\documentclass[prd,preprintnumbers,showkeys,eqsecnum,showpacs]{revtex4}
\usepackage{bm} 
\usepackage{graphicx}
\usepackage{hyperref}
\newcommand{\be}{\begin{eqnarray}}
\newcommand{\ee}{\end{eqnarray}}
\newcommand{\beq}{\begin{equation}}
\newcommand{\eeq}{\end{equation}}
\begin{document}
\title{Chiral dynamics and peripheral transverse densities}
\author{C.~Granados}
\affiliation{Theory Center, Jefferson Lab, Newport News, VA 23606, USA}
\author{C.~Weiss}
\affiliation{Theory Center, Jefferson Lab, Newport News, VA 23606, USA}
\begin{abstract}
In the partonic (or light--front) description of relativistic systems the 
electromagnetic form factors are expressed in terms of frame--independent 
charge and magnetization densities in transverse space. This formulation 
allows one to identify the chiral components of nucleon structure as the 
peripheral densities at transverse distances $b = O(M_\pi^{-1})$ and 
compute them in a parametrically controlled manner. A dispersion relation 
connects the large--distance behavior of the transverse charge and 
magnetization densities to the spectral functions of the Dirac and Pauli 
form factors near the two--pion threshold at timelike $t = 4 M_\pi^2$, 
which can be computed in relativistic chiral effective field theory. 
Using the leading--order approximation we 
(a)~derive the asymptotic behavior (Yukawa tail) of the isovector
transverse densities in the ``chiral'' region $b = O(M_\pi^{-1})$ and the
``molecular'' region $b = O(M_N^2/M_\pi^3)$; (b)~perform the 
heavy--baryon expansion of the transverse densities; (c)~explain
the relative magnitude of the peripheral charge and magnetization densities 
in a simple mechanical picture; (d)~include $\Delta$ isobar intermediate 
states and study the peripheral transverse densities in the large--$N_c$ 
limit of QCD; (e)~quantify the region of transverse distances where the 
chiral components of the densities are numerically dominant; 
(f)~calculate the chiral divergences of the $b^2$--weighted moments of the 
isovector transverse densities (charge and anomalous magnetic radii) 
in the limit $M_\pi \rightarrow 0$ and determine their spatial support. 
Our approach provides a concise formulation of the spatial structure of 
the nucleon's chiral component and offers new insights into basic 
properties of the chiral expansion. It relates the information extracted 
from low--$t$ elastic form factors to the generalized parton distributions 
probed in peripheral high--energy scattering processes.
\end{abstract}
\preprint{JLAB-THY-13-1763}
\maketitle
\newpage
\tableofcontents
\newpage
\section{Introduction}
\label{sec:introduction}
Understanding the spatial structure of hadrons and their interactions
is one of the main objectives of modern strong interaction physics.
A space--time picture is needed not only to gain a more intuitive 
understanding of hadrons as extended systems, but also to enable the 
formulation of approximation methods taking advantage of the relevant 
distance scales. For non--relativistic quantum systems such as atoms 
in electrodynamics or nuclei in conventional many--body theory a 
space--time picture follows naturally from the Schr\"odinger wave 
function, resulting in a rich intuition based on concepts like the 
spatial size of configurations and the orbital motion of the constituents. 
For essentially relativistic systems such as hadrons the space--time picture 
is more complex, as the particle number can change due to vacuum 
fluctuations, the notion of wave function is reference frame--dependent, 
and constraints like crossing invariance and analyticity need to 
be satisfied. 

The light--front description of relativistic systems provides a framework 
in which it is possible to formulate a rigorous space--time picture.
One way to arrive at this description is to consider the system in
a frame where it moves which a large momentum and decouples from the
vacuum fluctuations (``infinite--momentum frame'') 
\cite{Weinberg:1966jm,Kogut:1969xa,Bjorken:1970ah,Gribov:1973jg}. 
Another, equivalent 
way is to view the system at fixed light--front time, which can be
done in any frame (``light--front quantization'') 
\cite{Dirac:1949cp,Leutwyler:1977vy,Lepage:1980fj}; see 
Ref.~\cite{Brodsky:1997de} for a review. Either way one obtains 
a closed quantum--mechanical system that can be described by a wave 
function, consisting of a coherent superposition of components with 
definite particle number, with simple transformation properties 
under Lorentz boosts. Most observables of interest, such as the matrix 
elements of current operators, can be expressed as overlap integrals 
of the wave functions in the initial and final state. The resulting 
space--time picture is frame--independent and preserves much of the 
intuition of non--relativistic physics. It is important to realize that
the light--front formulation of relativistic dynamics can be used not 
only when describing hadron 
structure in terms of the fundamental theory of QCD (where it matches 
with the conventional parton model), but also in effective theories 
based on hadronic degrees of freedom. The space--time picture available 
in this formulation can greatly help to elucidate the physical 
basis of the approximations made in such effective theories 
and quantify the limits of their applicability.

The most basic information about the spatial structure of the nucleon
comes from the transition matrix elements of conserved currents
(vector, axial vector) between nucleon states. They are parametrized 
by form factors depending on the invariant four--momentum transfer, $t$. 
In the light--front picture of nucleon structure, the Fourier transform of
the form factors describe the spatial distributions of charge and 
magnetization in the transverse 
plane \cite{Soper:1976jc,Burkardt:2000za,Burkardt:2002hr,Miller:2007uy};
see Ref.~\cite{Miller:2010nz} for a review. They represent the cumulative 
4--vector current seen by an observer at a transverse distance 
(or impact parameter) $b$ from the center of momentum (``line--of--sight 
densities'') and have an objective physical meaning. They are true spatial 
densities in the light--front wave functions of the system and, thanks to 
the frame independence of the latter, can be unambiguously related to 
other nucleon observables of interest. In particular, in the context of 
QCD the transverse densities correspond to a reduction of the generalized
parton distribution (or GPDs), 
which describe the transverse spatial distributions
of quarks, antiquarks and gluons in the nucleon
\cite{Burkardt:2000za,Burkardt:2002hr,Diehl:2002he}. The transverse charge 
and magnetization densities thus represent an essential tool for exploring 
the spatial structure of the nucleon as a relativistic system. 
Empirical densities have been obtained by Fourier--transforming the
elastic form factor data 
\cite{Miller:2007uy,Carlson:2007xd,Vanderhaeghen:2010nd,Venkat:2010by} 
and can be interpreted in terms of partonic structure of the nucleon 
or compared with dynamical model calculations; 
see Ref.~\cite{Miller:2010nz} for a review.

At large distances the behavior of strong interactions is governed by 
the spontaneous breaking of chiral symmetry. The associated Goldstone 
bosons -- the pions -- are almost massless on the hadronic scale, 
couple weakly to hadronic matter in the long--wavelength limit, 
and act as the longest--range carriers of the strong force. The resulting 
effective dynamics manifests itself in numerous distinctive phenomena
in low--energy $\pi\pi, \, \pi N$ and $NN$ interactions, as well as 
electromagnetic and weak processes. It can be studied systematically 
using methods of chiral effective field theory (chiral EFT, or 
chiral perturbation theory), in which one separates the dynamics 
at distances of the order $M_\pi^{-1}$ from that at typical 
hadronic distances, as represented e.g.\ by the inverse vector 
meson mass $M_V^{-1}$ \cite{Gasser:1983yg,Gasser:1984gg,%
Weinberg:1990rz,Weinberg:1991um}; see Ref.~\cite{Bernard:1995dp} 
for a review. A natural question is what this ``chiral dynamics'' implies 
for the transverse densities in the nucleon at distances of the order 
$b = O(M_\pi^{-1})$. This question has several interesting aspects, 
both methodological and practical, which make it a central problem 
of nucleon structure physics.

On the methodological side, the light--front formulation allows us to study 
how chiral dynamics plays out in the space--time picture appropriate for 
relativistic systems. It is important to note that in typical chiral 
processes the pion momenta are of the order of the pion mass, 
$k = O(M_\pi)$, i.e.\ the pion velocity is $v = O(1)$, so that chiral 
pions represent an essentially relativistic system. Methods from 
non-relativistic physics, such as the Breit frame density representation
of form factors, are not adequate for describing the spatial structure
of this system. In the light--front formulation the transverse distance 
$b$ has an objective physical meaning and acts as a new parameter justifying 
the chiral expansion. The peripheral transverse densities at 
$b = O(M_\pi^{-1})$ represent clean chiral observables free of 
short--distance contributions. They exhibit ``Yukawa tails'' similar 
to the classic results from non--relativistic $NN$ interactions, but
their interpretation is not restricted to the non--relativistic limit.
Generally, the possibility to study well--defined spatial densities 
rather than integral quantities (charge radii, magnetic moments and radii,
etc.) provides many new insights into basic properties of the chiral
expansion. For example, it allows us to study the spatial support of the 
chiral divergences of the charge and magnetic radii and provides 
a new perspective on the convergence of the heavy--baryon expansion 
for nucleon form factors.

The spatial view enabled by the transverse densities also sheds
new light on the role of the intrinsic non--chiral hadronic size
in chiral processes. The EFT describes the dynamics of 
the pion field at momenta $O(M_\pi)$ by an effective Lagrangian,
in which the non--chiral degrees of freedom --- e.g.\ the bare nucleon 
in processes with baryons \cite{Gasser:1987rb,Bernard:1992qa} --- 
are introduced as pointlike sources.
Their finite physical size is encoded in the pattern of higher--order 
coupling constants and counter terms appearing in loop calculations 
\cite{Ecker:1988te}. While efficiently implementing the separation 
of scales, this formulation does not convey an immediate sense of 
the spatial size of the hadrons 
involved in chiral processes. The spatial representation in the
light--front formulation clearly reveals the non-chiral size of the 
participating hadrons. This allows one to quantify the size of chiral 
and non--chiral contributions to nucleon observables and connect the 
couplings of the chiral Lagrangian with other measures of the hadron size. 

On the practical side, the chiral periphery of the transverse densities
represents an element of nucleon structure that can be computed from
first principles and included in a comprehensive parametrization.
The chiral periphery influences the behavior of the form factors
at very low spacelike momentum transfers 
$|t| \lesssim 10^{-2} \, \textrm{GeV}^2$ (see Ref.~\cite{Strikman:2010pu} 
for a preliminary assessment). It affects extrapolation 
of the form factor data to $t = 0$ and comparison with the 
charge radii measured in atomic physics experiments, and could 
possibly be studied in dedicated experiments.
Another interesting aspect is the connection of the transverse charge
and magnetization densities with the peripheral nucleon GPDs. The latter
could be probed in peripheral hard high--energy processes which directly
resolve the quark/gluon content of the nucleon's chiral periphery
\cite{Strikman:2003gz,Strikman:2009bd}.

In this article we perform a comprehensive study of the peripheral 
transverse charge and magnetization densities in the nucleon using methods 
of dispersion analysis and chiral EFT. We establish the parametric
regimes in the transverse distance, develop a practical method for 
calculating the peripheral densities, compute the chiral components 
of the charge and magnetization densities using leading--order chiral EFT, 
discuss their formal properties within the chiral expansion (heavy--baryon 
expansion, parametric order of charge and magnetization density, chiral 
divergences of moments), include $\Delta$ isobar intermediate states 
and explore the peripheral densities in the large--$N_c$ limit of QCD, 
and quantify the spatial region where the chiral component is 
numerically relevant. 

The main tool used in our study is a dispersion representation of the
transverse charge and magnetization densities, which expresses them as 
dispersion integrals of the imaginary parts (or spectral functions) of 
the Dirac and Pauli form factors in the timelike region $t > 0$. 
The large--distance behavior of the isovector densities is governed by 
the spectral functions near the threshold at $t = 4 M_\pi^2$, and the 
chiral expansion of the densities can be obtained directly from that of 
the spectral functions in this region \cite{Gasser:1987rb,%
Bernard:1996cc,Becher:1999he,Kubis:2000zd,Kaiser:2003qp}.
The dispersion representation of transverse densities offers many 
practical advantages. The dispersion integral for the 
densities converges exponentially at large $t > 0$ and effectively 
extends over masses in a range $\sqrt t - 2 M_\pi = O(b^{-1})$, such 
that the transverse distance $b$ acts as the physical parameter 
justifying the chiral expansion. The dispersion representation
allows one to compute the peripheral transverse densities using 
well--established methods of Lorentz--invariant relativistic chiral EFT,
even though the quantities computed have a partonic interpretation.
It greatly simplifies the chiral EFT calculations, as only the spectral 
functions need to be computed using $t$--channel cutting rules. 
The dispersion representation also allows one to combine chiral and 
non--chiral contribution to the transverse densities in a consistent 
manner; the latter result from the higher--mass states in the spectral 
function, particularly the $\rho$ meson resonance, and can be 
modeled phenomenologically. Using the dispersion representation
we study several aspects of the peripheral transverse densities
in the nucleon:

(a) \textit{Large--distance behavior of transverse densities.}
We analyze the asymptotic behavior of the transverse densities at large
distances on general grounds. In the dispersion representation it is 
directly related to the behavior of the spectral functions of the form 
factors near the threshold at $t = 4 M_\pi^2$. It is well--known that
the spectral functions in this region are essentially influenced by 
a subthreshold singularity on the unphysical sheet, whose presence is 
required by the general analytic properties of the $\pi N$ scattering 
amplitude \cite{Frazer:1960zza,Frazer:1960zzb,Hohler:1976ax}. 
The distance of this singularity from threshold is 
$M_\pi^4/M_N^2$ and thus anomalously small on the chiral 
scale, $M_\pi^2$. It implies the existence of two parametric regimes 
of the transverse densities: regular ``chiral'' distances 
$b = O(M_\pi^{-1})$, and anomalously large ``molecular'' distances, 
$b = O(M_N^2/M_\pi^3)$. They exhibit different asymptotic
behavior and require dedicated approximation methods. The structure
of the peripheral densities is thus much richer than that of a single
``Yukawa tail.'' A similar phenomenon was observed in the two--pion 
exchange contribution to the low--energy $NN$ interaction in
nonrelativistic chiral EFT \cite{Robilotta:1996ji,Robilotta:2000py};
see Ref.~\cite{Epelbaum:2005pn} for a review.

(b) \textit{Heavy--baryon expansion of transverse densities.}
We derive the heavy--baryon expansion (i.e., the power expansion 
in $M_\pi / M_N$) of the transverse charge and magnetization densities in 
the chiral region $b = O(M_\pi^{-1})$ and study its practical usefulness. 
In our approach it is directly obtained from the heavy--baryon expansion 
of the spectral functions near threshold, which was studied in detail in 
Refs.~\cite{Bernard:1996cc,Becher:1999he,Kubis:2000zd,Kaiser:2003qp}. 
The subthreshold singularity in the spectral 
functions limits the convergence of the heavy--baryon expansion. Even so, 
a very satisfactory heavy--baryon expansion of the peripheral
charge and magnetization densities is obtained, which can be used for 
numerical evaluation at all practically relevant distances. 

(c) \textit{Charge vs.\ magnetization density.}
We compare the transverse charge and magnetization densities in the 
nucleon's chiral periphery at $b = O(M_\pi^{-1})$. It is shown 
that the spin--independent and spin--dependent components of the
4--vector current matrix element, which are directly related to the charge 
and magnetization densities \cite{Miller:2010nz}, are of the same order 
in the parameter $M_N/M_\pi$. Moreover, the absolute value of the 
spin--dependent current density is found to be bounded by the
spin--independent density. Both observations can naturally be explained 
in an intuitive ``mechanical'' picture of the chiral $\pi N$ component of the
nucleon's light--cone wave function producing the peripheral densities.
It shows how the particle--based light--front formulation can illustrate 
basic properties of chiral dynamics that are not obvious in the 
general field--theoretical formulation. A detailed exposition of the
mechanical picture will be given in a forthcoming article, where we 
study the time evolution of chiral processes and express the peripheral 
charge and magnetization densities as overlap integrals of the light--front 
wave functions of the chiral $\pi N$ system \cite{inprep}.

(d) \textit{Intermediate $\Delta$ isobars and large--$N_c$ limit of QCD.}
We calculate the effect of $\Delta$ isobar intermediate states
on the nucleon's transverse densities at large distances. 
Intermediate $\Delta$ states pose a challenge for the traditional 
chiral expansion of integral quantities, as the $N\Delta$ mass difference 
represents a non--chiral scale that is numerically not far from the 
physical pion mass. In our coordinate--space approach we focus on the 
two--pion contribution to the densities at distances $b = O(M_\pi^{-1})$ 
and can include the $\Delta$ in a natural manner, as a modification
(new singularity) of the $\pi N$ scattering amplitude describing the 
coupling of the two--pion $t$--channel state to the nucleon. In this way 
we study the interplay of $N$ and $\Delta$ states in the transverse
densities at fixed $b = O(M_\pi^{-1})$, with the $N\Delta$ mass splitting
an unrelated external parameter. Inclusion of the 
$\Delta$ is important for practical reasons, as the $\pi N \Delta$ 
coupling is large and results in substantial contribution to the density 
at intermediate distances $b \sim 1-2 \, \textrm{fm}$. It is even more 
important theoretically, to ensure the proper scaling behavior of the 
transverse densities in the large--$N_c$ limit of QCD
\cite{Witten:1979kh,Dashen:1993jt,Dashen:1994qi}.
We show that in large--$N_c$ limit the $N$ and $\Delta$ 
contributions to the isovector charge density at $b = O(M_\pi^{-1})$ 
cancel each other in leading order of the $1/N_c$ expansion, bringing about
the correct $N_c$--scaling required by QCD. In the isovector magnetization
density the $N$ and $\Delta$ contributions add and give a large--$N_c$ 
value that is $3/2$ times the density from intermediate $N$ alone,
as expected on general grounds; see Ref.~\cite{Cohen:1996zz} for
a review. These results show that the two--pion
components of the transverse densities obtained in our approach 
obey the general $N_c$--scaling laws and can be regarded as legitimate
approximations to peripheral nucleon structure in large--$N_c$ QCD.

(e) \textit{Region of dominance of chiral component.} We quantify the
region of transverse distances where the chiral component of the nucleon
densities becomes numerically dominant. The spatial view of the nucleon,
combined with the dispersion representation of the transverse densities,
provides a framework that allows us to address this question in a 
transparent and physically motivated manner. Non--chiral contributions
to the transverse densities arise from higher--masss states in the spectral 
functions, particularly the vector meson states, and can be added to the 
chiral near--threshold contribution without double counting. 
Using a simple parametrization of the higher--mass states in terms of 
vector meson poles we show that the chiral component of the isovector 
transverse densities becomes numerically dominant only at surprisingly
large distances $b \gtrsim 2\, \textrm{fm}$. 
More generally, our coordinate--space approach provides a novel way of 
identifying the chiral component of nucleon structure, for the purpose
of either theoretical calculations or experimental probes.

(f) \textit{Chiral divergences of moments.} The $b^2$--weighted integrals
(moments) of the transverse charge and magnetization densities are 
proportional to the derivatives of the Dirac and Pauli form factors at 
$t = 0$ and represent the analog of the traditional charge and magnetic 
radii in the 2--dimensional partonic picture of spatial nucleon structure. 
These quantities exhibit chiral divergences in the limit 
$M_\pi \rightarrow 0$. We verify that the moments of our peripheral
densities at $b = O(M_\pi^{-1})$ reproduce the well--known universal
chiral divergences of the nucleon's charge and magnetic 
radii \cite{Gasser:1987rb}.
This also allows us to determine the spatial support of the chiral
divergences. It is seen that the chiral logarithm of the transverse 
charge radius results from the integral over a broad range of 
distances $b_0 \ll b \ll 1/M_\pi$ ($b_0$ represents a short--distance
cutoff), while the power--like divergence of the magnetic radius 
comes from distances $b \sim M_\pi^{-1}$. These findings connect our
approach with the usual chiral EFT studies of the pion mass dependence
of integral quantities and illustrate its spatial structure.

The plan of this paper is as follows. In Sec.~\ref{sec:densities}
we summarize the basic properties of the transverse densities associated
with the nucleon's electromagnetic form factors and discuss their 
space--time interpretation, in particular the relation between the
magnetization density and the physical spin--dependent current density.
We then describe the dispersion representation of the transverse densities 
and its usage, discuss the behavior of the spectral functions near 
threshold based on general principles, and introduce the parametric
regions of transverse distances. In Sec.~\ref{sec:chiral}
we calculate the chiral component of the transverse densities and 
perform a detailed analysis of its properties. We summarize the
chiral Lagrangian and the basics of the dispersive approach to chiral 
EFT and present a $t$--channel cutting rule that permits efficient 
calculation of the spectral functions from the chiral EFT Feynman 
diagrams (Appendix~\ref{app:cutting}). We study the spectral functions 
near threshold and numerically evaluate the transverse densities. 
We derive the heavy--baryon expansion of the densities in the chiral 
region, $b = O(M_\pi^{-1})$, and study its convergence numerically.
Explicit analytic expressions for the densities are obtained and 
evaluated in terms of special functions (Appendix~\ref{app:heavy}).
We also derive the asymptotic behavior of the density in the molecular 
region, $b = O(M_N^2/M_\pi^3)$, and give explicit formulas.  
We then compare the relative magnitude of the charge and magnetization
densities in the nucleon's periphery and explain it in a simple mechanical 
picture. Finally, we discuss the physical significance of the contact terms
appearing in the chiral EFT calculation, and their relation to the
form of the $\pi NN$ vertex in the chiral Lagrangian (axial vector
vs.\ pseudoscalar coupling). In Sec.~\ref{sec:delta} we
calculate the peripheral densities arising from $\Delta$ intermediate states
and evaluate them numerically. We then discuss the general large--$N_c$
scaling behavior of the transverse densities in QCD, and show that
the two--pion component of the peripheral densities, including both $N$
and $\Delta$ intermediate states, obeys the general large--$N_c$
scaling laws. In Sec.~\ref{sec:region} we quantify the region of 
transverse distances where the chiral component of the charge and 
magnetization densities becomes numerically dominant. Using a simple 
parametrization of higher--mass states in the spectral functions in terms 
of vector meson poles, we compare the chiral and non--chiral contributions 
to the transverse densities at different distances $b$. 
In Sec.~\ref{sec:moments} we study the chiral divergences of the 
$b^2$--weighted moments of the transverse densities. We show that our 
results for the peripheral densities reproduce the universal
chiral divergences of the nucleon's charge and magnetic radii
(i.e., the slope of the Dirac and Pauli form factors) and discuss the 
spatial support of the chiral divergences in our picture. 
A summary of our main conclusions and an outlook on further studies 
are presented in Sec.~\ref{sec:summary}.

An overview of the properties of the peripheral transverse charge density 
and their phenomenological implications was given already in 
Ref.~\cite{Strikman:2010pu}. In the present article we offer a detailed
exposition of the theoretical framework, extend the calculations to 
the Pauli form factor and the magnetization density, and explore several 
new aspects of the chiral component of transverse densities
(heavy--baryon expansion, mechanical interpretation, 
spatial support of chiral divergences).

In this paper we study the chiral component of the transverse charge and 
magnetization densities using the established Lorentz--invariant formulation 
of chiral EFT, taking advantage of the analytic properties of the form
factors. The partonic or light--front picture will be invoked only
for the interpretation of the densities, not as a framework for
actual calculations, and readers not familiar with these aspects
should be able to follow the presentation. It is, of course, possible to
calculate the chiral component of the densities directly in a
partonic picture, using the infinite--momentum frame or light--front
time--ordered perturbation theory. In this formulation the densities 
are expressed as overlap integrals of the peripheral $\pi N$ light--cone 
wave functions of the physical nucleon, which are calculable directly 
from the chiral Lagrangian. This formulation will reveal several new 
aspects, such as the role of orbital angular momentum in chiral counting,
the longitudinal structure of the configurations contributing to the
densities at given $b$, and the connection with chiral contributions 
to the nucleon's parton densities and high--energy scattering processes. 
We shall explore this formulation in a following article and
address all pertinent questions there \cite{inprep}.

In the present study we use chiral EFT in the leading--order approximation
to evaluate the transverse densities in the chiral region. The leading--order
densities do not depend on an explicit short--distance cutoff, involve only
a few basic parameters, and have a transparent physical structure. 
Our intention 
here is to discuss the properties of the peripheral densities at this level 
and compare them to the non--chiral densities generated by higher--mass states
in the spectral function. We comment on the places where higher--order 
effects are seen to be explicitly important; e.g., in the magnetization
density in the molecular region. We emphasize that the basic framework 
presented here (space--time picture, dispersion representation) is by no 
means limited to the leading--order approximation and could be explored 
in higher--order calculations as well. Higher--order calculations of the 
spectral functions of the nucleon form factors have been performed in
relativistic \cite{Kubis:2000zd} and heavy--baryon chiral EFT 
\cite{Bernard:1996cc,Kaiser:2003qp} and could be adapted for our purposes.
This extension, however, requires new physical considerations regarding 
the regularization of chiral loops in coordinate space and will be left 
to a future study.
\section{Transverse charge and magnetization densities}
\label{sec:densities}
\subsection{Definition and interpretation}
\label{subsec:definition}
The transition matrix element of the electromagnetic current
between nucleon (proton, neutron) states with three--momenta 
$\bm{p}_1$ and $\bm{p}_2$ and spin quantum numbers 
$\sigma_1$ and $\sigma_2$ can be parametrized as
\beq
\langle \bm{p}_2, \sigma_2 | J^\mu (x) | \bm{p}_1, \sigma_1 \rangle\
\;\; = \;\; \bar u_2 \left[ \gamma^\mu F_1(t) - 
\frac{\sigma^{\mu\nu} \Delta_\nu}{2 M_N} F_2(t) \right] u_1 \, 
e^{i\Delta x} ,
\label{me_general}
\eeq
where the nucleon momentum states are normalized according to the relativistic
convention, $\langle \bm{p}_2 | \bm{p}_1 \rangle
= 2 (p_1)^0 \delta^{(3)}(\bm{p}_2 - \bm{p}_1)$.
Here $u_1 \equiv u(\bm{p}_1, \sigma_1)$ and $u_2 \equiv u(\bm{p}_1, \sigma_1)$
are the nucleon bispinors, normalized to $\bar u_1 u_1 = \bar u_2 u_2 = 2 M_N$,
and $\sigma^{\mu\nu} \equiv \frac{1}{2}\left[ \gamma^\mu, \gamma^\nu\right]$.
The 4--momentum transfer is denoted by
\beq
\Delta \;\; \equiv \;\; p_2 - p_1 ,
\eeq
and the dependence of the matrix element on the space--time point $x$
follows from translational invariance. The functions $F_1$ and $F_2$ are 
known as the Dirac and Pauli form factors and depend on the invariant
momentum transfer,
\beq
t \;\; \equiv \;\; \Delta^2 ,
\eeq
with $t < 0$ (spacelike momentum transfer) in the physical region for 
electromagnetic scattering.
Equation~(\ref{me_general}) applies to either proton or neutron states.
The value of the Dirac form factor at zero momentum transfer is given
by the total charge of the nucleon,
\beq
F_1^p(0) \;\; = \;\; 1, \hspace{2em} F_1^n(0) \;\; = \;\; 0; 
\eeq
the value of the Pauli form factor by the anomalous magnetic moment,
\beq
F_2^p(0) \;\; = \;\; \kappa_p, \hspace{2em} F_2^n(0) \;\; = \;\; \kappa_n, 
\eeq
whose empirical values are $\kappa_p = 1.79$ and $\kappa_n = -1.91$.
Experimental knowledge of the nucleon form factors at finite $t < 0$ 
is reviewed in Ref.~\cite{Perdrisat:2006hj}; for a discussion of the most 
recent data see Refs.~\cite{Cates:2011pz,Bernauer:2013tpr}
and references therein. For theoretical analysis 
it is convenient to consider the isoscalar and isovector combinations 
of form factors \footnote{In Ref.~\cite{Strikman:2010pu} we 
considered the difference of proton and neutron form factors without 
a factor $1/2$. In the present article we follow the standard convention 
for the isoscalar and isovector form factors with the factor $1/2$.}
\be
F_1^{S, V}(t) &\equiv& {\textstyle\frac{1}{2}} [F_1^p(t) \pm F_1^n(t)], 
\hspace{2em} \textrm{etc.}
\label{F_V_S}
\ee
which are normalized such that 
\beq
F_1^{S, V}(0) \; = \; 1/2 ,
\hspace{2em}
F_2^{S, V}(0) \; = \; {\textstyle\frac{1}{2}} (\kappa_p \pm \kappa_n) .
\label{S_V_def}
\eeq

The form factors are Lorentz--invariant functions and can be analyzed
independently of any reference frame. Their space--time interpretation,
however, requires choosing a specific reference frame. In the context
of the light--front or partonic description of nucleon structure it
is natural to represent the form factors as the Fourier transform 
of certain 2--dimensional spatial densities. Choosing a frame such
that the spacelike momentum transfer lies is in the $xy$ (or transverse)
plane,
\beq
\Delta^\mu \;\; \equiv \;\; (\Delta^0, \Delta^x, \Delta^y, \Delta^z) 
\;\; = \;\; (0, \bm{\Delta}_T, 0), 
\hspace{2em} 
\bm{\Delta}_T \; = \; (\Delta^x, \Delta^y), 
\hspace{2em} t \; = \; -\bm{\Delta}_T^2
\eeq
and defining a conjugate coordinate variable 
as \footnote{Because the vector $\bm{b}$ is defined only in transverse space
and does not appear as the transverse component of a 4--vector, we omit
the usual label $T$ denoting transverse vectors.}
\beq
\bm{b} \;\; \equiv \;\; (b^x, b^y)
\eeq
one writes \cite{Miller:2007uy,Miller:2010nz}
\beq
F_{1, 2}(t = -\bm{\Delta}_T^2) \;\; = \;\; \int d^2 b \; 
e^{i \bm{\Delta}_T \bm{b}} \; \rho_{1, 2} (b) .
\label{rho_def}
\eeq
The functions $\rho_{1, 2}(b)$ are called the transverse charge and 
anomalous magnetization density (or simply ``magnetization density,'' for
short); their precise physical meaning will be elaborated in the following.
Their names refer to the obvious property that the 
spatial integral of the densities, i.e., the Fourier integral 
Eq.~(\ref{rho_def}) at $\bm{\Delta}_T = 0$, returns the form factors 
at $t = 0$, and thus the total charge and anomalous magnetic moment 
of the nucleon,
\be
\int d^2 b \; \rho_1^{S, V} (b) &=& \textstyle{\frac{1}{2}}, 
\\[1ex]
\int d^2 b \; \rho_2^{S, V} (b) &=& \textstyle{\frac{1}{2}} 
(\kappa_p \pm \kappa_n) , 
\ee
Because of rotational invariance in the transverse plane, 
the densities are functions only of the modulus $b \equiv |\bm{b}|$.
The transverse densities can be obtained from the form factor as 
\be
\rho_{1, 2} (b) &=& \int \frac{d^2\Delta}{(2\pi)^2}
\; e^{-i \bm{\Delta}_T \bm{b}}  \; F_{1, 2}(t = -\bm{\Delta}_T^2) 
\label{rho_fourier}
\\
&=& \int\limits_0^\infty \frac{d\Delta_T}{2\pi} 
\; \Delta_T \; J_0 (\Delta_T b) \; 
F_{1, 2} (t = -\Delta_T^2 ) ,
\label{rho_fourier_radial}
\ee
where $\Delta_T \equiv |\bm{\Delta}_T|$. In the last step we have performed 
the integral over the angle between the transverse vectors, and $J_0$ 
denotes the Bessel function.

%
% FIGURE
%
\begin{figure}
\begin{tabular}{ll}
\parbox[c]{.28\textwidth}{
\includegraphics[width=.25\textwidth]{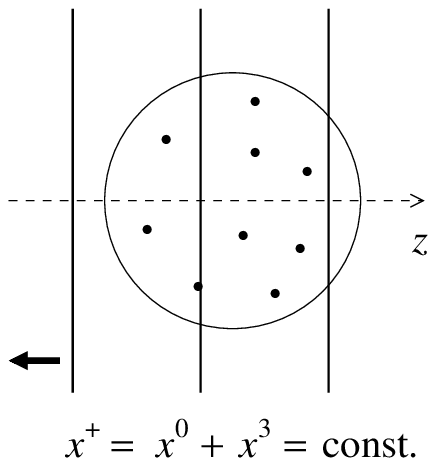}}
\hspace{.1\textwidth}
&
\parbox[c]{.26\textwidth}{
\includegraphics[width=.23\textwidth]{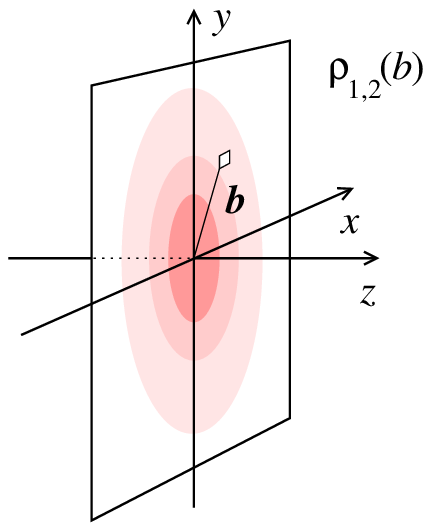}}
\\[-1ex]
(a) & (b)
\end{tabular}
\caption[]{(a) Light--front view of a relativistic system.
(b) Transverse densities in the nucleon. The function
$\rho_1(b)$ describes the spin--independent part of the
expectation value of the $J^+$ current in a nucleon state
localized at the transverse origin, Eq.~(\ref{J_plus_spin_indep});
the function $(2 M_N)^{-1} \, (b^x/b) \, \partial\rho_2(b)/\partial b$ the  
spin--dependent part in a nucleon polarized in the 
positive $y$--direction, Eq.~(\ref{J_plus_spin_dep}).}
\label{fig:interpretation}
\end{figure}
The physical interpretation of the 2--dimensional densities refers
to the light--front or partonic picture of nucleon structure and 
has been extensively discussed in the literature \cite{Soper:1976jc,%
Burkardt:2000za,Burkardt:2002hr,Miller:2007uy,Carlson:2007xd,Miller:2010nz}; 
here we only
summarize the main points. In the light--front picture one considers
the evolution of a relativistic system in light--front time 
$x^+ \equiv x^0 + x^3 = x^0 + z$, as corresponds to clocks synchronized by a 
light--wave traveling through the system in the $z$--direction
(see Fig.~\ref{fig:interpretation}a). 
Particle states such as the nucleon are characterized by their 
light--cone momentum $p^+ \equiv p^0 + p^z$ and transverse momentum
$\bm{p}_T \equiv (p^x, p^y)$, and $p^- \equiv p^0 - p^z$
plays the role of the energy, with $p^- = (p_T^2 + M_N^2)/p^+$.
One is generally interested in the ``plus'' component of the nucleon
current, which possesses a simple interpretation in dynamical models.
In a frame where the momentum transfer to the nucleon is in the 
transverse direction, 
\beq
\Delta^\pm \; = \; 0, \hspace{2em}
\bm{\Delta}_T \; = \; \bm{p}_{2T} - \bm{p}_{1T} \; \neq \; 0,
\eeq
the matrix element Eq.~(\ref{me_general}) takes the form
\be
\langle p^+, \bm{p}_{T2}, \lambda_2 | \, J^+ (x) \, | p^+, \bm{p}_{T1}, 
\lambda_1 \rangle\
&=& \bar u_2 \left[ \gamma^+ F_1(t) \; + \; 
\frac{\sigma^{+i} \Delta_T^i}{2 M_N} F_2(t) \right] u_1 \, 
e^{-i\bm{\Delta}_T \bm{x}_T} 
\nonumber \\[1ex]
&& (t \; \equiv \; - \bm{\Delta}_T^2) ,
\label{me_lf}
\ee
where now the momentum states are normalized as
$\langle p_2^+, \bm{p}_{T2} | p_1^+ , \bm{p}_{T1} \rangle
= 2 p_1^+ \, \delta(p_2^+ - p_1^+) 
\delta^{(2)}(\bm{p}_{T2} - \bm{p}_{T1})$. The polarization states of the 
initial and final nucleon can be defined in several ways and are usually
chosen as helicity eigenstates, with $\lambda_{1, 2} = \pm$ denoting
the helicities. An explicit representation of the
corresponding 4--spinors can be obtained by applying a Lorentz boost
to rest--frame spinors polarized in the $z$--direction, and is given 
by \cite{Leutwyler:1977vy}
\beq
u_1 \;\; \equiv \;\; u(p^+, \bm{p}_{1T}, \lambda_1 ) \;\; = \;\; 
\frac{\sqrt{2}}{\sqrt{p^+}} \left( p^+ + \gamma^0 M_N + \gamma^0 \bm{\gamma}_T
\bm{p}_{1T} \right) \frac{\gamma^-\gamma^+}{4}
\left( \begin{array}{c} \chi (\lambda_1 ) \\[1ex] 0 \end{array} \right) ,
\label{spinor_boosted}
\eeq
and similarly for $u_2$. Here $\chi (\lambda = \pm)$ are rest 
frame 2--spinors for polarization in the positive and negative 
$z$--direction. The transition matrix element then falls into two
structures, a ``spin--independent'' one proportional to
\beq
\delta ({\lambda_2 , \lambda_1}) \;\; = \;\; 
\chi^\dagger (\lambda_2 ) \chi (\lambda_1) ,
\eeq
which contains the Dirac form factor, and a ``spin--dependent'' one 
proportional to the vector
\beq
\bm{S} ({\lambda_2 , \lambda_1}) \;\; \equiv \;\; \chi^\dagger ({\lambda_2})
({\textstyle\frac{1}{2}} \bm{\sigma}) \chi ({\lambda_1}) ,
\label{S_vector}
\eeq
which contains the Pauli form factor.

To describe the transverse spatial structure of the nucleon one defines 
nucleon states in the transverse coordinate representation, corresponding 
to nucleons with a transverse center--of--momentum localized at given 
points $\bm{x}_{1T}$ and $\bm{x}_{2T}$, 
as \footnote{The proper mathematical definition
of the transversely localized nucleon states uses wave packets of
finite width and takes the limit of zero width at the end of
the calculation. The simplified derivation presented here,
using states ``normalized to a delta function,'' is legitimate
as long as one keeps $\bm{x}_{T2} \neq \bm{x}_{T1}$ until the 
end of the calculation.}
\be
| \bm{x}_{1T} \rangle &\equiv& \int \frac{d^2 p_{1T}}{(2\pi)^2} \;
e^{-i \bm{p}_{1T}\bm{x}_{1T}}\; | \bm{p}_{1T} \rangle ,
\\
\langle \bm{x}_{2T} | &\equiv& \int \frac{d^2 p_{2T}}{(2\pi)^2} \;
e^{i \bm{p}_{2T}\bm{x}_{2T}}\; \langle \bm{p}_{2T} | ,
\ee
which are normalized such that $\langle \bm{x}_{2T} | \bm{x}_{1T} \rangle
= \delta^{(2)}(\bm{x}_{2T} - \bm{x}_{1T})$. We now consider the matrix 
element of the current at light--front time $x^+ = 0$ and position 
$x^- = 0$, and a transverse position $\bm{x}_T$, between 
such transversely localized nucleon states with (arbitrary) 
longitudinal momentum $p^+$. Using Eqs.~(\ref{me_lf}) 
and (\ref{spinor_boosted}) it is 
straightforward to show that the spin--independent part of 
the matrix element of $J^+$ is given by
\be
\lefteqn{\langle p^+, \bm{x}_{2T}, \lambda_2 | 
\; J^+ (x^\pm = 0, \bm{x}_T) \; | 
p^+, \bm{x}_{1T}, \lambda_1 \rangle_{\text{\scriptsize spin-indep.}}} &&
\nonumber \\[2ex]
&=& [ 2p^+ \, \delta^{(2)}(\bm{x}_{2T} - \bm{x}_{1T}) ] \;
\delta (\lambda_2 , \lambda_1) \;
\int\frac{d^2\Delta}{(2\pi )^2} \; e^{-i \bm{\Delta}_T 
(\bm{x}_T - \bm{x}_{1T})}
\; F_1(-\bm{\Delta}_T^2)
\\[1ex]
&=& [ \ldots ] \;
\delta (\lambda_2 , \lambda_1) \;
\rho_1(\bm{x}_T - \bm{x}_{1T}) .
\label{me_imf_charge}
\ee
The factor in brackets results from the normalization of the nucleon states.
One sees that the function $\rho_1 (b)$ of Eq.~(\ref{rho_def}) describes 
the spin--independent part of the current in the nucleon, with 
\beq
\bm{b} \;\; \equiv \;\; \bm{x}_T - \bm{x}_{1T}
\label{b_from_x}
\eeq
defined as the displacement from the transverse center--of--momentum of 
the nucleon. In short, for a nucleon localized at the origin, 
$\bm{x}_{1T} = 0$, the spin--independent current at transverse 
position $\bm{x}_T = \bm{b}$ is (see Fig.~\ref{fig:interpretation}b)
\beq
\langle J^+ (\bm{b}) \rangle_{\text{\scriptsize spin-indep.}}
\;\; = \;\; \rho_1(b) .
\label{J_plus_spin_indep}
\eeq
Likewise, the spin--dependent part of the matrix element 
of $J^+$ is given by
\be
\lefteqn{\langle p^+, \bm{x}_{2T}, \lambda_2 | 
\; J^+ (x^\pm = 0, \bm{x}_T) \; | 
p^+, \bm{x}_{1T}, \lambda_1 \rangle_{\text{\scriptsize spin-dep.}}} &&
\nonumber \\[2ex]
&=& [ \ldots ] \; (-i)
\int\frac{d^2\Delta}{(2\pi )^2} \; e^{-i \bm{\Delta}_T 
(\bm{x}_T - \bm{x}_{1T})}
\; \frac{\bm{S}(\lambda_2, \lambda_1)}{M_N} 
\cdot (\bm{e}_z \times \bm{\Delta}_T) \;
F_2(-\bm{\Delta}_T^2) 
\\[1ex]
&=& [ \ldots ] \; \frac{\bm{S}(\lambda_2, \lambda_1)}{M_N} 
\cdot \left( \bm{e}_z \times 
\frac{\partial}{\partial \bm{x}_T} \right) 
\; \rho_2 (\bm{x}_T - \bm{x}_{1T}) ,
\label{me_imf_current}
\ee
where $\bm{S}(\lambda_2, \lambda_1)$ is the spin vector of the 
transition defined in Eq.~(\ref{S_vector}), and $\bm{e}_z$ the unit vector
in the $z$--direction. Thus, the ``crossed'' gradient of the
function $\rho_2 (b)$ of Eq.~(\ref{rho_def}) describes the spin--dependent 
current measured by an observer at a displacement $\bm{b}$ from the 
center--of--momentum of the nucleon. In Eq.~(\ref{me_imf_current})
the nucleon polarization states are characterized by the $z$--component
of the spin in the rest frame, $\lambda_{1, 2}$, 
cf.\ Eq.~(\ref{spinor_boosted}). If instead we prepared initial and final
nucleon state with definite spin in the $y$--direction and the same 
projection for both, the spin vector in 
Eq.~(\ref{me_imf_current}) would be replaced by
\beq
\bm{S}(\lambda_2, \lambda_1) \;\; \rightarrow \;\;
S^y \bm{e}_y 
\hspace{2em} \text{(nucleon polarized in $y$--direction)} ,
\eeq
where $S^y = \pm 1/2$ is the spin projection on the $y$--axis.
For a nucleon localized at the origin and polarized in the
$y$--direction, the spin--dependent current at a transverse position 
$\bm{b}$ is thus (see Fig.~\ref{fig:interpretation}b)
\be
\langle J^+ (\bm{b}) \rangle_{\text{\scriptsize spin-dep.}}
&=& (2S^y) \, 
\frac{\partial}{\partial b^x} 
\left[ \frac{\rho_2(b)}{2 M_N} \right]
\;\; = \;\; (2S^y) \, 
\frac{b^x}{b} \; 
\frac{\partial}{\partial b} 
\left[ \frac{\rho_2(b)}{2 M_N} \right]
\;\; = \;\; (2S^y) \, \cos\phi \; \widetilde\rho_2 (b) ,
\label{J_plus_spin_dep}
\ee
where $\cos\phi \equiv b^x/b$ is the cosine of the azimuthal angle and
\be
\widetilde\rho_2 (b) &\equiv& \frac{\partial}{\partial b} 
\left[ \frac{\rho_2(b)}{2 M_N} \right] .
\label{rho_2_tilde_def}
\ee
Now the term ``spin--dependent'' can be understood to mean that part 
of the current which changes sign when the transverse nucleon polarization 
is reversed. We shall refer to the function $\widetilde \rho_2$ as 
the ``spin--dependent current density,'' keeping in mind that the 
actual spin--dependent current matrix element involves also the 
polarization $(2 S^y)$ and the geometric factor $\cos\phi$.
Note that for a given spin orientation the spin--dependent current 
changes sign between positive (``right,'' when looking at the 
nucleon from $z = + \infty$) and negative (``left'')
values of $b^x$, as would be the case for a convection current
due to rotational motion around the $y$--axis. 
Finally, the total current in a nucleon polarized in the 
$y$--direction is then, in the same short--hand notation as used
above,
\be
\langle J^+ (\bm{b}) \rangle
&=&
\langle J^+ (\bm{b}) \rangle_{\text{\scriptsize spin-indep.}}
\;  + \; \langle J^+ (\bm{b}) \rangle_{\text{\scriptsize spin-dep.}}
\\[2ex]
&=& \rho_1 (b) \; + \; (2 S^y) \; \cos\phi \; \widetilde\rho_2 (b) .
\label{J_plus_total}
\ee
This expression, together with Eq.~(\ref{rho_2_tilde_def}),
concisely summarizes the physical significance of the transverse 
densities introduced as the 2--dimensional Fourier transforms 
of the invariant form factors, Eq.~(\ref{rho_def}). 
We shall use it to develop a simple mechanical interpretation of the
chiral component of the transverse densities below 
(see Sec.~\ref{subsec:charge_vs_current}).

The light--front interpretation of the nucleon
current matrix elements described here assumes only that the momentum
transfer to the nucleon is in the transverse direction, 
$\Delta^\pm = 0$ and $\bm{\Delta}_T \neq 0$,
but does not depend on the value of the nucleon's 
longitudinal momentum $p^+$. As such 
it is valid for any $p^+$, including the rest frame where $p^+ = M_N$. 
In Sec.~\ref{subsec:charge_vs_current} we shall use the rest frame
to obtain a simple interpretation of the relative order--of--magnitude 
of the chiral components of the charge and magnetization densities. 
Alternatively, one may consider the limit $p^+\rightarrow \infty$,
where the description sketched here coincides with the conventional
parton picture of nucleon structure (``infinite--momentum frame'').

In the present study we refer to the light--front representation 
of the transverse densities only for their interpretation; the actual 
calculations of the chiral component are carried out at the level 
the invariant form factors, without specifying a reference frame. 
For this purpose we may think of the transverse densities defined 
by Eq.~(\ref{rho_def}) as just a particular functional transform of the 
invariant form factors, i.e., an equivalent mathematical representation of the 
information contained in these functions. We shall return to the 
light--front picture only at the end, when interpreting the results 
of our calculation. The power of transverse densities is precisely 
that they connect the invariant form factors with the light--front 
picture of nucleon structure and can be accessed from both sides.

In dynamical models where the nucleon has a composite structure, the 
transverse densities Eq.~(\ref{rho_def}) can be represented as overlap 
integrals of the frame--independent light--cone wave functions of 
the system. With the momentum transfer chosen such that $\Delta^\pm = 0$
and $\bm{\Delta}_T \neq 0$ the current cannot produce particles but
simply ``counts'' the charge and current of the constituents in the 
various configuration of the wave functions. It is possible to compute 
the chiral component of transverse densities directly in this formulation,
using light--front time--ordered perturbation theory; this approach
will be explored in a subsequent article \cite{inprep}.
\subsection{Dispersion representation}
\label{subsec:dispersion}
Much insight into the behavior of the transverse densities can be gained 
by making use of the analytic properties of the nucleon form factors as
functions of the invariant momentum transfer. The form factors $F_{1, 2}(t)$
are analytic functions of $t$, with singularities (branch cuts, poles) on
the positive real axis. They correspond to processes in which a current
with timelike momentum converts to a hadronic state coupling to the nucleon, 
which may occur below the physical threshold for nucleon--antinucleon 
($N\bar N$) pair production. The principal cut in the physical sheet of the 
form factor starts at the squared mass of the lowest hadronic state, 
the two--pion state, $t = 4 M_\pi^2$, and runs to $t = + \infty$. 
Assuming that the form factors vanish at $|t| \rightarrow \infty$, 
as expected from the power behavior implied by perturbative QCD 
(with logarithmic modifications) and supported by present experimental 
data, the form factors satisfy an unsubtracted dispersion relation,
\beq
F_{1, 2} (t) \;\; = \;\; 
\int\limits_{4M_\pi^2}^\infty \frac{dt'}{t' - t} 
\; \frac{\textrm{Im}\, F_{1, 2}(t' + i0)}{\pi} .
\label{dispersion}
\eeq
It expresses the form factors as integrals over their imaginary parts 
on the principal cut, also known as the spectral functions. In the region 
below the $N\bar N$ threshold, $t' < 4 M_N^2$, which dominates the integral 
Eq.~(\ref{dispersion}) at all values of $t$ of interest,
the spectral function cannot be measured directly in conversion experiments
and can only be calculated using theoretical methods (dispersion theory,
chiral EFT) or determined empirically from fits to form factor 
data \cite{Hohler:1976ax,Belushkin:2006qa}. Even so, this representation 
of the form factor turns out to be extremely useful for the theoretical
analysis of transverse densities. Substituting Eq.~(\ref{dispersion}) in 
Eq.~(\ref{rho_fourier}) and carrying out the Fourier integral, one 
obtains a dispersion (or spectral) representation of the transverse
densities of the form \cite{Strikman:2010pu} 
\beq
\rho_{1, 2} (b) \;\; = \;\; \int\limits_{4M_\pi^2}^\infty \frac{dt}{2\pi} 
\; K_0(\sqrt{t} b) \; \frac{\textrm{Im}\, F_{1, 2} (t + i0)}{\pi} ,
\label{rho_dispersion}
\eeq
where $K_0$ denotes the modified Bessel function and we have dropped 
the prime on the integration variable $t$. This representation has several 
interesting mathematical properties. Because of the exponential decay of 
the modified Bessel function at large arguments,
\beq
K_0 (\sqrt{t} b) \;\; \sim \;\; \sqrt{\frac{\pi}{2}}
\; \frac{e^{-\sqrt{t} b}}{(\sqrt{t} b)^{1/2}}
\hspace{2em} (\sqrt{t} b \; \gg \; 1),
\label{K0_asymptotic}
\eeq
the dispersion integral for the density 
converges exponentially at large $t$, in contrast
to the power--like convergence of the original integral for the form 
factor, Eq.(\ref{dispersion}) \footnote{Use of a subtracted 
dispersion relation in Eq.~(\ref{rho_fourier}) would lead to an expression for 
$\rho_{1, 2}(b)$ which differs from Eq.~(\ref{rho_dispersion}) only by a term 
$\propto \delta^{(2)}(\bm{b})$. Subtractions therefore have no influence 
on the dispersion representation of the transverse density at finite $b$.
In this sense the dispersion representation Eq.~(\ref{rho_dispersion}) 
is similar to the Borel transform used to eliminate polynomial terms 
in QCD sum rules \cite{Shifman:1978bx,Shifman:1978by}.}. 
Equation~(\ref{rho_dispersion})
thus corresponds to integrating over the spectral function with an 
exponential filter of width $1/b$ applied to the energy $\sqrt{t}$.
Significant numerical suppression happens already inside the range
$\sqrt{t} \lesssim 1/b$ and determines the \textit{absolute} magnitude of 
the resulting density; the important point is that the contribution from 
larger energies in the integral are \textit{relatively} suppressed 
compared to those inside the range with exponential strength
(see Refs.~\cite{Miller:2010tz,Miller:2011du} for a 
detailed discussion). In this sense the transverse distance $b$ acts as an
external parameter that allows one to ``select'' energies in the range 
$\sqrt{t} \lesssim 1/b$ in the spectral functions of the form factors.

The spectral representation Eq.~(\ref{rho_dispersion}) is particularly
suited to the study of the asymptotic behavior of the transverse densities
at large distances. Generally, any singularity (pole or branch cut) 
in the form factors at a squared mass $t = \mu^2$, which contributes 
to the imaginary parts $\textrm{Im} \, F_{1, 2}(t + i0)$, 
produces densities which asymptotically decay as
\beq
\rho_{1, 2}(b)_{\text{singularity at} \; \mu^2} 
\;\; \sim \;\; P_{1, 2}(b) \; e^{-\mu b} 
\hspace{2em} (b \rightarrow \infty) ,
\label{large_b_general}
\eeq
where $P_{1, 2}$ are functions with power--like asymptotic behavior.
The rate of exponential decay is governed by the position of the
singularity alone; the pre-exponential factor $P_{1, 2}$ depends on the
strength of the singularity and the variation of the spectral functions 
over the relevant range of integration (which may involve other mass 
scales besides $\mu$) and has to be determined by detailed calculation.
Equation~(\ref{large_b_general}) expresses the traditional notion of 
the range of an ``exchange mechanism'' in the spatial representation
of nucleon structure through transverse densities.

Here we are interested in the transverse densities in the chiral periphery,
at distances of the order $b \sim M_\pi^{-1}$. In the context of the
spectral representation Eq.~(\ref{rho_dispersion}) the densities at
such distances are determined by the behavior of the spectral function 
near the two--pion threshold, $t = 4 M_\pi^2$; more precisely, at
masses
\beq
t - 4 M_\pi^2 \;\; \sim \;\; \textrm{few} \; M_\pi^2 .
\eeq
Physically, this corresponds to chiral processes in which the current 
operator couples to the nucleon by exchange of two ``soft'' pions, with 
momenta $|\bm{k}_{1, 2}| \sim \textrm{few} \, M_\pi$ in the nucleon 
rest frame (details will be given below).
The two--pion cut in the nucleon form factor has isovector quantum numbers
and contributes with different sign to the proton and neutron.
In our theoretical analysis we therefore focus on the isovector
combination of the form factors and the transverse densities,
\beq
\rho^V (b) \;\; \equiv \;\; {\textstyle\frac{1}{2}} [\rho^p (b) - \rho^n(b)] .
\eeq
In the isoscalar density the chiral contribution starts with three--pion
exchange and is numerically irrelevant at all distances of interest
(see Refs.~\cite{Miller:2011du} for a phenomenological analysis).

The spectral representation of Eq.~(\ref{rho_dispersion}) offers many
practical advantages for the study of the chiral component of the 
transverse densities. First, it relates the chiral component to the 
isovector spectral function near threshold, which possesses a rich 
structure (see Sec.~\ref{subsec:spectral}) that expresses itself in
the densities and can be exhibited in this way. The calculation
of the spectral function in chiral EFT is particularly simple and
can be performed very efficiently using $t$--channel cutting rules.
The chiral and heavy--baryon expansions of the spectral functions
have been studied extensively in the literature
\cite{Gasser:1987rb,Bernard:1996cc,Becher:1999he,Kubis:2000zd,Kaiser:2003qp},
and these results can directly be imported into the study of 
transverse densities. Second, the spectral representation
allows us to combine chiral and non--chiral contributions to the
transverse densities in a consistent manner. The latter arise from
higher--mass states in the spectral functions, particularly the 
$\rho$ meson in the isovector channel. The total spectral function 
can be constructed such that the chiral EFT result is used only in 
the near--threshold region $t - 4 M_\pi^2 \sim \textrm{few} \; M_\pi^2$,
where the chiral expansion is manifestly valid, and the higher--mass 
region is parametrized empirically. In this way the chiral and non--chiral
components can be added without double--counting and compared quantitatively
as functions of $b$.

In the following we 
use the the spectral representation Eq.~(\ref{rho_dispersion})
as a tool to calculate the chiral component of the transverse densities
in chiral EFT. It is worth noting that this representation has many 
applications beyond this specific purpose. It can be used to quantify the 
vector meson contribution in the nucleon's transverse 
densities \cite{Miller:2011du}, 
and to construct the transverse charge density in the pion from precise 
data of the timelike form factor obtained in $e^+e^-$ annihilation 
experiments \cite{Miller:2010tz}. It can also be extended to other
nucleon form factors and corresponding densities, such as the form 
factors of the energy--momentum tensor and the ``generalized form factors''
defined by the moments of the nucleon GPDs.
\subsection{Spectral functions near threshold}
\label{subsec:spectral}
The isovector transverse densities in the chiral periphery are determined 
by the spectral functions of the nucleon form factors in the vicinity of 
the two--pion threshold at $t = 4 M_\pi^2$. Before turning to the chiral 
EFT calculations it is worth reviewing the analytic
structure of the form factor near threshold as it follows from general 
considerations \cite{Frazer:1960zza,Frazer:1960zzb,Hohler:1976ax}. 
In particular, this explains the nature of the subthreshold 
singularity at $t = 4 M_\pi^2 - M_\pi^4/M_N^2$, 
which defines the parametric regimes in the analysis of the transverse 
densities and determines the convergence of the chiral expansion.

\begin{figure}
\begin{tabular}{ll}
\parbox[c]{.24\textwidth}{
\includegraphics[width=.24\textwidth]{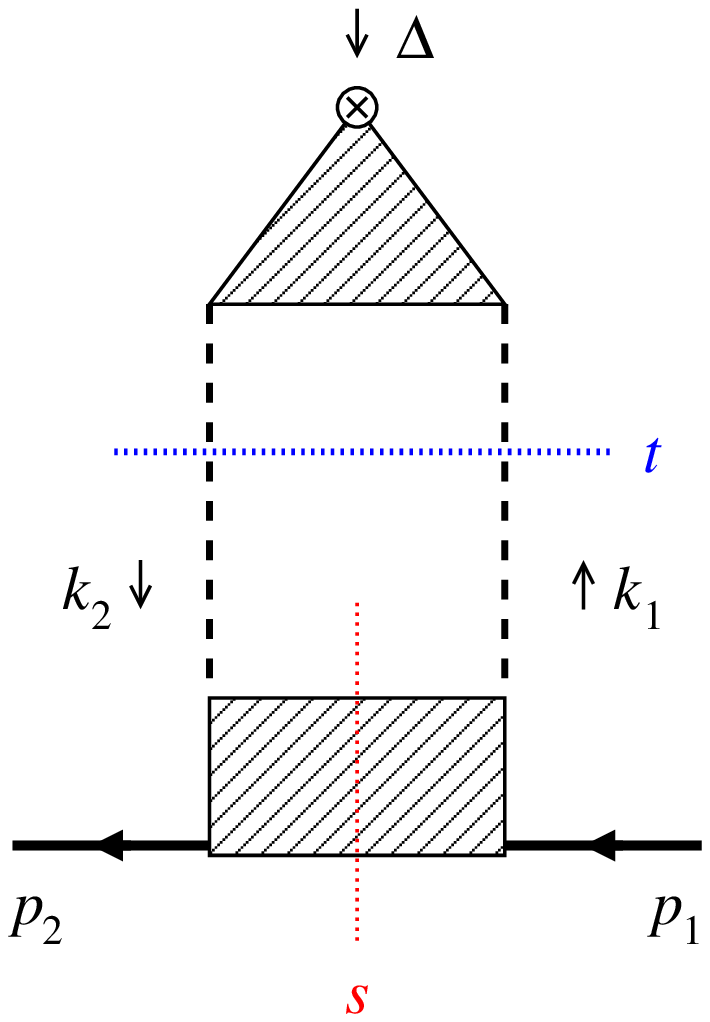}}
\hspace{.1\textwidth}
&
\parbox[c]{.32\textwidth}{
\includegraphics[width=.32\textwidth]{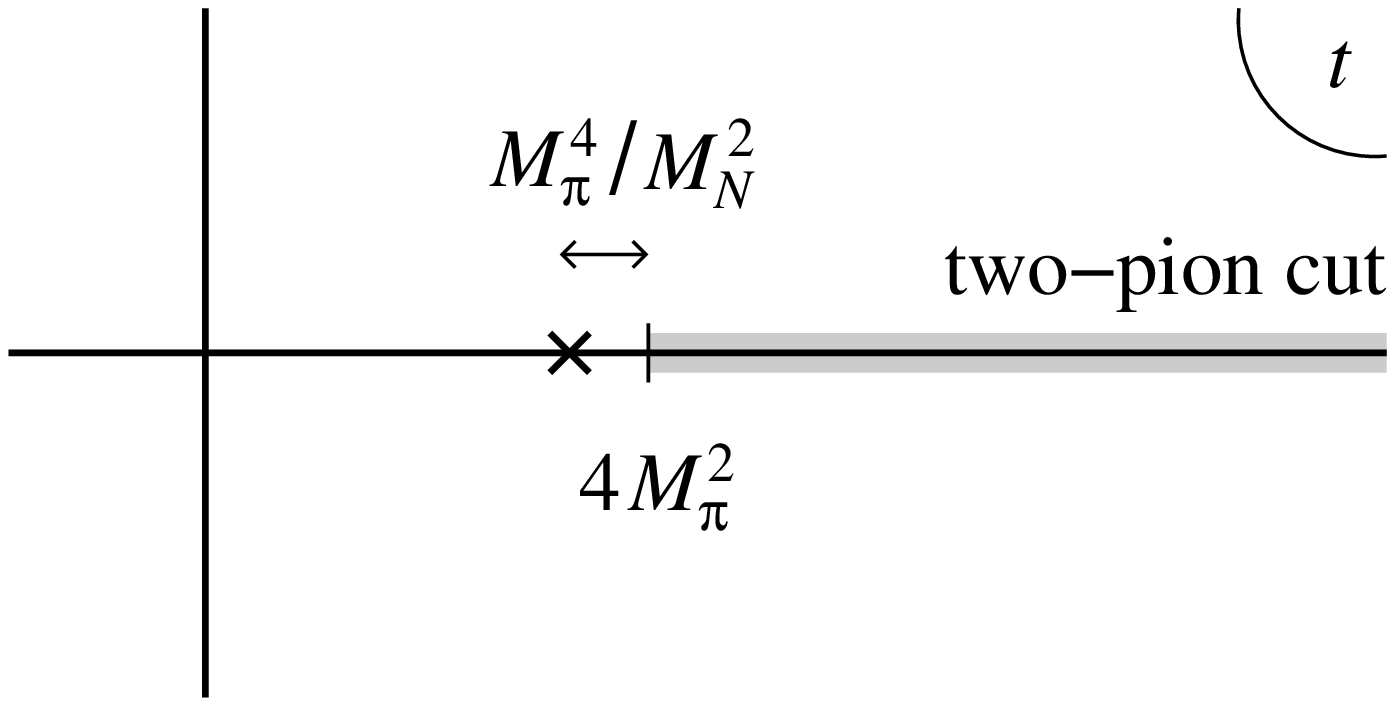}}
\\[-1ex]
(a) & (b)
\end{tabular}
\caption[]{(a) Virtual processes generating the two--pion cut in the
nucleon form factor. The triangle denotes the timelike pion form factor, 
the rectangle the full $\pi N$ scattering amplitude in the region 
$t > 4 M_\pi^2$.
(b) Analytic structure of the nucleon form factor in the vicinity 
of the two--pion threshold $t = 4 M_\pi^2$.
The cross denotes the subthreshold singularity on the unphysical sheet,
resulting from the intermediate nucleon pole in the $\pi N$ amplitude 
in the virtual process [see drawing (a)].}
\label{fig:twopi}
\end{figure}

The spectral functions at $t = 4 M_\pi^2 + \textrm{few} \, M_\pi^2$
result from virtual processes in which the current couples to the 
nucleon by conversion to a two--pion state of mass $\sqrt{t}$
(see Fig.~\ref{fig:twopi}a). The coupling of this system to the nucleon 
is described by the $\pi N$ scattering amplitude, which at $t < 0$ can 
be determined in $\pi N$ scattering experiments but is evaluated here 
in the region $t > 0$. The analytic 
structure of the $\pi N$ scattering amplitude implies the existence of 
certain singularities on the unphysical sheet of the nucleon form factor,
below the principal cut starting at $t = 4 M_\pi^2$
(see Fig.~\ref{fig:twopi}b). 
They occur because for certain values of $t$ the invariant mass of 
the $s$--channel intermediate state of the $\pi N$ scattering process 
can reach the value of physical baryon masses (specifically, the $N$ 
and $\Delta$), where the $\pi N$ scattering amplitude has a pole.
This can be seen most easily in the center--of--mass frame of the 
$t$--channel process of production of the two--pion system by the 
electromagnetic current. Let $p_{1, 2}$ be the 4--momenta of the 
initial and final nucleon, and $k_{1, 2}$ those of the two pions.
Introducing the average nucleon and pion 4--momenta and their 
difference,
\beq
P \; \equiv \; {\textstyle\frac{1}{2}} (p_1 + p_2) ,
\hspace{2em}
k \; \equiv \; {\textstyle\frac{1}{2}} (k_1 + k_2) ,
\hspace{2em}
\Delta \; \equiv \;  p_2 - p_1 \; = \; k_2 - k_1 ,
\eeq
we express the individual 4--momenta as $p_{1, 2} = P \mp \Delta/2$
and $k_{1, 2} = k \mp \Delta/2$. The mass shell conditions for the 
initial and final nucleon 4--momenta imply
\be
P\Delta &=& 0 ,
\label{PDelta}
\\
P^2 &=& M_N^2 - t/4 ,
\label{P2}
\ee
where $t = \Delta^2$.
%
% FIGURE
%
The spectral function corresponds to the process of Fig.~\ref{fig:twopi}a
with on--shell external nucleons but values of $t > 4 M_\pi^2$, for which the
current can produce a two--pion state. In this state also the pion 4--momenta
are on mass--shell, and in addition to Eq.~(\ref{PDelta}) and (\ref{P2})
one has the relations
\be
k\Delta &=& 0 ,
\label{kDelta}
\\
k^2 &=& M_\pi^2 - t/4 .
\label{k2}
\ee
The $t$--channel center--of--mass (or CM) frame is defined as the frame 
in which the 4--momentum of the current, which is the total 4--momentum 
of the pion pair, has components
\beq
\Delta^\mu \;\; = \;\; (\sqrt{t}, 0, 0, 0),
\label{Delta_cm}
\eeq
where $t > 0$. Because of Eq.~(\ref{PDelta}) the average nucleon momentum 
$P$ in this frame has only spatial components, and we choose it to point 
in the $z$--direction,
\beq
P^\mu \;\; = \;\; \left(0, 0, 0, P^z \right) ,
\label{P_cm}
\eeq
where the component $P^z$ is determined by Eq.~(\ref{P2}) as
\beq
P^z \;\; = \;\; \left\{ 
\begin{array}{lll}
{\textstyle\sqrt{t/4 - M_N^2}} 
&= \sqrt{-P^2} \hspace{2em} & t > 4 M_N^2, \\[1ex]
i \textstyle{\sqrt{M_N^2 - t/4}} &= i\sqrt{P^2} & t < 4 M_N^2 .
\end{array} \right.
\label{P3_cm}
\eeq
In the near--threshold region $t = 4 M_\pi^2 + \textrm{few} \, M_\pi^2$
we need to use the lower expression, where the value of $P^z$ is imaginary. 
Note that the sign of the imaginary part 
of $P^z$ in the region $t < 4 M_N^2$ follows from the analytic continuation 
of the expression for $t > 4 M_N^2$ with the prescription 
$t \rightarrow t + i0$. In sum, the choice of 4--vectors 
Eqs.~(\ref{Delta_cm})--(\ref{P3_cm}) satisfies the
invariant constraints Eqs.~(\ref{PDelta})--(\ref{P2}) 
for any value of $t > 0$. 

Further in the CM frame, Eq.~(\ref{kDelta}) requires that the average 
pion 4--momentum $k$ have components
\beq
k^\mu \; = \; \left(0, \bm{k} \right),
\eeq
and the modulus of the 3--momentum is determined by Eq.~(\ref{k2}) as
\beq
|\bm{k}| \; = \; {\textstyle\sqrt{t/4 - M_\pi^2}} \; \equiv \;\; k_{\rm cm}
\label{k_cm}
\eeq
and referred to as the pion CM momentum. Here we assume that 
$t > 4 M_\pi^2$; the values of $k_{\rm cm}$ below threshold
are obtained by analytic continuation with $t \rightarrow t + i0$. 
Denoting the polar angle of the pion momentum by $\theta$, we have
\beq
k^z \; = \; k_{\rm cm} \, \cos\theta ,
\hspace{2em}
kP \; = \; -i k_{\rm cm} \sqrt{P^2} \cos\theta .
\eeq
The two--pion contribution to the spectral functions of the electromagnetic 
form factors at $t > 4 M_\pi^2$ is now given by the product of the
invariant amplitudes for the $\textrm{current} \rightarrow \pi\pi$
and the $\pi \pi \rightarrow N \bar N$ transitions, integrated over the
solid angle of the pion CM momentum $\bm{k}$ (Fig.~\ref{fig:twopi}a). 
Because of $t$--channel angular momentum conservation the angular 
dependence of the $\textrm{current} \rightarrow \pi\pi$ amplitude in 
the CM frame is $\sim \cos\theta$, and the integral projects the
$\pi \pi \rightarrow N \bar N$ amplitude on the $J = 1$ partial 
wave (P wave). The well--known result is
\cite{Frazer:1960zza,Frazer:1960zzb,Hohler:1976ax}
\beq
\frac{1}{\pi} \textrm{Im} \, F_{1, 2} (t) \;\; = \;\; 
\frac{k_{\rm cm}^3}{\pi\sqrt{t}} \; F_\pi^\ast (t) \; \Gamma_{1, 2} (t) ,
\label{partial_wave}
\eeq
where $F_\pi^\ast (t)$ is the (complex--conjugate) pion form factor
and $\Gamma_{1, 2} (t)$ the $\pi \pi \rightarrow N \bar N$ partial
wave amplitude \cite{Hohler:1974ht}.

Equation~(\ref{partial_wave}) describes the spectral functions of
the form factors, which are real functions defined in the physical region 
of the $t$--channel process, $t > 4 M_\pi^2$. The behavior of the complex
form factors themselves can be studied in a very similar manner, by 
interpreting Eq.~(\ref{partial_wave}) as a discontinuity of the complex
function which can be analytically continued. The net result is that the 
singularities of the $\pi\pi \rightarrow N\bar N$ partial--wave amplitude 
are ``transmitted'' to the form factors
\cite{Frazer:1960zza,Frazer:1960zzb,Hohler:1976ax}. Specifically, at a given
value of $t$ and $\cos\theta$, the squared invariant mass of the 
$s$--channel intermediate state in the $\pi N$ invariant amplitude is
(see Fig.~\ref{fig:twopi}a) 
\be
s &\equiv& (p_1 -k_1)^2 \;\; = \;\; (p_2 - k_2)^2 \;\; = \;\; (k - P)^2
\nonumber
\\[1ex]
&=& -k_{\rm cm}^2 - t/4 + 2 i k_{\rm cm} \sqrt{P^2} \cos\theta + M_N^2 .
\label{s_pole}
\ee
The $\pi N$ invariant amplitude has singularities at the values of $s$ 
corresponding to physical intermediate states; in particular the nucleon
pole at $s = M_N^2$. Upon integration over $\cos\theta$, it produces
branch cut singularities in the partial wave amplitudes $\Gamma_{1, 2} (t)$
on the unphysical sheet of $t$. The start of the cut (the position of
the branch point) coincides with the end points of the angular integration
and is thus determined by the condition $s(t, \cos\theta = \pm 1) = M_N^2$, 
or
\beq
-k_{\rm cm}^2 - t/4 \;\; = \;\; \pm 2 i k_{\rm cm} \sqrt{P^2} . 
\label{pole_condition}
\eeq
Taking the square of both sides, and substituting the 
expressions Eqs.~(\ref{P2}) and (\ref{k_cm}) for $P^2$ and $k_{\rm cm}$,
this becomes
\beq
(t/2 - M_\pi^2)^2 \;\; = \;\; - (t - 4 M_\pi^2) (M_N^2 - t/4) ,
\eeq
the solution of which is
\beq
t \;\; = \;\; 4 M_\pi^2 \; - \; \frac{M_\pi^4}{M_N^2} 
\;\; \equiv \;\; t_{\rm sub} .
\label{t_sub}
\eeq
In sum, the form factors as analytic functions of $t$ have a 
branch cut on the unphysical sheet, starting at the value given 
by Eq.~(\ref{t_sub}), which corresponds to the intermediate nucleon 
state in the $\pi N$ scattering amplitude going on mass 
shell (see Fig.~\ref{fig:twopi}b) 
\cite{Frazer:1960zza,Frazer:1960zzb,Hohler:1976ax}. The presence 
of this subthreshold singularity can be established on general grounds;
it can also be seen explicitly in the relativistic chiral EFT results 
quoted below.

A point of great importance is that the distance of the subthreshold
singularity from the threshold is small on the scale of $M_\pi^2$:
\beq
t_{\rm sub} \; - \; 4 M_\pi^2 \;\; = \;\; \epsilon^2 \, M_\pi^2 ,
\label{t_sub_epsilon}
\eeq
where
\beq
\epsilon \;\; \equiv \;\; \frac{M_\pi}{M_N} .
\label{epsilon}
\eeq
The ratio $\epsilon$ is a small parameter in both the chiral and the 
heavy--baryon limit. The spectral functions of the isovector form 
factors thus exhibit structure on two different scales. Looking at them
on the ``coarse'' scale, $t = O(M_\pi^2)$, one sees them rising
from the threshold at $t = 4 M_\pi^2$ and varying on average with 
a characteristic scale $\sim M_\pi^2$. Looking at the functions near 
threshold on the ``fine'' scale, $t - 4 M_\pi^2 = O(\epsilon^2 M_\pi^2)$, 
one sees a variation with characteristic scale $\epsilon^2 M_\pi^2$,
caused by the closeness of the subthreshold singularity.

The presence of a singularity close to the physical threshold affects 
the convergence of the chiral expansion of the spectral function near 
threshold \cite{Bernard:1996cc,Becher:1999he,Kubis:2000zd,Kaiser:2003qp}.
For instance, one immediately sees that a naive expansion 
of $\textrm{Im} \, F_{1, 2}(t)$ in powers of the pion CM momentum
$k_{\rm cm}$ would converge only in the parametrically small region 
$t - 4 M_\pi^2 < \epsilon^2$, or $k_{\rm cm} < \epsilon / 2$,
and thus produce unnaturally large expansion coefficients growing like
inverse powers of $\epsilon$. This situation generally requires the use
of different expansion schemes in different parametric regions of 
$t$; uniform approximations can be obtained by matching the
different expansions \cite{Becher:1999he}.

The nucleon pole in the $\pi N$ scattering amplitude is special in that
it produces a subthreshold singularity extremely close to the threshold,
which strongly influences the behavior of the spectral function above
threshold. Higher mass $\pi N$ resonances give rise to further 
subthreshold singularities of the form factor, which, however, lie 
farther away from threshold. Below we consider the $\Delta$ isobar
at $M_\Delta = 1.23 \, \textrm{GeV}$, which couples strongly to the
$\pi N$ channel and becomes degenerate with the $N$ in the large--$N_c$
limit of QCD. For this state the pole condition $s = M_\Delta^2$ 
becomes [cf.\ Eq.~(\ref{pole_condition})]
\beq
-k_{\rm cm}^2 - t/4 - M_\Delta^2 + M_N^2 
\;\; = \;\; \pm 2 i k_{\rm cm} \sqrt{P^2} ,
\eeq
whose solution is
\beq
t \;\; = \;\; 4 M_\pi^2 - \frac{(M_\Delta^2 - M_N^2 + M_\pi^2)^2}{M_\Delta^2}
\;\; \equiv \;\; t_{{\rm sub}, \Delta} 
\label{t_sub_delta}
\eeq
[the expression reduces to Eq.~(\ref{t_sub}) if one sets $M_\Delta = M_N$].
One sees that this subthreshold singularity is removed from threshold
by a distance in $t$ that does not tend to zero in the chiral limit 
$M_\pi \rightarrow 0$. Numerically, with the physical $\pi, N$ and $\Delta$
masses, the distance from threshold is $0.022\, M_\pi^2$ for the $N$ 
and $0.43 \, M_\pi^2$ (or 20 times larger) for the $\Delta$ singularity, 
showing clearly the qualitative difference between the $N$ pole and 
higher--mass $\pi N$ resonances.
\subsection{Parametric regions of transverse distance}
\label{subsec:parametric}
In the context of our dispersion analysis of transverse densities,
the ``two--scale'' structure of the spectral function near threshold
defines the parametric regions of the transverse distance $b$ at which 
we aim to compute the densities. Again, it is useful to establish this
connection on general grounds, before turning to the actual chiral 
expansion of the functions.

In the dispersion integral Eq.~(\ref{rho_dispersion}) the distance
$b$ effectively controls the region of $t$--channel masses over which
the spectral function is integrated. To make this more explicit,
we substitute the asymptotic expression Eq.~(\ref{K0_asymptotic}) for the
modified Bessel function; the deviations between the exact function
and the asymptotic approximation are not important for the parametric
estimates made here. We obtain
\beq
\rho_{1, 2} (b) \;\; = \;\; e^{-2 M_\pi b} \;
\int\limits_{4M_\pi^2}^\infty dt \; 
\frac{e^{-(\sqrt{t} - 2 M_\pi) b}}{(8 \pi \sqrt{t}b)^{1/2}}
\; \frac{\textrm{Im}\, F_{1, 2} (t + i0)}{\pi} .
\label{rho_dispersion_preexp}
\eeq
We have extracted the exponential factor $\exp (-2 M_\pi b)$ from the
integral, so that the remaining integral represents the pre-exponential 
factor $P_{1, 2}(b)$ in 
the general asymptotic form Eq.~(\ref{large_b_general}).
In Eq.~(\ref{rho_dispersion_preexp}) the exponential function under
the integral restricts the integration to masses $\sqrt{t}$ for which
\beq
(\sqrt{t} - 2 M_\pi) b \;\; = \;\; O(1) .
\eeq
We can therefore distinguish two parametric regions in $b$. 
\begin{itemize}
\item[(a)] In the region
\beq
b \; = \; O(M_\pi^{-1})
\hspace{2em}  \textrm{(``chiral distances'')}
\label{b_chiral}
\eeq
the integral of Eq.~(\ref{rho_dispersion_preexp}) extends over masses
in the region
\beq
\sqrt{t} - 2 M_\pi \; = \; O(M_\pi) ,
\hspace{2em}
\textrm{or}
\hspace{2em}
t - 4 M_\pi^2 \; = \; O(M_\pi^2) ,
\eeq
with no additional restriction to values near threshold. 
The $t$--channel pion CM momenta are of the order
\beq
k_{\rm cm} \; = \; O(M_\pi) ,
\eeq 
which is the domain usually associated with chiral dynamics.
\item[(b)] In the region
\beq
b \; = \; O(\epsilon^{-2} M_\pi^{-1}) \; = \; O(M_N^2/M_\pi^3)
\hspace{2em}  \textrm{(``molecular distances'')}
\label{b_molecular}
\eeq
[$\epsilon = M_\pi / M_N$, cf.\ Eq.~(\ref{epsilon})]
the integral over masses is restricted to the near--threshold
region
\beq
\sqrt{t} - 2 M_\pi \; = \; O(\epsilon^2 M_\pi) ,
\hspace{2em}
\textrm{or}
\hspace{2em}
t - 4 M_\pi^2 \; = \; O(\epsilon^2 M_\pi^2) .
\eeq
The distance of $t$ from threshold is comparable to that of the 
subthreshold singularity from threshold, Eq.~(\ref{t_sub_epsilon}),
so that the behavior of the spectral function is essentially 
influenced by the subthreshold singularity. The pion CM momenta are 
now of the order
\beq
k_{\rm cm} \; = \; O(\epsilon M_\pi) ,
\label{k_cm_molecular}
\eeq 
corresponding to the $t$--channel system moving non--relativistically 
with velocity $v = k_{\rm cm}/M_\pi = O(\epsilon)$. 
\end{itemize}

We refer to the parametric domain of Eq.~(\ref{b_molecular}) as the 
molecular region, as the typical transverse distances between the pion 
and the initial/final nucleon are much larger than the Compton wavelength 
of the pion. At the physical pion and nucleon mass $\epsilon \approx 1/7$, 
so that such distances can numerically be as large as 
$10^2 \, \textrm{fm}$. Since the densities decay with an overall 
exponential factor of $\exp (-2 M_\pi b)$, they are extremely small at
such large distances. The molecular region of the nucleon's transverse 
densities is therefore mostly of theoretical interest. However, the
existence of this regime in coordinate space affects the magnitude 
of higher $b^2$--weighted moments of the densities, which are proportional
to higher derivatives of the form factors at $t = 0$, and thus may in
principle have observable consequences.

The parametric classification of distances, Eqs.~(\ref{b_chiral}) and
Eqs.~(\ref{b_molecular}), can be established on general grounds, starting
from the scales governing the behavior of the spectral function.
In Sec.~\ref{subsec:heavy_baryon} we show that the invariant chiral EFT
result bears out this general structure and perform the heavy--mass
expansion of the densities in the different parametric regions.
We note that the existence of a regime of anomalously large 
distances $\sim M_N^2/M_\pi^3$
is not specific to the isovector transverse charge and magnetization densities
but common to all nucleon observables governed by $t$--channel exchange 
of two pions, which are sensitive to the subthreshold singularities of
the $\pi N$ scattering amplitude. A similar phenomenon has been observed 
in the two--pion exchange contribution to the low--energy $NN$ 
interaction, where it can be expressed in terms of the large--distance 
behavior of the 3--dimensional 
$NN$ potential \cite{Robilotta:1996ji,Robilotta:2000py};
see Ref.~\cite{Epelbaum:2005pn} for a review.
\section{Peripheral densities from chiral dynamics}
\label{sec:chiral}
\subsection{Two--pion spectral functions}
\label{subsec:two_pion}
We now want to calculate the chiral component of the transverse densities
in the nucleon within the framework laid out in Sec.~\ref{sec:introduction}.
We use the leading--order chiral EFT results for the spectral functions of 
the form factors to compute the peripheral densities from the dispersion 
integral Eq.~(\ref{rho_dispersion_preexp}) and study their properties
in the parametric regions identified in Sec.~\ref{subsec:parametric}.
In view of the essential role of analyticity we employ the relativistic 
formulation of chiral EFT with baryons, which generates amplitudes with
the correct analytic structure in the form of Feynman diagrams with 
relativistic propagators; the heavy--baryon limit will be investigated
by expanding the explicit expressions obtained in the relativistic
formulation. 

The spectral functions of the nucleon form factors have been studied
extensively both in the relativistic and the heavy--baryon formulations
of chiral EFT \cite{Gasser:1987rb,Bernard:1996cc,Becher:1999he,%
Kubis:2000zd,Kaiser:2003qp} (see e.g.\ Ref.~\cite{Kaiser:2003qp} for a 
discussion of the literature), and we can use these results for 
our purposes. For several reasons it will be useful to revisit the 
leading--order relativistic calculation and summarize the essential 
steps here. First, the spectral functions can be computed very 
efficiently using $t$--channel cutting rules; this method can easily be 
extended to $\Delta$ intermediate states (see Sec.~\ref{sec:delta}) 
and to form factors of other operators (energy--momentum tensor, 
GPD moments) that will be calculated in a future study. 
Second, we need the explicit expressions of the
Feynman integrals for the partonic interpretation of our results and
future comparison with the light--front approach. In particular,
the physical origin of the contact term in the chiral EFT result for
the spectral function of $F_1$ is best understood at the level of the 
original Feynman integrals and was not discussed in this form before. 
Third, we present a very compact representation of the leading--order 
chiral EFT results that can easily be used for numerical analysis.

In the relativistic formulation of chiral EFT with 
nucleons \cite{Becher:1999he} the leading--order chiral Lagrangian 
is given by $\mathcal{L}^{(1)}_\chi = \mathcal{L}^{(1)}_N + 
\mathcal{L}^{(1)}_\pi$, 
where $\mathcal{L}^{(1)}_\pi$ is the usual chiral Lagrangian of 
the pion field, while $\mathcal{L}^{(1)}_N$ describes the dynamics of 
the nucleon field and its coupling to the pion and is of the form
\be
\mathcal{L}^{(1)}_N &=& 
\bar\psi [ i (\hat \partial + \hat \Gamma) 
- M_N ] \psi \; + \; {\textstyle\frac{1}{2}} g_A  
\bar\psi\hat u \gamma_5 \psi .
\label{chiral_lagrangian_orig}
\\[1ex]
\Gamma_\mu &\equiv & 
{\textstyle\frac{1}{2}} [ U^{-1/2} , \partial_\mu (U^{1/2}) ] ,
\\[1ex]
u_\mu &\equiv& i \, U^{-1/2} \, (\partial_\mu U) \, U^{-1/2} ,
\\[1ex]
U &\equiv& \exp [i\bm{\pi}\cdot\bm{\tau}/{F_\pi}] ,
\hspace{2em}
U^{\pm 1/2} \; = \; 
\exp [\pm i\bm{\pi}\cdot\bm{\tau}/(2 F_\pi )] ,
\label{U}
\ee
where $\hat\partial \equiv \partial_\mu \gamma^\mu$ etc.
Here $\psi$ is the Dirac field of the nucleon, and $\pi^a (a = 1, 2, 3)$
the chiral pion field. In Eq.~(\ref{chiral_lagrangian_orig})
$g_A$ denotes the nucleon axial vector coupling and $F_\pi$ the 
pion decay constant; at leading order these parameters are taken at
their physical (tree--level) values $g_A = 1.26$ and 
$F_\pi = 93\, \textrm{MeV}$.
In the calculation of the leading--order isovector spectral functions
one needs the pion--nucleon coupling to second
order in the pion field. Expanding Eq.~(\ref{chiral_lagrangian_orig})
in powers of the pion field one obtains
\be
\mathcal{L}^{(1)}_N &=& \bar\psi (i\hat\partial - M_N) \psi
\; - \; \frac{g_A}{2 F_\pi} \bar\psi \gamma_\mu \gamma_5 \tau^a \psi 
\; \partial_\mu \pi^a 
\; - \; \frac{1}{4F_\pi^2} \bar\psi \gamma_\mu \tau^a \psi \;
\epsilon^{abc} \pi^b \partial_\mu \pi^c .
\label{lag}
\ee
The second term on the right--hand side of Eq.~(\ref{lag}) 
describes a Yukawa--type $\pi NN$ 
coupling (three--point vertex). We note that the axial vector coupling 
used here is equivalent to the conventional pseudoscalar $\pi NN$
coupling for on--shell nucleons; namely
\be
- \frac{g_A}{2 F_\pi} 
\bar u_2 i \hat\Delta \gamma_5 \tau^a u_1
&=& 
\frac{g_A M_N}{F_\pi} 
\bar u_2 i \gamma_5 \tau^a u_1
\;\; \equiv \;\; g_{\pi NN} \bar u_2 i \gamma_5 \tau^a u_1
\label{axial_pseudoscalar}
\ee
between nucleon spinors $u_1 \equiv u(p_1)$ and $\bar u_2 \equiv \bar u(p_2)$
with $\Delta = p_2 - p_1$. The identification of the pseudoscalar coupling 
constant of Eq.~(\ref{axial_pseudoscalar}) is precisely the 
Goldberger--Treiman relation for the nucleon's axial current matrix element.
The third term in Eq.~(\ref{lag}) describes a local $\pi\pi NN$ coupling
(four--point vertex). Its appearance is due to the specific representation 
of the nucleon fields adopted in Eq.~(\ref{chiral_lagrangian_orig}), 
and the coupling constant is fixed by chiral symmetry and does not involve 
any free parameter. The vertex couples the isovector--vector current
of the nucleon field to that of the pion field. 

%
% FIGURE
%
\begin{figure}
\includegraphics[width=.5\textwidth]{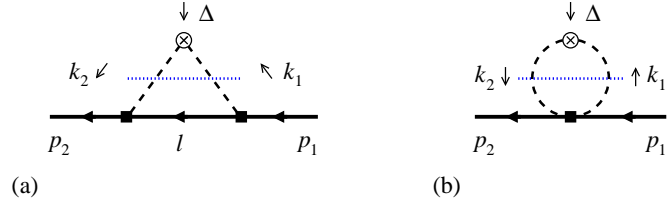}
\caption[]{The Feynman diagrams describing the leading--order chiral
contributions to the two--pion cut of the isovector nucleon form factor.
The dotted line indicates the Cutkosky cut.}
\label{fig:diag}
\end{figure}
The calculation of the spectral functions starts from the matrix element 
of the electromagnetic current between nucleon states. In general the 
electromagnetic current operator of the effective chiral theory consists 
of the currents of the pion and nucleon fields and contributions resulting 
from their pointlike interactions. We are interested only in the spectral 
functions of the isovector form factors in the region 
$t = 4M_\pi^2 + \textrm{few}\; M_\pi^2$, which results from processes
in which the current couples to the nucleon through two--pion exchange. 
In leading order these are given by the two Feynman diagrams 
of Fig.~\ref{fig:diag}, where the current is the leading--order 
isovector current of the pion field,
\beq
J_\pi^{\mu, \, (1)} \;\; = \;\; \epsilon^{3ab} \pi^a \partial^\mu \pi^b .
\eeq
Other diagrams appearing at the same order only contribute to the 
two--nucleon cut of the spectral function [which gives a 
short--distance contribution to the density at $b = O(M_N^{-1})$] or
modify the real part of the nucleon vertex function current, but do 
not contribute to the two--pion cut; this simplification is a major 
advantage of the dispersive approach. The contributions to the isovector 
current matrix element resulting from the diagrams of Fig.~\ref{fig:diag} 
can be computed using standard rules of Lorentz--invariant perturbation 
theory, and one obtains
\be
\langle N_2 | \, J^\mu (0) \, | N_1 \rangle_{\pi\pi \; {\rm cut}} 
&=& \frac{i g_A^2}{F_\pi^2} \;
\int\!\frac{d^4 k}{(2\pi)^4} \; 
\; \frac{[\bar u_2 \hat k_2 \gamma_5 (\hat l + M_N) \hat k_1 \gamma_5 u_1]
\, k^\mu}
{(k_1^2 - M_\pi^2 + i0) (k_2^2 - M_\pi^2 + i0) (l^2 - M_N^2 + i0)} 
\label{triangle_momentum_1}
\\[1ex]
&+& \frac{i}{F_\pi^2} \;
\int \! \frac{d^4 k}{(2\pi )^4} 
\; \frac{(\bar u_2 \hat k u_1) \, k^\mu}{(k_1^2 - M_\pi^2 + i0)
(k_2^2 - M_\pi^2 + i0)} .
\label{contact_momentum}
\ee
The label ``$\pi\pi$ cut'' indicates that we retain only the 
diagrams contributing to the two--pion cut.
The first integral, Eq.~(\ref{triangle_momentum_1}), results from 
diagram Fig.~\ref{fig:diag}a with the 
$\pi NN$ three--point vertex; 
the second, Eq.~(\ref{contact_momentum}), from diagram 
Fig.~\ref{fig:diag}b with the $\pi\pi NN$ four--point vertex
(or contact term). In both diagrams the pion 4--momenta 
are decomposed as
\beq
k_{1, 2} \;\; = \;\; k \mp \Delta/2 ,
\eeq
and the average momentum $k$ was chosen as integration variable.
In Eq.~(\ref{contact_momentum}) we have dropped terms in the integrand
which integrate to zero because of the symmetry of the integrand with 
respect to reflections $k \rightarrow - k$. In 
Eq.~(\ref{triangle_momentum_1})
\be
l &\equiv& p_1 - k_1 \;\; = \;\; p_2 - k_2 \;\; = \;\; P - k
\ee
is the 4--momentum of the intermediate nucleon, with $P = (p_1 + p_2)/2$
the average nucleon momentum. The expression of this diagram can be 
simplified further. Namely, the integral in Eq.~(\ref{triangle_momentum_1})
contains a term in which the pole of the intermediate
nucleon propagator cancels, and which is of the same structure as
the integral of Eq.~(\ref{contact_momentum}).
Making use of the anticommutation relations between the gamma matrices 
and the Dirac equation for the nucleon spinors, one can rewrite the 
bilinear form in Eq.~(\ref{triangle_momentum_1}) as
\be
\bar u_2 \hat k_2 \gamma_5 (\hat l + M_N) \hat k_1 \gamma_5 u_1 
&=& \bar u_2 \left[  - 2 M_N (l^2 - M_N^2) - (l^2 - M_N^2) \hat k
- 4 M_N^2 \hat k \right] u_1 .
\label{bilinear_simplified}
\ee
The first term in the bracket on the right--hand side 
integrates to zero because the integrand 
is antisymmetric under $k \rightarrow - k$, and can be dropped.
The second term leads to an integral of the same form as 
Eq.~(\ref{contact_momentum}) and can be combined with 
Eq.~(\ref{contact_momentum}), effectively changing the coefficient
of the contact term resulting from diagram Fig.~\ref{fig:diag}b as
\beq
\frac{1}{F_\pi^2} \;\; \rightarrow \;\; 
\frac{1 - g_A^2}{F_\pi^2} .
\label{contact_combined}
\eeq
The appearance of the combination $1 - g_A^2$ here is not accidental 
but has a deeper physical meaning, as is explained 
in Sec.~\ref{subsec:contact}. 
The third term in Eq.~(\ref{bilinear_simplified}) represents the 
genuine ``non--contact'' contribution from the diagram Fig.~\ref{fig:diag}a,
corresponding to an intermediate state with a propagating nucleon.

The tensor integrals in Eq.~(\ref{triangle_momentum_1}) and 
(\ref{contact_momentum}) can be reduced to scalar integrals
with the help of standard projection formulas. Using the Dirac equation
to convert the resulting bilinear forms $\bar u_2 \ldots u_1$ to those 
of the right--hand side of Eq.~(\ref{me_general}), one obtains
the chiral contribution to the isovector Dirac and Pauli form factors 
in terms of invariant integrals as
\be
F_1^V(t)_{\pi\pi \; {\rm cut}} 
&=& 
\frac{4 M_N^2 g_A^2}{F_\pi^2} \, I_1 (t)
\; + \; 
\frac{1 - g_A^2}{F_\pi^2}  \, I_{\rm cont}(t) ,
\label{chiral_F_1}
\\
F_2^V(t)_{\pi\pi \; {\rm cut}} 
&=& \frac{4 M_N^2 g_A^2}{F_\pi^2} \, I_2 (t) ,
\label{chiral_F_2}
\ee
where
\be
I_{1, 2} &\equiv & -i \int\!\frac{d^4 k}{(2\pi )^4} \;
\frac{N_{1, 2}}
{(k_1^2 - M_\pi^2 + i0) (k_2^2 - M_\pi^2 + i0) (l^2 - M_N^2 + i0)} ,
\label{I_12}
\\[1ex]
I_{\rm cont} &\equiv& 
\phantom{-} i \int\!\frac{d^4 k}{(2\pi )^4} \;
\frac{N_{\rm cont}}{(k_1^2 - M_\pi^2 + i0)
(k_2^2 - M_\pi^2 + i0)} ,
\label{I_cont}
\\[1ex]
N_1
&\equiv& \frac{1}{P^2} \left\{
-\frac{t}{8} \left[ k^2 - \frac{(k\Delta)^2}{\Delta^2} \right]
\; + \; \left( M_N^2 + \frac{t}{8} \right) 
\frac{(kP)^2}{P^2} \right\} ,
\\[1ex]
N_2 &\equiv& -\frac{1}{2} \left[ - k^2 + 3 \frac{(kP)^2}{P^2} 
+ \frac{(k\Delta)^2}{\Delta^2} \right] \frac{M_N^2}{P^2} ,
\\[1ex]
N_{\rm cont} &\equiv& \frac{1}{3} 
\left[ k^2 - \frac{(k\Delta)^2}{\Delta^2} \right] .
\ee
For the spectral functions we need only the imaginary part of the 
invariant integrals Eqs.~(\ref{I_cont}) 
and Eqs.~(\ref{I_12}) above the two--pion 
threshold $t > 4 M_\pi^2$. The imaginary part can be computed very
efficiently using the $t$--channel cutting rule given in
Appendix~\ref{app:cutting}. We go to the $t$--channel CM frame
described in Sec.~\ref{subsec:spectral}, where the external 4--momenta
have components [cf.\ Eqs.~(\ref{Delta_cm})--(\ref{P3_cm})]
\beq
\Delta^\mu \; = \; (\sqrt{t}, 0, 0, 0), 
\hspace{2em}
P^\mu \; = \; (0, 0, 0, i \sqrt{P^2}).
\eeq
The on--shell constraints Eq.~(\ref{k_constraint}) restrict the 
integration momentum in this frame to
\beq
k^\mu \; = \; (0, \bm{k}), \hspace{2em}
|\bm{k}| \; = \; k_{\rm cm} ,
\eeq
where $k_{\rm cm}$ is defined in Eq.~(\ref{k_cm}).
It is straightforward to express the invariants in 
Eqs.~(\ref{I_cont}) and (\ref{I_12}) in terms of these vector components; 
specifically, the intermediate nucleon denominator in 
Eq.~(\ref{I_12}) becomes [cf.\ Eq.~(\ref{s_pole})]
\be
l^2 - M_N^2 &=&  -A + i B \cos\theta ,
\\[1ex]
A &\equiv & t/2 - M_\pi^2 ,
\\[1ex]
B &\equiv&  2 k_{\rm cm} \sqrt{P^2} .
\ee
Applying Eq.~(\ref{cutkosky_theta}) the imaginary parts then become
elementary phase space integrals over the polar angle of the 
pion $t$--channel CM momentum, $\cos \theta$. Performing the
integrals, one readily obtains \footnote{For brevity we omit the
infinitesimal imaginary part of the argument $t$ when quoting
explicit expressions of the spectral function and write 
$\textrm{Im} \, F (t) \equiv \textrm{Im} \, F (t + i0)$.
This convention will be applied throughout the following
text and figures.}
\be
\frac{1}{\pi} \, \textrm{Im} \, F_1^V (t)
&=& \frac{M_N^2 g_A^2 A^2}{(4\pi F_\pi)^2 (P^2)^{5/2} 
\sqrt{t}}
\left[ -\frac{t}{8} x^2 \arctan x
\; + \; \left( M_N^2 + \frac{t}{8} \right) ( x - \arctan x) 
\right] 
\label{Im_F1_noncontact}
\\[1ex]
&+& \frac{2 (1 - g_A^2) k_{\rm cm}^3}
{3 (4\pi F_\pi)^2 \sqrt{t}} 
\label{Im_F1_contact}
\\[2ex]
\frac{1}{\pi} \, \textrm{Im} \, F_2^V (t)
&=& \frac{M_N^4 g_A^2 A^2}{2 (4\pi F_\pi)^2 (P^2)^{5/2} 
\sqrt{t}}
\left[ x^2 \arctan x \; - \; 3 (x - \arctan x) \right] ,
\label{Im_F2}
\\[2ex]
x &\equiv & x(t) \;\; \equiv \;\; 
\frac{B}{A}
\;\; = \;\;
\frac{2 \sqrt{t/4 - M_\pi^2} \sqrt{M_N^2 - t/4}}{t/2 - M_\pi^2} 
\label{x}
\\[2ex]
&& \left[ k_{\rm cm} \, = \, {\textstyle\sqrt{t/4 - M_\pi^2}}, \;
P^2 \, = \, M_N^2 - t/4, \;
A \, = \, t/2 - M_\pi^2, \; B = 2 k_{\rm cm} \sqrt{P^2} \right] .
\nonumber
\ee
Equations~(\ref{Im_F1_noncontact})--(\ref{x}) represent the leading--order 
result for the isovector spectral functions of the nucleon's Dirac
and Pauli form factor in relativistic chiral 
EFT \cite{Gasser:1987rb,Bernard:1996cc,Kubis:2000zd,Kaiser:2003qp} 
and are our starting point for the study of the chiral component of 
the transverse charge
and magnetization densities. Despite their compact form the expressions 
of Eqs.~(\ref{Im_F1_noncontact})--(\ref{x}) contain very rich structure, 
which will be exhibited in the following.

The leading--order chiral result for the spectral functions
Eqs.~(\ref{Im_F1_noncontact})--(\ref{x}) embodies the general analytic
structure of the form factors near threshold described in 
Sec.~\ref{subsec:spectral}. First, one sees that the subthreshold singularity
Eq.~(\ref{t_sub}) is encoded in the inverse tangent function;
it has branch point singularities at complex values of the argument
\beq
x \; = \; \pm i ,
\eeq
which correspond to the value of $t$ given by Eq.~(\ref{t_sub}). 
The presence of these singularities restricts the power series expansion
of the function in $x$ around $x = 0$ to the region $|x| < 1$. Second,
we note that the expressions in Eqs.~(\ref{Im_F1_noncontact})--(\ref{x}) 
are not singular at $t = 4 M_N^2$; the inverse powers of 
$\sqrt{P^2} = \sqrt{M_N^2 - t/4}$ appearing in the prefactors are 
compensated by the vanishing of the expressions in the brackets for 
$x \rightarrow 0$. Physically this is obvious, as the chiral contribution
given by diagrams Fig.~\ref{fig:diag}a and b does not know about the
$N\bar N$ production threshold.

%
% FIGURE
%
\begin{figure}
\begin{tabular}{ll}
\includegraphics[width=.45\textwidth]{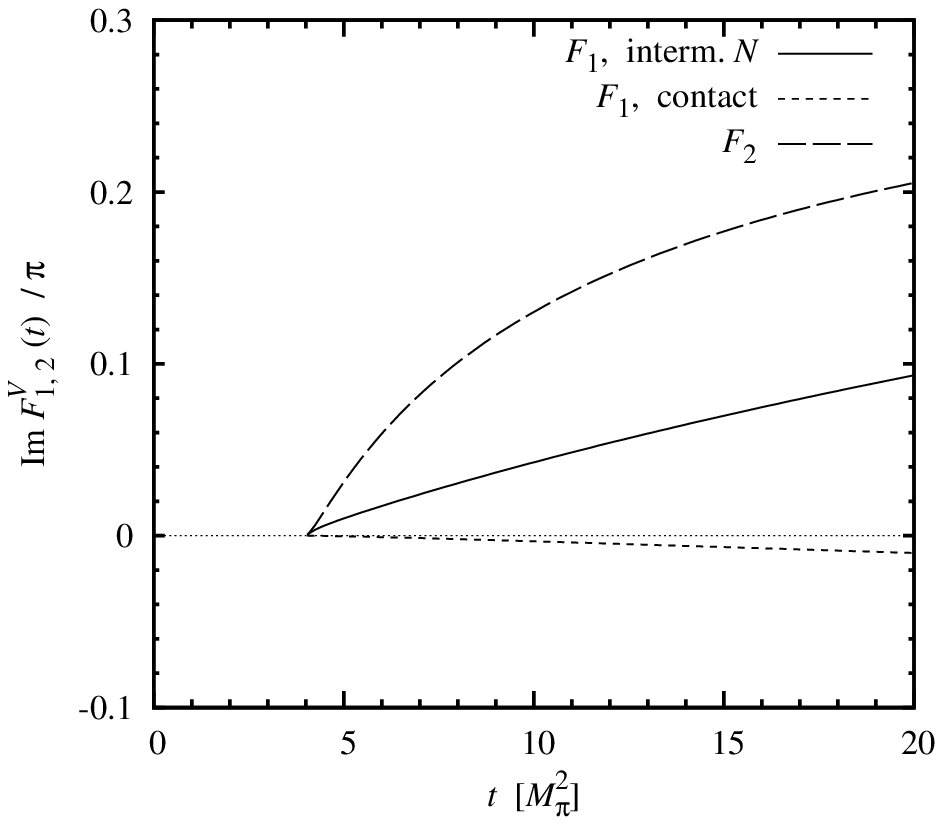}
&
\includegraphics[width=.45\textwidth]{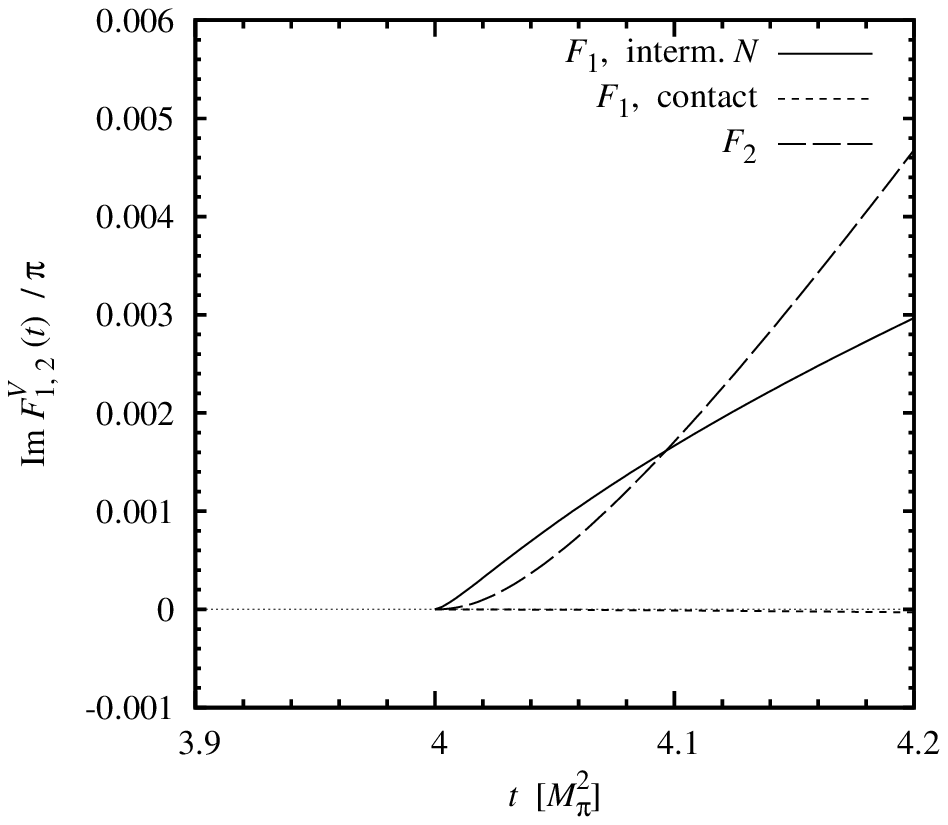}
\\[-2ex]
(a) & (b)
\end{tabular}
\caption[]{Leading--order spectral functions of the nucleon's
isovector Dirac and Pauli form factors, $\textrm{Im} \, F_{1, 2}^V(t)/\pi$,
Eqs.~(\ref{Im_F1_noncontact})--(\ref{x}). The variable $t$ is given 
in units of $M_\pi^2$. Panel (a) shows the
functions over the entire chiral region $t \sim \textrm{few} \, M_\pi^2$, 
panel (b) the behavior in the near--threshold region.
Solid lines: $F_1$, intermediate nucleon part, 
Eq.~(\ref{Im_F1_noncontact}).
Dotted lines: $F_1$, contact term, Eq.~(\ref{Im_F1_contact}).
Dashed lines: $F_2$, Eq.~(\ref{Im_F2}).}
\label{fig:ft}
\end{figure}
The numerical results for the chiral spectral functions is shown in
Fig.~\ref{fig:ft}a and b. Panel (a) shows the functions over the 
entire chiral region $t \sim \textrm{few} \, M_\pi^2$, panel (b) 
the behavior in the near--threshold region. Several features are
worth noting. First, Fig.~\ref{fig:ft}a shows that most of the
spectral function in the chiral region comes from the intermediate
nucleon part of diagram Fig.~\ref{fig:diag}a, 
Eq.~(\ref{Im_F1_noncontact}); 
the combined contact term resulting from diagram Fig.~\ref{fig:diag}b
and the non--propagating part of Fig.~\ref{fig:diag}a, 
Eq.~(\ref{Im_F1_contact}), accounts only for $<10\%$ in the region shown here.
Second, at non--exceptional values $t \sim \textrm{few} \, M_\pi^2$
the spectral function of the Pauli form factor $\textrm{Im}\, F_2(t)$ 
is several times larger than that of the Dirac form factor 
$\textrm{Im}\, F_1(t)$ (see Fig.~\ref{fig:ft}a). 
However, at values of $t$ close to threshold the pattern reverses,
and $\textrm{Im}\, F_2(t)$ vanishes faster than $\textrm{Im}\, F_1(t)$
(see Fig.~\ref{fig:ft}b). Third, in the near--threshold region
both spectral functions show a rapid change of behavior over a 
range $t - 4 M_\pi^2 \ll M_\pi^2$. This can be traced back to the 
``unnaturally small'' scale $M_\pi^4/M_N^2$ present in the 
distance of the subthreshold singularity from threshold, 
Eq.~(\ref{t_sub_epsilon}), and will be investigated further
in the context of the heavy--baryon expansion in 
Sec.~\ref{subsec:heavy_baryon}.
\subsection{Chiral component of transverse densities}
\label{subsec:chiral_densities}
%
% FIGURE
%
\begin{figure}
\includegraphics[width=.45\textwidth]{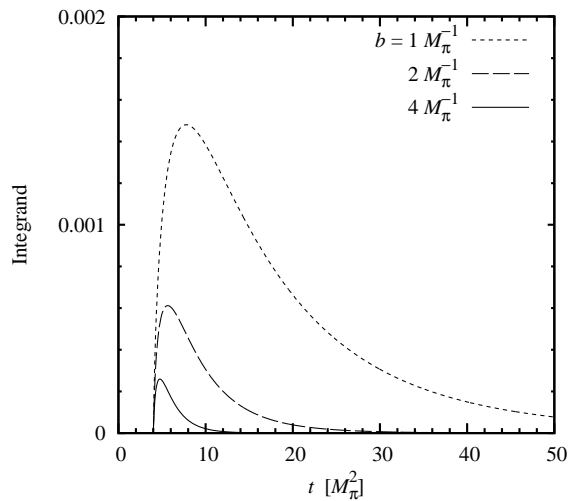}
\caption[]{Integrand of the dispersion integral for the isovector transverse
charge density, Eq.~(\ref{rho_dispersion_preexp}) [with the overall
exponential factor $\exp (-2 M_\pi b)$ extracted], for various
values of $b$.}
\label{fig:weighted}
\end{figure}
Using the leading--order result for the two--pion spectral functions
Eqs.~(\ref{Im_F1_noncontact})--(\ref{x}) we can now calculate the
chiral component of the transverse densities with the help of the
dispersion representation Eq.~(\ref{rho_dispersion}). Before
computing the integral we first want study the numerical distribution 
of strength in the integrand and how it varies when 
changing the distance $b$. Figure~\ref{fig:weighted} shows the
integrand of Eq.~(\ref{rho_dispersion_preexp}),
defining the pre-exponential factor in the charge density $\rho_1(b)$, 
for several values of $b$
in the chiral region $b \sim \textrm{few} \, M_\pi$. One clearly 
sees the exponential suppression of large masses $\sqrt{t}$.
At $b = 1 \, M_\pi^{-1}$ the integral still extends over a broad region
of $t$ including values up to $\sim 50 \, M_\pi^2 \approx 1 \, \textrm{GeV}^2$
where the chiral expansion can not be trusted. 
At $b = 2 \, M_\pi^{-1}$ the region of integration has shrunk
to values $\lesssim 20 \, M_\pi^2$; at $b = 4 \, M_\pi^{-1}$ it
shrinks further to values $\lesssim 10 \, M_\pi^2$. 
This shows quantitatively how the transverse distance $b$ determines 
the range of masses over which the spectral function is integrated.
Similar distributions are found in the integral for the magnetization 
density $\rho_2$. We conclude that the chiral components of the transverse 
densities can reliably be calculated starting from 
$b \gtrsim 2 \, M_\pi^{-1} \approx 3 \, \textrm{fm}$. 
Note that this corresponds to rather large distances on the hadronic scale.

\begin{figure}
\begin{tabular}{ll}
\includegraphics[width=.45\textwidth]{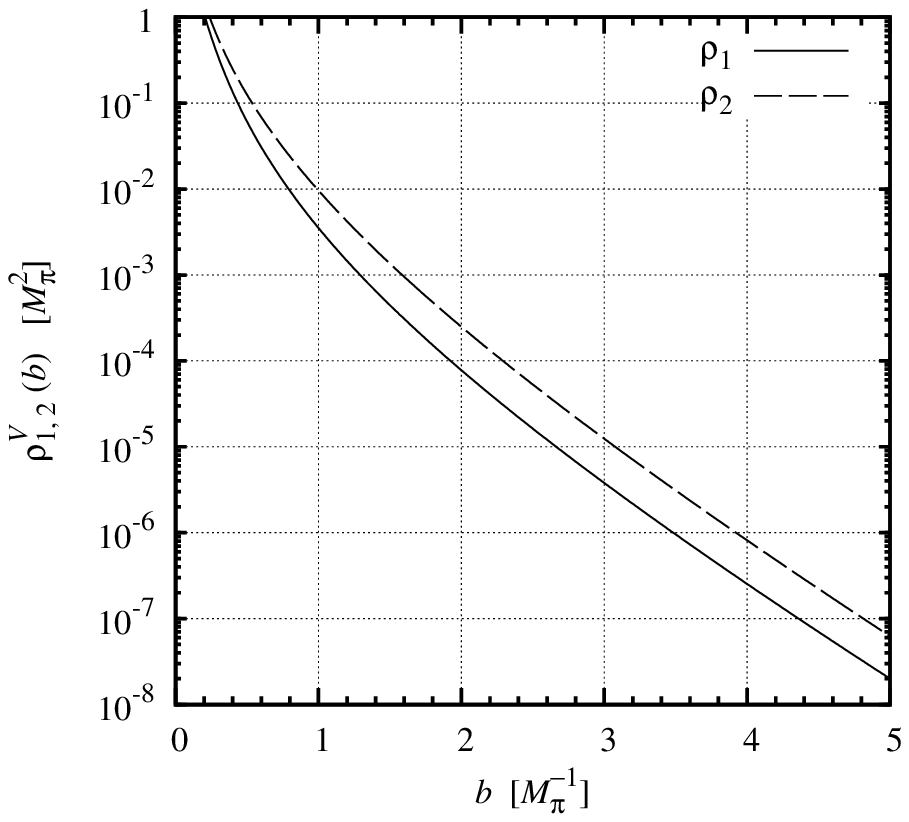}
&
\includegraphics[width=.45\textwidth]{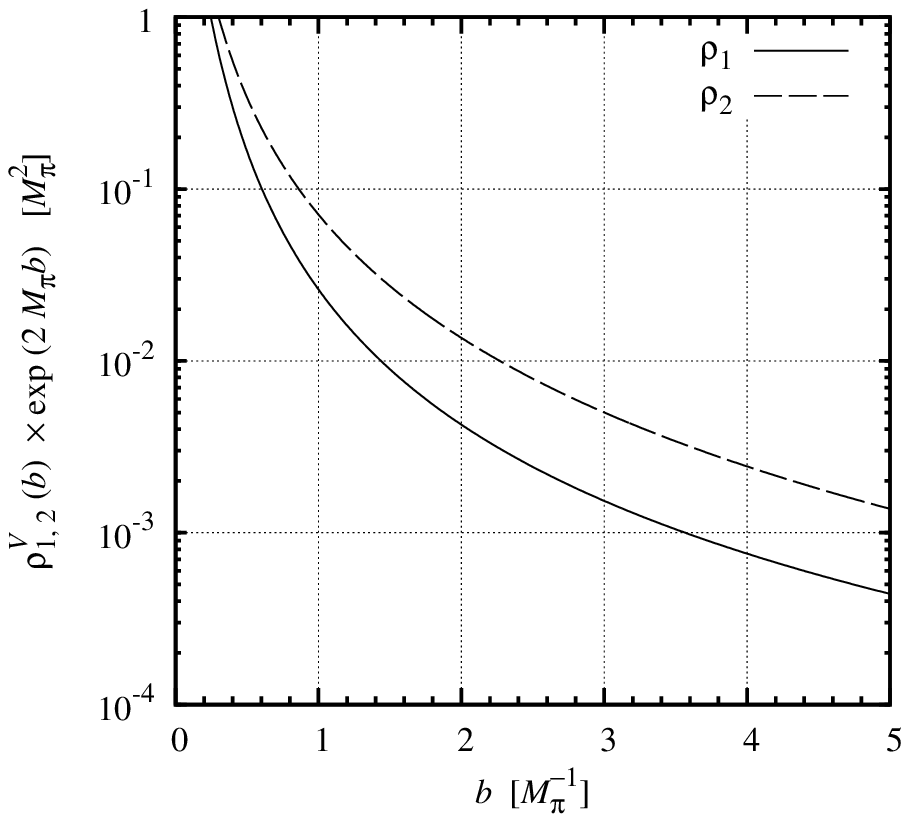}
\\[-2ex]
(a) & (b)
\end{tabular}
\caption[]{Leading--order chiral component of the nucleon's isovector 
transverse charge and magnetization densities
$\rho_{1, 2}^V(b)$, as functions of $b$. Plot (a) shows the true
densities, plot (b) the dependence on $b$ after extracting the 
exponential factor $\exp (- 2 M_\pi b)$ [the functions shown in
this plot are the pre-exponential factors $P_{1, 2}(b)$ in
the general asymptotic expression Eq.~(\ref{large_b_general})]. 
The distance $b$ is given in units
of $M_\pi^{-1}$, the densities in units of $M_\pi^2$.}
\label{fig:rho_b}
\end{figure}

The chiral components of the isovector charge and magnetization densities
obtained from the dispersion integral are shown in Fig.~\ref{fig:rho_b}
as functions of $b$. Fig.~\ref{fig:rho_b}a shows the full densities, 
Fig.~\ref{fig:rho_b}b the dependence on $b$ after extracting the 
exponential factor $\exp (- 2 M_\pi b)$, i.e., the pre-exponential 
factors $P_{1, 2}(b)$ in the general asymptotic expression 
Eq.~(\ref{large_b_general}). One sees that the densities drop
very rapidly with increasing $b$. The decrease is substantially
faster than the exponential fall--off $\sim \exp (-2 M_\pi b)$ 
required by the position of the two--pion threshold 
(see Fig.~\ref{fig:rho_b}b). This behavior is due to the non--trivial 
structure of the $\pi N$ scattering amplitude near threshold, 
particularly the subthreshold nucleon singularity, which brings in 
an additional scale in the form of the distance $M_\pi^4/M_N^2$,
Eq.~(\ref{t_sub}).
%
% FIGURE
%

In our numerical study of the chiral periphery here we have used the 
leading--order chiral result for the spectral functions as given by
Eqs.~(\ref{Im_F1_noncontact})--(\ref{x}). It is known that next--to--leading
order corrections increase the magnitude of the spectral functions
by $\lesssim 40\%$ in the near--threshold region $t < 10 \, M_\pi^2$
\cite{Kaiser:2003qp}. These corrections could easily be incorporated in
our numerical analysis but would not change our overall conclusions. 
In the following study of general properties of the large--$b$ densities
(heavy--baryon expansion, large--$b$ asymptotics) we shall continue to 
use the leading--order approximation, where the spectral functions are 
given by the compact expressions Eqs.~(\ref{Im_F1_noncontact})--(\ref{x}), 
and simple analytic formulae for the densities can be obtained.
\subsection{Heavy--baryon expansion}
\label{subsec:heavy_baryon}
We now consider the heavy--baryon expansion of the chiral component of
the nucleon's transverse densities. This expansion is interesting from
a theoretical point of view, as it separates the unrelated physical scales 
of the nucleon and pion mass and simplifies the interpretation
of the expressions. It is also interesting as a practical tool, as it 
provides us with analytic approximations to the densities that may
be used for numerical evaluation. 

In the context of our study of transverse densities we understand the 
heavy--baryon limit as the limit $M_N \rightarrow \infty$ at \textit{fixed} 
pion mass $M_\pi$ and a \textit{fixed} value of the cutoff mass scale.
Physically, this corresponds to the situation that the basic range of the 
chiral fields carrying charge and magnetization remains fixed, while the 
source producing them becomes heavy. We investigate this regime 
by taking the heavy--baryon limit of the leading--order relativistic 
chiral EFT results for the spectral functions and the resulting densities; 
how the resulting densities could be reproduced or improved in a
suitable variant of heavy--baryon chiral EFT remains an interesting
problem for further study.

The behavior of the spectral function near threshold is dominated by the
subthreshold singularity at a distance $M_\pi^4/M_N^2 = \epsilon^2 M_\pi^2$ 
from the threshold, Eq.~(\ref{t_sub_epsilon}). As shown in 
Sec.~\ref{subsec:parametric}, this distance defines two parametric 
regimes in $t > 4 M_\pi^2$, which are sampled in the dispersion integral 
for the density in different parametric regions of $b$. The heavy--baryon 
limit corresponds to the situation that the subthreshold singularity 
approaches the physical threshold, $\epsilon \rightarrow 0$. This clearly
has different implications in the different parametric regions of $t$
(or $b$), and we have to consider the heavy--baryon limit separately
in the two regions.

\textit{Chiral region.} In the region of distances 
$b = O(M_\pi^{-1})$ the dispersion integral extends over values
of $t$ for which $t - 4 M_\pi^2 = O(M_\pi^2)$, or $k_{\rm cm} = O(M_\pi)$. 
We thus need to carry out the heavy--mass expansion of the 
spectral function for such non--exceptional values of $t$.
The presence of the subthreshold singularity implies that this
expansion is non--uniform and diverges near threshold.
In the chiral result Eqs.~(\ref{Im_F1_noncontact})--(\ref{x})
the heavy--baryon expansion in this region of $t$ corresponds
to the limit
\beq
x \;\; = \;\; \frac{2 k_{\rm cm} \sqrt{P^2}}{A} \;\; = \;\;
\frac{2 \sqrt{t/4 - M_\pi^2} \sqrt{M_N^2 - t/4}}{t/2 - M_\pi^2}
\;\; \rightarrow \;\; \infty ,
\eeq
and we can simplify the expressions by substituting the asymptotic 
series for the inverse tangent function,
\beq
\arctan x \;\;\ = \;\; \frac{\pi}{2} \; - \; \frac{1}{x} \; + \; 
\frac{1}{3 x^3} \;  + O\left(\frac{1}{x^5}\right) 
\hspace{2em} (x \rightarrow \infty).
\eeq
This formally results in a series in inverse powers of $M_N$. However, 
these are accompanied by inverse powers of the CM momentum 
$k_{\rm cm} = \sqrt{t/4 - M_\pi^2}$, which vanishes at threshold.
It causes the series to diverge near threshold, as expected. 
To get approximations to the densities we perform the expansion 
up to the last order at which the terms are still 
integrable over $t$ in the dispersion integral 
Eq.~(\ref{rho_dispersion}), namely terms with inverse 
powers $k_{\rm cm}^{-1}$. Furthermore, when expanding 
Eqs.~(\ref{Im_F1_noncontact})--(\ref{x}) in inverse powers of $M_N$, 
we must also expand the factors $\sqrt{P^2} = \sqrt{M_N^2 - t/4}$ 
in powers of $t/M_N^2$ and consistently take into account the factors of 
$t$ and $M_N^2$ in the expressions. In this way we obtain
\be
\frac{1}{\pi} \, \textrm{Im} \, F_1^V (t)
&=& 
\frac{g_A^2}{(4\pi F_\pi)^2 \sqrt{t}}
\left[ 2 A k_{\rm cm} - \frac{\pi  (2 A^2 + k_{\rm cm}^2 t)}{4 M_N}
+ \frac{A (A^2 + 3 k_{\rm cm}^2 t)}{2 M_N^2 k_{\rm cm}} 
\right.
\nonumber
\\[1ex]
&-& \left. \frac{3\pi t (4 A^2 + k_{\rm cm}^2 t)}{32 \, M_N^3} 
+ O\left(\frac{M_\pi^4}{M_N^4}\right) \right]
\nonumber
\\[1ex]
&+& \frac{2 (1 - g_A^2) k_{\rm cm}^3}
{3 (4\pi F_\pi)^2 \sqrt{t}} ,
\label{F_1_heavy}
\\[3ex]
\frac{1}{\pi} \, \textrm{Im} \, F_2^V (t)
&=& \frac{g_A^2}{(4\pi F_\pi)^2 \sqrt{t}}
\left[ \pi M_N k_{\rm cm}^2 - 4 A k_{\rm cm}
+ \frac{3 \pi  (2 A^2 + k_{\rm cm}^2 t)}{8 M_N}
- \frac{2 A (A^2 + 3 k_{\rm cm}^2 t)}{3 M_N^2 k_{\rm cm}} 
\right.
\nonumber
\\[1ex]
&+& \left. \frac{15\pi t (4 A^2 + k_{\rm cm}^2 t)}{128 \, M_N^3} 
+ O\left(\frac{M_\pi^4}{M_N^4}\right) \right]
\label{F_2_heavy}
\\[3ex]
&& 
\left[ t = O(M_\pi^2), \; A = t/2 - M_\pi^2 = O(M_\pi^2), \;
k_{\rm cm} = \sqrt{t/4 - M_\pi^2} = O(M_\pi) \right] .
\nonumber
\ee
In both expressions the terms $O(M_\pi^4 / M_N^4)$ involve inverse 
powers $k_{\rm cm}^{-3}$, which are no longer integrable over $t$.
In the Dirac spectral function the ``useful'' part of the
series consists of four terms; in the Pauli spectral function 
it consists of five terms. The results Eqs.~(\ref{F_1_heavy}) 
and (\ref{F_2_heavy}) show several interesting features. 
First, one sees that in the chiral region the Pauli spectral 
function is parametrically larger than the Dirac one,
\beq
\frac{\textrm{Im} \, F_2^V (t)}{\textrm{Im} \, F_1^V (t)}
\;\; = \;\; O\left(\frac{M_N}{M_\pi}\right) 
\hspace{2em} [t = O(M_\pi^2)] .
\label{F_ratio_chiral}
\eeq
This enhancement carries over to the densities and implies that
\beq
\frac{\rho_2^V (b)}{\rho_1^V (b)}
\;\; = \;\; O\left(\frac{M_N}{M_\pi}\right) 
\hspace{2em} [b = O(M_\pi^{-1})] .
\label{rho_ratio_chiral}
\eeq
The physical interpretation of this finding will be discussed in
Sec.~\ref{subsec:charge_vs_current}. Second, we see that the successive 
terms in the series in $1/M_N$ have alternating sign. This is a necessary
consequence of the fact that these terms involve positive powers
of $t$ (or $k_{\rm cm}$), which causes them to grow rapidly at large $t$, 
while the spectral functions themselves grow only very modestly 
with increasing $t$ (see Fig.~\ref{fig:ft}a). There are thus large
cancellations between successive terms at larger values of $t$,
limiting the usefulness of the series as a numerical approximation.

%
% FIGURE
%
\begin{figure}
\begin{tabular}{ll}
\includegraphics[width=.45\textwidth]{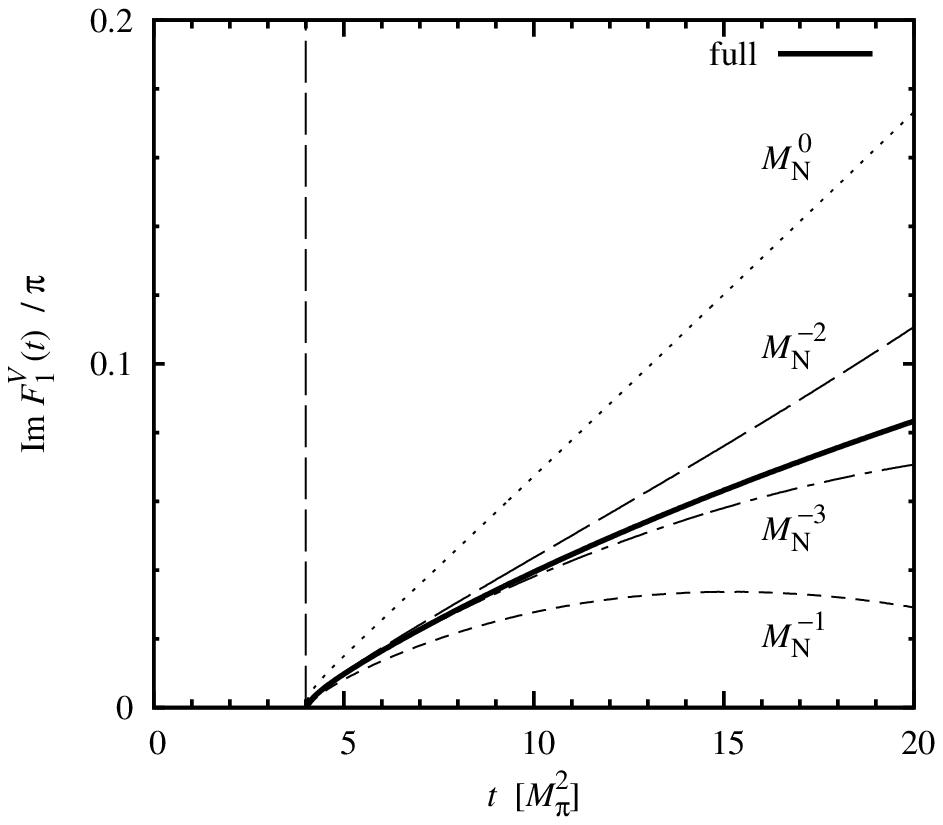}
&
\includegraphics[width=.45\textwidth]{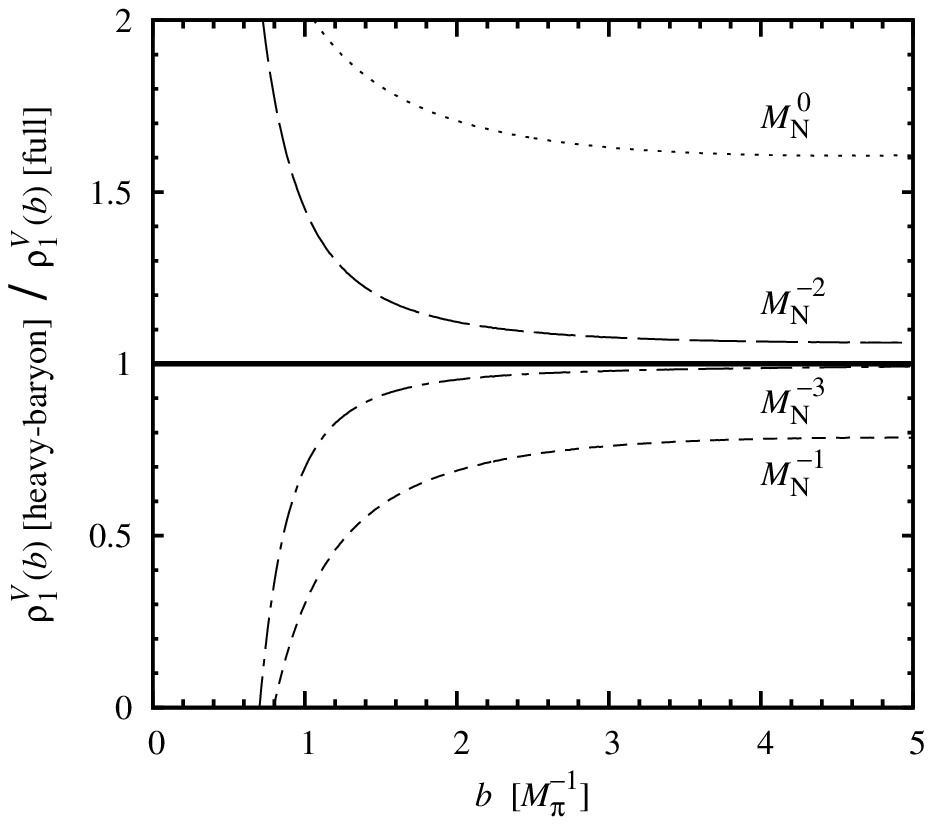}
\\[-2ex]
(a) & (c)
\\[1ex]
\includegraphics[width=.45\textwidth]{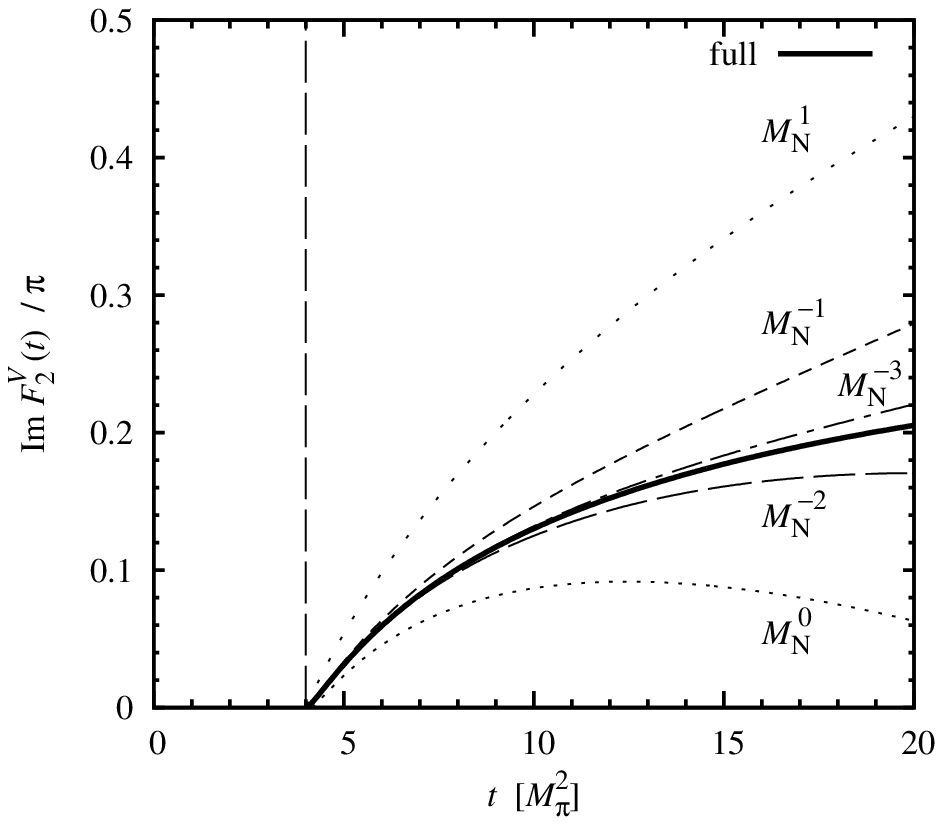}
&
\includegraphics[width=.45\textwidth]{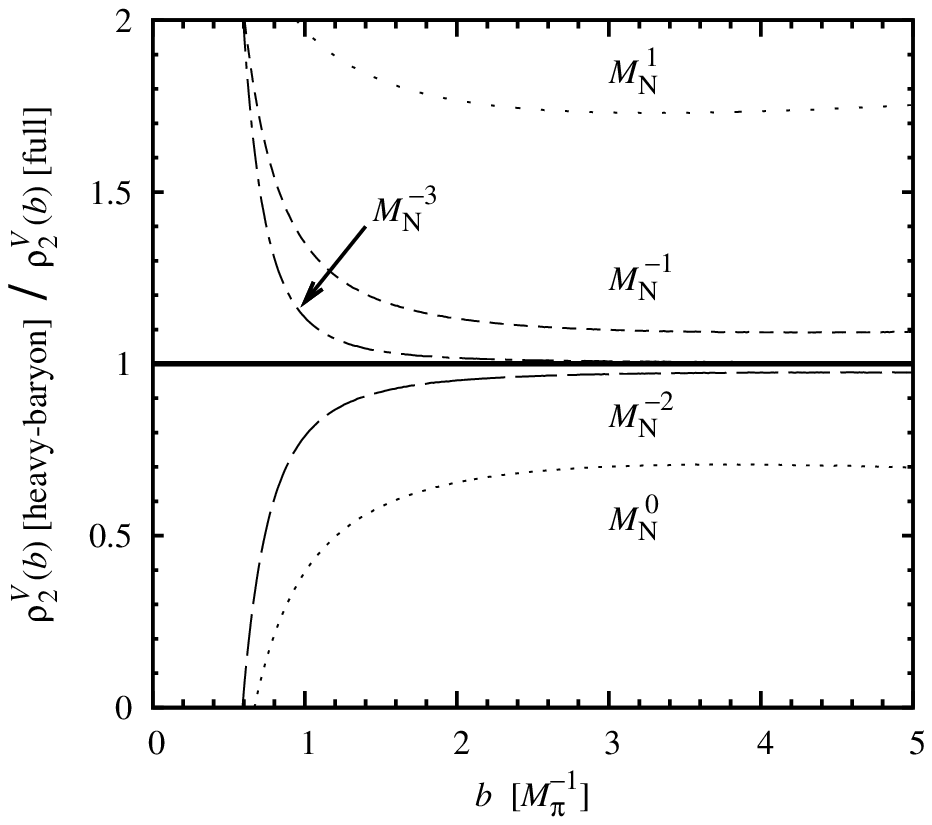}
\\[-2ex]
(b) & (d)
\end{tabular}
\caption[]{(a, b) Heavy--baryon expansion of the leading--order 
isovector Dirac and Pauli spectral functions, 
$\textrm{Im} \, F_{1, 2}^V(t)/\pi$, in the chiral region $t = O(M_\pi^2)$,
Eqs.~(\ref{F_1_heavy}) and (\ref{F_2_heavy}). The thick solid
lines show the full unexpanded expressions,
Eqs.~(\ref{Im_F1_noncontact})--(\ref{x}). The broken lines
show the heavy--baryon series of Eqs.~(\ref{F_1_heavy}) and (\ref{F_2_heavy}),
summed up to (and including) terms of the oder indicated by the labels 
above or below the curves. Starting from order $M_N^{-2}$ the heavy--baryon
series diverges near the threshold $t = 4 M_\pi^2$; the details of
the near--threshold behavior are not visible on the scale at which
the functions are plotted here. \\[0ex] 
(c, d) Heavy--baryon expansion of the leading--order isovector transverse 
charge and magnetization densities, $\rho_{1, 2}^V (b)$, 
in the chiral region $b = O(M_\pi^{-1})$.
The plots show the ratio of the densities obtained with the heavy--baryon
expansion of given order, Eqs.~(\ref{F_1_heavy}) and (\ref{F_2_heavy}),
to those obtained from the full expressions, 
Eqs.~(\ref{Im_F1_noncontact})--(\ref{x}).}
\label{fig:heavy}
\end{figure}
The numerical convergence of the heavy--baryon expansion of the
leading--order chiral component of the spectral functions is
shown in Fig.~\ref{fig:heavy}a and b. The thick solid lines show the full 
expressions Eqs.~(\ref{Im_F1_noncontact})--(\ref{x}); the broken lines
show the series of Eqs.~(\ref{F_1_heavy}) and (\ref{F_2_heavy}),
summed up to (and including) terms of the oder indicated by the 
labels above or below the curves. One sees that the alternating
signs of the successive terms cause the series to converge slowly.
With the 4 terms up to order $M_N^{-3}$ the Dirac spectral function 
is approximated with an accuracy of $\sim 15\%$ over the range 
$t < 20 \, M_\pi^2$, excluding the near--threshold region where the
series diverges (see Fig.~\ref{fig:heavy}a). The Pauli spectral function 
is approximated by $\sim 10\%$ by the 5 terms up to order $M_N^{-3}$ over 
the same region (see Fig.~\ref{fig:heavy}b).

The heavy--baryon expansion of the transverse densities in the chiral
region $b = O(M_\pi^{-1})$ is obtained by substituting the series
Eqs.~(\ref{F_1_heavy}) and (\ref{F_2_heavy}) into the dispersion 
integral Eq.~(\ref{rho_dispersion}). Thanks to the exponential convergence 
of the integral at large $t$ the series for the spectral function can be 
integrated over $t$ term--by--term; the only restriction comes from the
divergence of the expansion near threshold $t = 4 M_\pi^2$, which limits 
the order of the expansion in $1/M_N$, as explained above. The resulting 
contributions to the density can be expressed in terms of standard 
integrals over the modified Bessel function and computed analytically 
(see Appendix~\ref{app:heavy}). The quality of the numerical approximation
to the densities is shown in Fig.~\ref{fig:heavy}c and d. The plots show
the relative accuracy of the approximation; i.e., the
ratio of the heavy--baryon expansion of the density (up to a given order) 
to the full result obtained by integrating the unexpanded expressions
Eqs.~(\ref{Im_F1_noncontact})--(\ref{x}). With the maximum number of 
terms up to order $M_N^{-3}$ an approximation of $<20\%$ ($< 15 \%$)
accuracy is achieved for $\rho_1$ ($\rho_2$) at all distances 
$b > 1\, M_\pi^{-1}$; the accuracy improves significantly 
at distances $b \sim 2-3 \, M_\pi^{-1}$. Note that the heavy--baryon 
expansion breaks down both at small $b$, because of the increasing 
sensitivity to large $t$, 
where the expansion for the spectral function converges poorly; 
and at large $b$, where values of $t$ close to threshold $t = 4 M_\pi^2$
become important (see below). Still, it provides a very decent numerical 
approximation to the density over most of the practically relevant range of 
distances $b = \textrm{few} \, M_\pi^{-1}$.

\textit{Molecular region.} In the region of anomalously large distances 
$b = O(\epsilon^{-2} M_\pi^{-1})$ the dispersion integral extends over
the near--threshold region $t - 4 M_\pi^2 = O(\epsilon^2 M_\pi^2)$,
or $k_{\rm cm} = O(\epsilon M_\pi)$
[$\epsilon = M_\pi / M_N$, cf.\ Eq.~(\ref{epsilon})]. 
In this region the spectral function
is under the influence of the subthreshold singularity at a distance
$\epsilon^2 M_\pi^2$ from threshold and exhibits a non--trivial variation 
over the relevant $t$--range. In the heavy--baryon limit 
\beq
\epsilon \;\; \rightarrow \;\; 0 ,
\eeq
so that the width of the relevant $t$--range becomes small. When carrying 
out the heavy--baryon expansion we must distinguish between ``slow'' 
functions of $t$, which vary only over the range $t - 4 M_\pi^2 \sim M_\pi^2$,
and ``fast'' functions, which exhibit a variation of order unity over
the range $t - 4 M_\pi^2 \sim \epsilon M_\pi^2$: the former can be expanded
around the threshold, $t = 4 M_\pi^2$, while the latter must be
retained as live functions in the dispersion integral. In this sense
we can replace in Eqs.~(\ref{Im_F1_noncontact})--(\ref{x}) the slow functions 
by
\be
P^2 \; = \; {\textstyle\sqrt{M_N^2 - t/4}} &\rightarrow& M_N,
\\[.5ex]
\sqrt{t} &\rightarrow& 2 M_\pi ,
\\[.5ex]
t/2 - M_\pi^2 &\rightarrow& M_\pi^2 ,
\ee
while $x \equiv x(t)$ is a fast function and becomes
\beq
x(t) 
\;\; \rightarrow \;\; \frac{2 k_{\rm cm}}{\epsilon M_\pi} \;\;  \equiv \;\; 
x_0 (t)
\hspace{2em} [ t - 4 M_\pi^2 = O(\epsilon^2 M_\pi^2), \, 
k_{\rm cm} = O(\epsilon M_\pi) ] .
\label{x_0}
\eeq
Note that $x_0 = O(1)$ in the region considered here. 
To leading order in $\epsilon$ the spectral functions then become
\be
\frac{1}{\pi} \, \textrm{Im} \, F_1^V (t)
&=& \frac{g_A^2 M_\pi^2 \, \epsilon}{2 (4\pi F_\pi)^2} 
( x_0 - \arctan x_0 ) ,
\label{Im_F1_molecular}
\\
\frac{1}{\pi} \, \textrm{Im} \, F_2^V (t)
&=& \frac{g_A^2 M_\pi^2 \, \epsilon}{4 (4\pi F_\pi)^2}
\left[ (x_0^2 + 3) \arctan x_0 - 3 x_0 \right] ,
\label{Im_F2_molecular}
\ee
where $x_0 \equiv x_0(t)$. The contact 
term in the Dirac spectral function, Eq.~(\ref{Im_F1_contact}),
is of order $\epsilon^3$ and can be neglected in this region.
Note that the Dirac and Pauli spectral functions are of the same
parametric order in the near--threshold region considered here;
in contrast to their behavior in the chiral region $t - 4 M_\pi^2 = 
O(M_\pi^2)$, Eq.~(\ref{F_ratio_chiral}), where the Pauli spectral
function is parametrically larger. This implies that at
distances $b = O(\epsilon^{-2} M_\pi^{-1})$ the densities $\rho_1(b)$ 
and $\rho_2(b)$ are of the same order and will be discussed further
in Sec.~\ref{subsec:charge_vs_current}.

When calculating the dispersion integral for the densities, 
Eq.~(\ref{rho_dispersion}), we note that the region of molecular 
distances $b = O(\epsilon^{-2} M_\pi^{-1})$ corresponds to values 
\beq
\sqrt{t} b \;\; = \;\; O(\epsilon^{-2}) ,
\eeq
where the modified Bessel function can be replaced by its
leading asymptotic form for large arguments, Eq.~(\ref{K0_asymptotic});
higher inverse powers of $\sqrt{t} b$ in the pre-exponential factor
of the modified Bessel function would give rise to higher powers of 
$\epsilon$ upon integration
over $t$. Furthermore, in leading order in $\epsilon$ 
we can replace the slowly varying function $\sqrt{t}$ multiplying
the exponential by its value at threshold, $2 M_\pi$
(in the exponent, where $\sqrt{t}$ is multiplied by $b$, we have to
retain it as is). 
It is convenient to use the CM momentum $k_{\rm cm}$ 
as integration variable, in terms of which $t = 4 k_{\rm cm}^2 + M_\pi^2$.
In leading order of $\epsilon$ the dispersion integral 
Eq.~(\ref{rho_dispersion}) then becomes
\be
\rho_{1, 2}(b) &=& 
\frac{1}{\sqrt{16 \pi M_\pi b}} \;
\int\limits_{4M_\pi^2}^\infty dt \; e^{-\sqrt{t} b}
\left[ 1 + O(\epsilon) \right] \, 
\ldots
\\
&=& \frac{2}{\sqrt{\pi M_\pi b}} \;
\int\limits_{0}^\infty dk_{\rm cm} \; k_{\rm cm} \; 
\exp \left[ -2 M_\pi b \left( 1 + \frac{k_{\rm cm}^2}{M_\pi^2} \right)^{1/2} 
\right]
\, \ldots
\\
&=& \frac{2}{\sqrt{\pi M_\pi b}} \; 
\int\limits_{0}^\infty dk_{\rm cm} \; k_{\rm cm} \; 
\exp \left[ - 2 M_\pi b - \frac{b k_{\rm cm}^2}{M_\pi} + O(\epsilon^2 ) 
\right]
\, \ldots
\ee
where the ellipsis $\ldots$ stands for the simplified spectral densities
Eqs.~(\ref{Im_F1_molecular}) and (\ref{Im_F2_molecular}), and we have
schematically indicated higher--order terms in $\epsilon$ that were
subsequently neglected. In the last step we have expanded the
square root in the exponent in powers of $k_{\rm cm}/M_\pi$ and
retained only the first two terms; the next term $\sim b k_{\rm cm}^4$
would be of order $\epsilon^2$ and modify the pre-exponential factor
in the same way as the other terms neglected previously. 
Our reasoning here follows the logic of the saddle point approximation 
for exponential integrals with a large parameter in the exponent.
The resulting Gaussian integral over the CM momentum is readily
computed, and we obtain
\be
\rho_1 (b) &=& 
\frac{g_A^2 M_\pi^4 \; e^{-2 M_\pi b}}{2 \, (4\pi F_\pi)^2 \, (M_\pi b)^2} 
\left[ 1 - e^\lambda \, \lambda^{1/2} \, 
\Gamma ({\textstyle\frac{1}{2}}, \lambda ) \right] ,
\label{rho_1_molecular}
\\[2ex]
\rho_2 (b) &=& 
\frac{g_A^2 M_\pi^4 \; e^{-2 M_\pi b}}{2 \, (4\pi F_\pi)^2 \, (M_\pi b)^2} 
\left[ \left(\frac{1}{2 \lambda} + 1 \right) e^\lambda \, \lambda^{1/2} \, 
\Gamma ({\textstyle\frac{1}{2}}, \lambda ) - 1 \right]
\label{rho_2_molecular}
\\[2ex]
&& \left[ \lambda \, \equiv \, \epsilon^2 M_\pi b/4 \, = \, O(1), 
\; b = O(\epsilon^{-2} M_\pi^{-1}) \right] .
\ee
Here $\Gamma ({\textstyle\frac{1}{2}}, \lambda )$ denotes the incomplete
Gamma function. Equations~(\ref{rho_1_molecular}) and (\ref{rho_2_molecular})
describe the transverse charge and magnetization densities at molecular 
distances $b = O(\epsilon^{-2} M_\pi^{-1})$ to leading order in $\epsilon$
and have several noteworthy properties. First, since the densities
are functions of the variable 
$\lambda = \epsilon^2 M_\pi b/4 = M_\pi^3 b/(4 M_N^2)$, one sees
explicitly that the limit $b \rightarrow \infty$ and the 
heavy--baryon expansion $M_N \gg M_\pi$ do not commute,
as noted already in the general parametric analysis of 
Sec.~\ref{subsec:parametric}. Second, substituting the asymptotic 
expansion of the incomplete Gamma function,
\beq
e^\lambda \, \lambda^{1/2} \, \Gamma ({\textstyle\frac{1}{2}}, \lambda )
\;\; \sim \;\; 1 - \frac{1}{2 \lambda} + \frac{3}{4 \lambda^2} 
+ O(\lambda^{-3}),
\eeq
we obtain the asymptotic behavior of the leading--order chiral 
component of the charge and magnetization densities at large $b$ as
\be
\rho_1 (b) &\sim& \frac{g_A^2 M_N^2 M_\pi^2}{(4\pi F_\pi)^2 (M_\pi b)^3}
\, e^{-2 M_\pi b} ,
\label{rho_1_asymp}
\\[2ex]
\rho_2 (b) &\sim& \frac{4 g_A^2 M_N^4}{(4\pi F_\pi)^2 (M_\pi b)^4}
\, e^{-2 M_\pi b} .
\label{rho_2_asymp}
\ee
The spin--dependent current density Eq.~(\ref{rho_2_tilde_def}) 
at this accuracy is obtained by differentiating only the fast--varying 
exponential factor,
\be
\widetilde\rho_2 (b) &=& \frac{1}{2 M_N} \, 
\frac{\partial \rho_2}{\partial b} 
\;\; \sim \;\; - \frac{4 g_A^2 M_N^3 M_\pi}{(4\pi F_\pi)^2 
(M_\pi b)^4} \, e^{-2 M_\pi b} .
\label{rho_2_tilde_asymp}
\ee
Equations~(\ref{rho_1_asymp})--(\ref{rho_2_tilde_asymp})
are the asymptotic densities one would obtain by direct expansion
of the spectral functions Eqs.~(\ref{Im_F1_noncontact})--(\ref{x})
at threshold in powers of the CM momentum $k_{\rm cm}$
\footnote{The contact term in the Dirac spectral function,
Eq.~(\ref{Im_F1_contact}), gives a term of the same form as
Eq.~(\ref{rho_1_asymp}), but with a coefficient that is suppressed
by a factor $\epsilon^2 = M_\pi^2/ M_N^2$.},
\be
\frac{1}{\pi} \, \textrm{Im} \, F_1^V (t)
&\sim& \frac{4 g_A^2 M_N^2 \, k_{\rm cm}^3}{3 (4\pi F_\pi )^2 M_\pi^3} ,
\\[1ex]
\frac{1}{\pi} \, \textrm{Im} \, F_2^V (t)
&\sim& \frac{32 g_A^2 M_N^4 \, k_{\rm cm}^5}{15 (4\pi F_\pi )^2 M_\pi^7} 
\hspace{2em} (t \rightarrow 4 M_\pi^2) .
\label{F_2_threshold}
\ee
It is curious to
note that this leading asymptotic form would approximate the density
only in the region $\lambda \gg 1$, which corresponds to distances
\beq
b \;\; \gg \;\; \frac{4}{\epsilon^2 M_\pi} \;\; = \;\; 
\frac{4 M_N^2}{M_\pi^3} \;\; \approx \;\; 250 \, \textrm{fm}. 
\eeq
This shows how misleading it would be to infer the asymptotic behavior 
of the density from just the leading threshold behavior of the 
spectral function. We stress again that all densities discussed here 
are exponentially suppressed by the factor $\exp (-2M_\pi b)$, 
and that their behavior in the molecular region is mainly
of mathematical interest.

Our study of the molecular region here is limited to inspection of 
the leading--order chiral EFT results. We do not claim that 
Eqs.~(\ref{rho_1_asymp}) and (\ref{rho_2_asymp}) represent the 
``true'' asymptotic behavior of the transverse densities.
In fact, it is known that higher--order corrections to the 
Pauli spectral function in relativistic chiral EFT change its 
power behavior near threshold to (in our notation) \cite{Kubis:2000zd}
\beq
\frac{1}{\pi} \, \textrm{Im} \, F_2^V (t)
\;\; \sim \;\; \frac{4 M_N \, c_4 \, k_{\rm cm}^3}{3 (4\pi F_\pi )^2 M_\pi} 
\hspace{2em} (t \rightarrow 4 M_\pi^2) ,
\eeq
where $c_4 \approx 3.4\, \textrm{GeV}^{-1}$ is a low--energy constant 
in the second--order relativistic chiral Lagrangian, whose value is
determined from $\pi N$ scattering data \cite{Fettes:1998ud}. 
This is qualitatively different from the $k_{\rm cm}^5$ behavior
of the leading--order result, Eq.~(\ref{F_2_threshold}). 
It indicates that the chiral expansion converges non--uniformly at
molecular distances, and that resummation may be necessary
to obtain the true asymptotic behavior in this parametric region.
A resummation of the logarithmic terms of the chiral expansion 
was performed in Refs.~\cite{Kivel:2007jj,Kivel:2008ry,Perevalova:2011qi}
and shown to qualitatively change the large--$b$ behavior of the
pion GPD at small $x$ compared to fixed--order calculations
(see the discussion in Sec.~\ref{sec:summary} below). 
We emphasize that the existence of the molecular regime as such 
follows from the general analytic structure of the form factor near 
threshold (see Sec.~\ref{subsec:parametric}) is not conditional
on the convergence of the chiral expansion. Note also that the convergence
issue discussed here affects only the molecular region; in the chiral 
region the effect of higher--order corrections the spectral functions 
is only quantitative \cite{Kaiser:2003qp}, and it is legitimate to use
the leading--order approximation to study the densities.

\textit{Uniform approximation.} The spectral functions obtained from 
leading--order relativistic chiral EFT, 
Eqs.~(\ref{Im_F1_noncontact})--(\ref{x}),
embody the full analytic structure of the form factor near the two--pion 
threshold, as governed by the two scales $M_\pi^2$ and $\epsilon^2 M_\pi^2$.
While a systematic expansion in $\epsilon$ can be performed in the 
chiral region of $t$ (see above), it converges non--uniformly and is 
of limited value for practical purposes.
Following general arguments presented in Ref.~\cite{Becher:1999he}, 
a more useful uniform approximation to the spectral functions can be 
obtained by neglecting in Eqs.~(\ref{Im_F1_noncontact})--(\ref{x}) terms 
of order $t/M_N^2$, while leaving the position of the subthreshold 
singularity unchanged. This amounts to replacing 
\beq
\sqrt{P^2} \;\; = \;\; {\textstyle\sqrt{M_N^2 - t/4}}
\;\; \rightarrow \;\; M_N 
\eeq
and dropping the terms with factors $t/M_N^2$ in Eq.~(\ref{Im_F1_noncontact}).
With these simplifications the spectral functions become
\be
\frac{1}{\pi} \, \textrm{Im} \, F_1^V (t)
&=& 
\frac{g_A^2 (t/2 - M_\pi^2)^2}{(4\pi F_\pi)^2 M_N \sqrt{t}}
(x_1 - \arctan x_1) \;\; + \;\; 
\frac{2 (1 - g_A^2) k_{\rm cm}^3}{3 (4\pi F_\pi)^2 \sqrt{t}} ,
\label{Im_F1_simplified}
\\[2ex]
\frac{1}{\pi} \, \textrm{Im} \, F_2^V (t)
&=& \frac{g_A^2 (t/2 - M_\pi^2)^2}{2 (4\pi F_\pi)^2 M_N
\sqrt{t}}
\left[ (x_1^2 + 3) \arctan x_1 - 3 x_1 \right] ,
\label{Im_F2_simplified}
\\[2ex]
x_1 &\equiv & x_1(t) \;\; \equiv \;\; 
\frac{2 M_N k_{\rm cm}}{t/2 - M_\pi^2}
\;\; = \;\;
\frac{2 M_N \sqrt{t/4 - M_\pi^2}}{t/2 - M_\pi^2} .
\label{x_1}
\ee
Equations~(\ref{Im_F1_simplified})--(\ref{x_1}) approximate the 
full leading--order expressions Eqs.~(\ref{Im_F1_noncontact})--(\ref{x})
with an accuracy of $< 15\%$ for all $4 \, M_\pi^2 < t < 20 \, M_\pi^2$,
while fully preserving the analytic structure near threshold.
They summarize in compact form the entire information contained 
in the leading--order chiral component of the isovector spectral
functions. The uniform approximation to the Dirac spectral function,
Eqs.~(\ref{Im_F1_simplified}) and (\ref{x_1}), was used in the
numerical studies of the chiral component of the transverse charge 
density in Refs.~\cite{Strikman:2010pu,Miller:2011du}.
\subsection{Charge vs.\ magnetization density}
\label{subsec:charge_vs_current}
So far we studied the chiral components of the transverse charge
and anomalous magnetization densities, $\rho_1(b)$ and $\rho_2(b)$.
It is interesting to explore what these results imply for the
spin--independent and --dependent nucleon matrix elements of the 
plus component of the vector current operator, whose relation to the 
transverse densities is described in Sec.~\ref{subsec:definition}. 
This excursion leads us to an interesting positivity property of the 
chiral component of the transverse densities. It also suggests that the 
main results of our dispersion--based calculation of the peripheral 
transverse densities can be understood in a simple quantum--mechanical 
picture of $\pi N$ configurations in the nucleon's light--cone wave 
function in the rest frame. The details of this picture will be presented
in a subsequent article, where we study the chiral processes in
time--ordered perturbation theory \cite{inprep}.

Following Sec.~\ref{subsec:definition}, the expectation values of 
the light--cone plus component of the vector current, in a
nucleon state polarized transversely along the $y$--axis, are
\be
\langle J^+ (\bm{b}) \rangle_{\text{\scriptsize spin-indep.}}
&=& \rho_1(b) ,
\label{J_plus_indep}
\\[2ex]
\langle J^+ (\bm{b}) \rangle_{\text{\scriptsize spin-dep.}}
&=& (2 S^y) \, \cos\phi \, \widetilde\rho_2(b) ,
\label{J_plus_dep}
\ee
where $\widetilde \rho_2$ is defined in Eq.~(\ref{rho_2_tilde_def}). 
The result of the heavy--baryon expansion of the densities $\rho_1$ and 
$\rho_2$, Eq.~(\ref{rho_ratio_chiral}), now implies that
\be
\frac{\widetilde\rho_2^V (b)}{\rho_1^V (b)} &=& 
O\left(\frac{M_\pi^0}{M_N^0}\right)
\;\;  \equiv \;\; O(1)
\hspace{2em} [b = O(M_\pi^{-1})].
\label{rho_2_tilde_order}
\ee
Thus the spin--independent and --dependent parts of the current
expectation value are of the same order in the chiral expansion
at non--exceptional angles. It therefore seems natural to focus
on the function $\widetilde\rho_2$ rather than $\rho_2$ when 
discussing the chiral periphery.
%
% FIGURE
%
\begin{figure}
\begin{tabular}{ll}
\parbox[c]{.45\textwidth}
{\includegraphics[width=.45\textwidth]{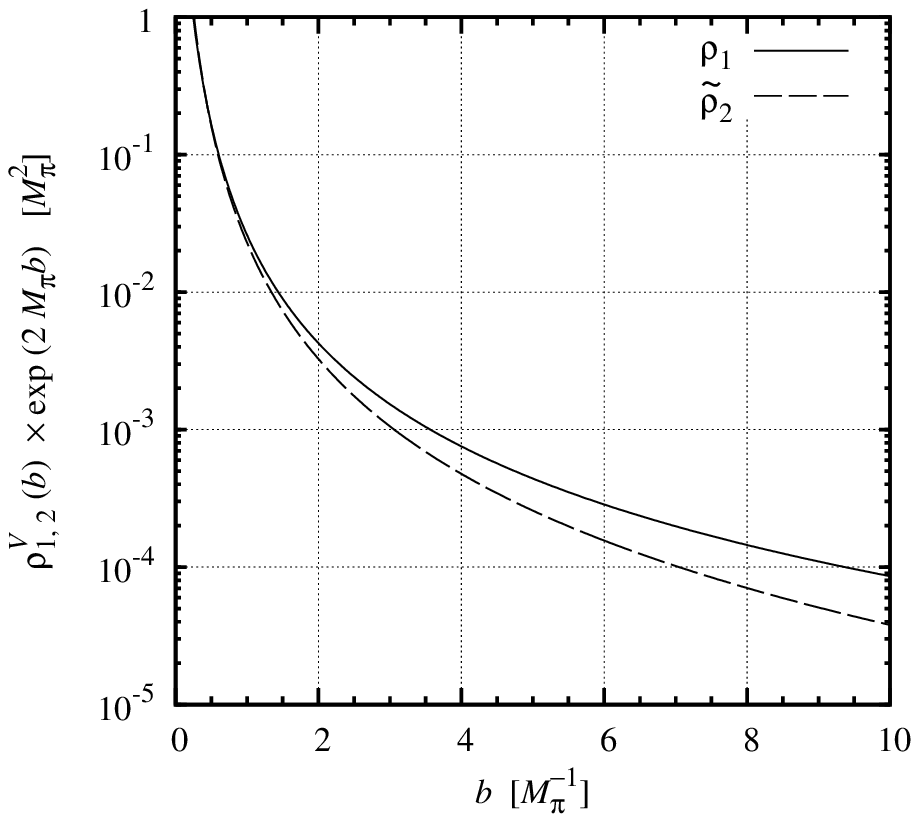}}
\hspace{.1\textwidth}
&
\parbox[c]{.32\textwidth}{\includegraphics[width=.32\textwidth]{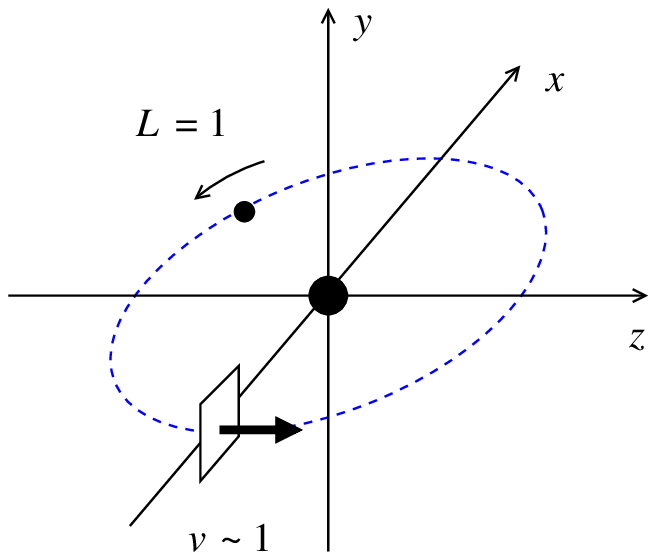}}
\\[-2ex]
(a) & (b)
\end{tabular}
\caption[]{(a) Comparison of the leading--order chiral component of the 
nucleon's isovector spin--independent current density $\rho_1^V(b)$ 
(solid line) and spin--dependent current density 
$\widetilde \rho_2^V(b)$ (dashed line). 
The plot shows the densities
with the exponential factor $\exp (- 2 M_\pi b)$ extracted
[the functions plotted correspond to the pre-exponential factor in 
Eq.~(\ref{large_b_general})]. The distance $b$ is given in units
of $M_\pi^{-1}$, the densities in units of $M_\pi^2$. (b) Mechanical
picture explaining the relation of the peripheral densities (details
see text). The nucleon state is polarized along the $y$--axis with 
$S^y = +1/2$. The peripheral densities are generated by components
of the light--cone wave function involving a peripheral pion with 
$L = 1$. At distances $b = O(M_\pi^{-1})$ the velocity of the pion 
is $v = O(1)$, and the current and charge density are of the same
order.}
\label{fig:rho_current}
\end{figure}

The numerical results for the densities $\rho_1(b)$ and $\widetilde\rho_2(b)$,
obtained from the leading--order chiral EFT result for the two--pion 
spectral functions, are compared in Fig.~\ref{fig:rho_current}a. 
One sees that at all distances the chiral spin--dependent current 
density $\widetilde\rho_2^V (b)$ is smaller in absolute value than 
the spin--independent density $\rho_1^V(b)$,
\beq
|\widetilde\rho_2^V (b)| \;\; \leq \;\; \rho_1^V(b) .
\label{inequality}
\eeq
At smaller (but parametrically still ``chiral'') distances 
$b \lesssim 2 \, M_\pi^{-1}$ they become practically equal in absolute value. 
The inequality Eq.~(\ref{inequality}) implies that the chiral 
result for the total expectation value of the plus component 
of the isovector current, given by the sum 
of Eqs.~(\ref{J_plus_indep}) and (\ref{J_plus_dep})
[cf.\ Eq.~(\ref{J_plus_total})], is positive
\be
\langle J^+ (\bm{b}) \rangle
&=& \rho_1^V (b) \;\; + \;\; (2 S^y) \, \cos\phi \, \widetilde\rho_2^V (b) 
\;\; \geq \;\; 0 .
\label{positivity}
\ee
We stress that Eqs.~(\ref{inequality}) and (\ref{positivity}) are numerical 
statements based on inspection of the leading--order chiral results,
and that we cannot claim that they hold under more general circumstances.

The two observations, Eqs.~(\ref{rho_2_tilde_order}) and (\ref{positivity}),
can be explained in a simple quantum--mechanical picture of 
peripheral nucleon structure.
Consider a nucleon with transverse spin $S^y = +1/2$ in the rest 
frame. The chiral component of the transverse densities at distances
$b = O(M_\pi^{-1})$ arises from virtual processes in which the nucleon 
fluctuates into a $\pi N$ system through the effective chiral interactions. 
Because pion emission flips the nucleon spin, the relevant configurations 
in the wave function have the pion moving with orbital angular momentum 
$L = 1$ (see Fig.~\ref{fig:rho_current}b). The momentum of the 
peripheral pion is $k_\pi = O(M_\pi)$, whence its velocity is 
$v = k_\pi / M_\pi = O(1)$; i.e., 
the motion of the pion is essentially relativistic.
Since the $\pi N$ interaction is pointlike on the scale $M_\pi^{-1}$,
we can regard the peripheral $\pi N$ system as non--interacting, and
use this simple model to infer the expectation value of the current
operator, including its light--cone component $J^+$. The spin--independent
part of $J^+$ is given by the charge density $J^0$ in the rest frame, 
the spin--dependent part by the current density $J^z$. By simple
geometry (see Fig.~\ref{fig:rho_current}b) the ratio of current to 
charge density in the system is given by the pion velocity,
and one obtains
\beq
\frac{|\langle J^+ (\bm{b}) \rangle_{\text{\scriptsize spin-dep.}}|}
{\langle J^+ (\bm{b}) \rangle_{\text{\scriptsize spin-indep.}}}
\;\; = \;\; \frac{|J^z|}{J^0} \;\; = \;\; v \;\; = \;\; O(1) 
\hspace{2em} [b = O(M_\pi^{-1})] ,
\label{ratio_mechanical}
\eeq
which naturally explains Eq.~(\ref{rho_2_tilde_order}). The positivity
condition Eq.~(\ref{positivity}) can be accounted for in a similar manner. 
To the extent that the peripheral $\pi N$ system can be regarded as 
non--interacting, the current at $b = O(M_\pi^{-1})$ should be 
proportional to the current produced by a free charged pion with 
four--momentum $k$, which is
\beq
\langle \pi (\bm{k}) | J^+ | \pi (\bm{k}) \rangle \;\ = \;\; 2 k^+
\;\ > \; 0 ,
\eeq
where the last relation is obtained because 
$k^0 = \sqrt{(k^z)^2 + \bm{k}_T^2 + M_\pi^2} \geq |k^z|$ for a free particle.
We emphasize that the mechanical picture presented here is just a 
heuristic tool, and that several aspects (role of relativity, 
non--interaction in the periphery) need to be clarified. A rigorous
particle--based interpretation of the peripheral densities can be
developed in the context of a time--ordered description of chiral processes 
and will be described elsewhere \cite{inprep}.

In the molecular region, $b = O(M_N^2/M_\pi^3)$, the asymptotic
behavior of the densities $\rho_1$ and $\rho_2$ is described by
Eqs.~(\ref{rho_1_molecular}) and (\ref{rho_2_molecular}). One can
easily see that this implies that the current matrix elements 
behave as
\be
\frac{\widetilde\rho_2^V (b)}{\rho_1^V (b)} &=& O\left(
\frac{M_\pi}{M_N} \right)
\hspace{2em} [b = O(M_N^2/M_\pi^3)];
\label{rho_2_tilde_order_molecular}
\ee
see also Eqs.~(\ref{rho_1_asymp}) and (\ref{rho_2_tilde_asymp}).
In this parametric region the spin--dependent part of the current 
matrix element is suppressed relative to the spin--independent one.
The numerical results of Fig.~\ref{fig:rho_current}a show that the 
ratio of the densities indeed decreases at distances
$b \gg M_\pi^{-1}$. This behavior again can be understood in the
mechanical picture of Fig.~\ref{fig:rho_current}b. The analysis of
Sec.~\ref{subsec:parametric} shows that in this region of distances 
the pion velocity becomes parametrically small, $v = O(M_\pi / M_N)$,
cf.\ Eq.~(\ref{k_cm_molecular}); using this result in the
model estimate Eq.~(\ref{ratio_mechanical}) one obtains exactly 
the parametric suppression Eq.~(\ref{rho_2_tilde_order_molecular})
\footnote{The molecular region corresponds to the extreme 
classically--forbidden range of motion of the peripheral pion in the
nucleon, as can be seen by the exponential suppression of the 
probability. A proper quantum--mechanical treatment of the motion
in this region can be developed using light--front wave functions
\cite{inprep}. It is interesting that the classical picture 
provides correct parametric estimates even though the motion is
essentially quantum--mechanical.}.
\subsection{Contact terms and pseudoscalar $\pi N$ coupling}
\label{subsec:contact}
The presence of a $\pi\pi NN$ contact term in the leading--order chiral 
EFT results for the isovector Dirac spectral function and transverse 
charge density is a matter that merits separate discussion. In fact, 
the compact expressions obtained in Sec.~\ref{subsec:chiral_densities}
shed new light on the physical interpretation of this structure.

The pion--nucleon contact couplings in the chiral Lagrangian 
Eq.~(\ref{chiral_lagrangian_orig}) encode the effect of internal structure 
of the nucleon which is not resolved by pions with moment $k = O(M_\pi)$.
In the isovector Dirac spectral function, the $\pi\pi NN$ contact 
term in the Lagrangian, exhibited explicitly in Eq.~(\ref{lag}),
induces a chiral process in which the two pions couple to the nucleon
locally on the scale $O(M_\pi^{-1})$, described by diagram 
Fig.~\ref{fig:diag}b. A local contribution of the same structure arises 
also from diagram Fig.~\ref{fig:diag}a, as a term in which the numerator
cancels the pole of the intermediate nucleon propagator. This results in
a net contact term with coefficient [cf.\ Eq.~(\ref{contact_combined})]
\beq
\frac{1 - g_A^2}{F_\pi^2}
\label{combination_contact}
\eeq
in the Dirac spectral function and transverse charge density. The appearance 
of the combination $1 - g_A^2$ here is very natural. For a pointlike 
(i.e., structureless) Dirac fermion the axial coupling is unity, 
$g_A = 1$, as can be seen trivially by computing the matrix element of 
the axial vector current between free--particle states. The combination 
Eq.~(\ref{combination_contact}) thus vanishes for a pointlike particle
and reflects the ``compositeness'' of the nucleon.
It would be interesting to explore the connection between $g_A > 1$ and 
the $\pi\pi NN$ contact coupling at a more microscopic level.
For example, using a composite model of nucleon structure, with pions 
coupling to quarks, one might be able to demonstrate explicitly that 
both effects arise from the same underlying 
dynamics \footnote{The finding that the 
net $\pi\pi NN$ contact term is proportional to $1 - g_A^2$, and thus 
to the compositeness of the nucleon, also resolves a more general 
issue that arises when using Dirac fermions to describe particles
with internal structure, as done in relativistic chiral EFT.
It has been argued that light--front time--ordered perturbation 
theory with Dirac fermions would give rise to $Z$--graphs (i.e., graphs 
with an $NN\bar N$ intermediate state), which should not contribute on 
physical grounds because they are strongly suppressed by the pion--nucleon 
form factor (S.~Brodsky, private communication). 
Our results show that the explicit contact term in the chiral
Lagrangian cancels most of this contribution, to the effect that the net 
result is proportional to $1 - g_A^2$, which reflects the composite nucleon
structure, as it should be. It shows how chiral invariance naturally 
arranges for the cancellation of contributions from high--mass intermediate 
states unrelated to true internal nucleon structure.}.

The presence of contact terms in the chiral Lagrangian is closely related 
to the form of the basic $\pi NN$ coupling adopted in formulating the
effective dynamics. The $\pi\pi NN$ contact term of Eq.~(\ref{lag})
is specific to the axial vector form of the $\pi NN$ coupling.
It is well--known that for on--shell nucleons this form is equivalent to 
the pseudoscalar form of the $\pi NN$ coupling, cf.\ 
Eq.~(\ref{axial_pseudoscalar}), if one identifies 
$g_{\pi NN} = M_N g_A/F_\pi$. [More generally, the axial--vector
chiral Lagrangian Eq.~(\ref{lag}), including the $\pi\pi NN$ contact term, 
can be obtained from a pseudoscalar Langrangian by performing a chiral 
rotation of the nucleon fields with the matrices $U^{\pm 1/2}$, Eq.~(\ref{U}).
The contact arises from the chiral rotation of the pseudoscalar kinetic 
term and therefore carries the universal coefficient $\sim 1/F_\pi^2$.]
It is interesting to note that we get the same result for the
intermediate nucleon (or ``non--contact'') part of the isovector 
Dirac spectral function, Eq.~(\ref{Im_F1_noncontact}), with the 
pseudoscalar and axial vector forms of the $\pi NN$ 
couplings \cite{Strikman:2010pu}. This part arises from the triangle 
graph Fig.~\ref{fig:diag}a with the intermediate nucleon propagator, after
separating out the off--shell terms in the numerator [keeping
only the third term in Eq.~(\ref{bilinear_simplified})], and one
can verify by explicit calculation that the result is the same 
as what one obtains with a pseudoscalar coupling given by
Eq.~(\ref{axial_pseudoscalar}). It shows that the difference between the
pseudoscalar and axial vector couplings is effectively contained in the
``net'' $\pi\pi NN$ contact term in the final result, 
Eq.~(\ref{Im_F1_contact}), which is again consistent with this term 
being proportional to $1 - g_A^2$ and reflecting the compositeness 
of the nucleon.

In the light--front formulation of chiral processes the contact terms
summarize the contributions from quasi--zero modes, in which the pion 
field carries a vanishing fraction of the nucleon's plus 
momentum \cite{Strikman:2010pu}. This can be shown explicitly by following 
the space--time--evolution of the chiral processes in time--ordered 
perturbation theory \cite{inprep}. It is also known that in time--ordered 
perturbation theory the different forms of the $\pi N$ coupling give 
apparently different results, as this formulation of relativistic dynamics
does not conserve four--momentum in intermediate states
\cite{Ji:2009jc,Burkardt:2012hk}. Our findings suggest that 
there is a natural connection between the two observations.
\section{Delta isobar and large--$N_c$ limit}
\label{sec:delta}
\subsection{Peripheral densities from $\Delta$ excitation}
\label{subsec:delta}
We now want to study the role of $\Delta$ isobar excitation in the 
nucleon's peripheral transverse charge and magnetization densities. 
While going beyond the domain of strictly chiral dynamics, inclusion of 
the $\Delta$ is important for practical as well as theoretical reasons.
First, the $\pi N \Delta$ coupling is large, and $N \rightarrow \pi \Delta$ 
transitions contribute significantly to the isovector spectral functions
at $t - 4 M_\pi^2 \sim \textrm{few} \, M_\pi^2$ and the transverse
densities at $b\sim \textrm{few} \, M_\pi^{-1}$. Second, with the
$\Delta$ we can see explicitly how the analytic structure of the form 
factor near the two--pion threshold changes in the case of a ``heavy'' 
intermediate state, and how the mass splitting affects the subthreshold 
singularity that plays such an important role in the amplitude
with the nucleon intermediate state. Third, and most important,
inclusion of the $\Delta$ is required to ensure the proper scaling
behavior of the peripheral densities in the large--$N_c$ limit of QCD.

The $N\Delta$ mass splitting $M_\Delta - M_N = 0.29 \, \textrm{GeV}$
represents a ``non--chiral'' mass scale that is numerically comparable
to the pion mass $M_\pi = 0.14 \, \textrm{GeV}$. Several schemes for
extending chiral EFT to include $\Delta$ degrees 
of freedom have been proposed, putting the $N\Delta$ mass splitting 
in some parametric relation to the pion mass. Our objectives here
are very specific and can be addressed without a fully developed
EFT of the $\Delta$. We want to estimate the
contribution of intermediate $\Delta$ states in the two--pion cut
of the isovector spectral function and the peripheral densities 
in a way that is consistent with the leading--order relativistic 
chiral EFT treatment of the nucleon of Sec.~\ref{sec:chiral}, and verify 
that the total result obeys the proper $N_c$--scaling. To this end we 
introduce the $\Delta$ as a relativistic point particle, with an 
empirical $\pi N \Delta$ coupling, and treat the $N\Delta$ mass 
splitting as a free parameter, with no defined relation to $M_\pi$; 
later, we let the masses and couplings scale according to their
large--$N_c$ behavior.

The spin--3/2 field of the $\Delta$ can be constructed by applying 
constraints to a four--vector bispinor field (Rarita--Schwinger 
formalism) \cite{LLIV}. Its Green function with four--momentum $l$ is
\beq
\frac{R_{\mu\nu}(l)}{l^2 - M_\Delta^2 + i0} ,
\label{green_delta}
\eeq
where the projector is explicitly given by
\be
R_{\mu\nu} (l) &\equiv& (\hat l + M_\Delta )
\left[ -g_{\mu\nu} + \frac{1}{3} \gamma_\mu \gamma_\nu
+ \frac{2}{3 M_\Delta^2} l_\mu l_\nu
- \frac{1}{3 M_\Delta} (l_\mu \gamma_\nu - \gamma_\mu l_\nu )
\right]
\label{rarita_schwinger_expanded}
\\
&=& (\hat l + M_\Delta) \left( -g_{\mu\nu} + 
\frac{l_\mu l_\nu}{M_\Delta^2}
\right)
\; - \; 
\frac{1}{3} 
\left( \gamma_\mu + \frac{l_\mu}{M_\Delta} \right)
(\hat l - M_\Delta)
\left( \gamma_\nu + \frac{l_\nu}{M_\Delta} \right)
\label{rarita_schwinger_factorized}
\ee
and obeys the constraints
\beq
\left.
\begin{array}{r}
l_\mu R_{\mu\nu}  \\[1ex]
R_{\mu\nu} l_\nu   \\[1ex]
(\hat l - M_\Delta) R_{\mu\nu} \\[1ex]
R_{\mu\nu} (\hat l - M_\Delta)
\end{array}
\right\}
\;\; \propto \;\; (l^2 - M_\Delta^2) ,
\label{rarita_schwinger_constraint}
\eeq
implying that in the corresponding contractions the pole of the
Green function is canceled. 
The $\pi N \Delta$ interaction is described by the Lagrangian
\be
\mathcal{L}_{\pi N \Delta} &=& \frac{i g_{\pi N \Delta}}{\sqrt{2} M_N}
\left[ 
\bar\psi_p \; \partial_\mu \pi^- \; \Psi_{\Delta ++}^\mu 
+ \sqrt{\frac{2}{3}} \; \bar\psi_p \; \partial_\mu \pi^0 \; 
\Psi_{\Delta +}^\mu
+ \frac{1}{\sqrt{3}} \; \bar\psi_p \; 
\partial_\mu \pi^+ \; \Psi_{\Delta 0}^\mu
\right. 
\nonumber
\\
&+& 
\left.
\bar\psi_n \; \partial_\mu \pi^+ \; \Psi_{\Delta -}^\mu 
+ \sqrt{\frac{2}{3}} \; \bar\psi_n \; \partial_\mu \pi^0 \; 
\Psi_{\Delta 0}^\mu
+ \frac{1}{\sqrt{3}} \; \bar\psi_n \; 
\partial_\mu \pi^- \; \Psi_{\Delta +}^\mu
\right] \;\; + \;\; \mbox{h.c.},
\label{L_pi_N_Delta}
\ee
where $\psi_{p, n}$ are the proton and neutron fields and 
$\Psi_{\Delta ++}^\mu$ etc.\ the $\Delta$ fields. 
The relative coefficients of the terms in 
Eq.~(\ref{L_pi_N_Delta}) are dictated by isospin invariance. 
Our definition of the coupling constant $g_{\pi N \Delta}$ corresponds 
to that of Ref.~\cite{Adkins:1983ya} (see 
Refs.~\cite{Strikman:2003gz,Strikman:2009bd} for comparison with
other conventions), and the empirical value of the coupling is 
$g_{\pi N \Delta} = 20.22$. We note that the introduction of the $\Delta$
into the effective Lagrangian could in principle result in, or require,
the addition of a $\pi\pi NN$ contact term of the same type as that already
present in the $\pi N$ Lagrangian; the physical meaning and implications
of such a term are discussed below.

%
% FIGURE
%
\begin{figure}
\includegraphics[width=.25\textwidth]{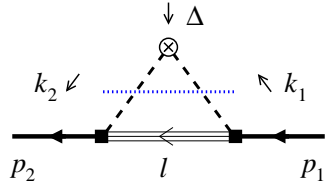}
\caption[]{Contribution of $\Delta$ isobar excitation to the 
two--pion cut of the isovector nucleon form factor. The labeling
of the 4--momenta is the same as for the intermediate 
$N$ diagram of Fig.~\ref{fig:diag}a.} 
\label{fig:diag_delta}
\end{figure}
With the coupling Eq.~(\ref{L_pi_N_Delta}) and the Green function
Eq.~(\ref{green_delta}) it is straightforward to calculate the
contribution of $N \rightarrow \Delta$ transitions to the two--pion
cut of the isovector spectral functions. The calculation closely
follows that for the intermediate nucleon state in 
Sec.~\ref{subsec:two_pion}, and we outline only the main steps.
The $\Delta$ contribution to the isovector current matrix element, 
as given by diagram Fig.~\ref{fig:diag_delta}, is 
\be
\langle N_2 | J^\mu (0) | N_1 \rangle_{\pi\pi \; {\rm cut}}
&=& - \frac{2 i g_{\pi N\Delta}^2}{3 M_N^2} 
\int\frac{d^4 k}{(2\pi )^4}
\frac{[\bar u_2 \, k_{2\alpha} R_{\alpha\beta} (l) k_{1\beta}\, u_1]\, 
k^\mu}{(k_2^2 - M_\pi^2 + i0)
(k_1^2 - M_\pi^2 + i0) (l^2 - M_\Delta^2 + i0)} ,
\label{triangle_momentum_delta}
\ee
where again $u_{1, 2}$ are the external nucleon bispinors and the labeling 
of the 4--momenta is the same as in the case of the intermediate 
nucleon diagram Fig.~\ref{fig:diag}a, Eq.~(\ref{triangle_momentum_1}). 
The bilinear form in the numerator contains terms
which vanish at $s \equiv l^2 = M_\Delta^2$ and result in integrals
of the same form as that obtained from the $\pi\pi NN$ contact
term, Eq.~(\ref{contact_momentum}). Making extensive use of the constraints 
Eq.~(\ref{rarita_schwinger_constraint}) [the result of the particular 
contractions has to be determined from Eqs.~(\ref{rarita_schwinger_expanded})
or (\ref{rarita_schwinger_factorized})], the kinematic relations between
the different momentum 4-vectors, and the Dirac equation for the
external nucleon spinors, we write the bilinear form as
\be
\bar u_2 \, k_{2\alpha} R_{\alpha\beta} (l) k_{1\beta}\, u_1
&=& \bar u_2 ( F + G \, \hat k) u_1 \; + \; 
(s - M_\Delta^2 ) \bar u_2 R_{\rm off} u_1 .
\label{R_delta_on_off}
\ee
The first term on the right--hand side remains non--zero on the 
baryon mass--shell. Here $F$ and $G$ denote scalar functions of
the invariants $t$ and $k_{1, 2}^2 = (k \mp \Delta/2)^2$,
\be
\left. \begin{array}{r} F(t, k_1^2, k_2^2) 
\\[2ex] G(t, k_1^2, k_2^2)  \end{array}\right\}
&\equiv & \left[ \frac{t}{2} - M_N^2 + 
\frac{(M_\Delta^2 + M_N^2 - k_2^2)
(M_\Delta^2 + M_N^2 - k_1^2)}{4 M_\Delta^2} \right]
\; \times \;  \left\{ \begin{array}{c} (M_\Delta + M_N) 
\\[2ex] (-1) \end{array}\right\}
\nonumber
\\[2ex]
&+& \frac{1}{3} 
\left( M_N + \frac{M_\Delta^2 + M_N^2 - k_2^2}{2 M_\Delta}\right)
\left( M_N + \frac{M_\Delta^2 + M_N^2 - k_1^2}{2 M_\Delta}\right)
\; \times \;  \left\{ \begin{array}{c} (M_\Delta - M_N) 
\\[2ex] 1 \end{array}\right\} .
\label{f0_f1_def}
\ee
Note that the functions are symmetric with respect to both
$k \rightarrow - k$ and $\Delta \rightarrow - \Delta$.
The second term in Eq.~(\ref{R_delta_on_off}) is an off--shell piece, with
\be
R_{\rm off} 
&\equiv & 
\frac{1}{3 M_\Delta^2}
\left[ M_\Delta^2 - M_N^2 + M_N M_\Delta + M_\pi^2 
+ \frac{1}{4} (s - M_\Delta^2) \right] (M_N + M_\Delta - \hat k)
\nonumber
\\[2ex]
&+& \frac{(s - M_\Delta^2 )}{12 \, M_\Delta^2} (M_N - M_\Delta - \hat k) ,
\ee
where $s = l^2 = (P - k)^2$. After inserting the decomposition
Eq.~(\ref{R_delta_on_off}) into Eq.~(\ref{triangle_momentum_delta}),
the tensor integrals are reduced to scalar integrals
with the help of standard projection formulas, making use of the symmetries
of the integrand. The resulting bilinear forms $\bar u_2 \ldots u_1$ are 
then converted to those of the right--hand side of Eq.~(\ref{me_general})
using the Dirac equation for the nucleon spinors. In this way we
obtain the $\Delta$ contribution to the isovector Dirac and Pauli 
form factors in terms of invariant integrals as
\be
F_1^V(t)_{\pi\pi \; {\rm cut}} &=& \frac{2 g_{\pi N\Delta}^2}{3 M_N^2} \, 
\left[ 
I_{\Delta 1} (t) \; + \; I_{\Delta 1\, {\rm cont}} (t) \right] ,
\\[2ex]
F_2^V(t)_{\pi\pi \; {\rm cut}} &=& \frac{2 g_{\pi N\Delta}^2}{3 M_N^2} \,
\left[ I_{\Delta 2} (t) \; + \; I_{\Delta 2 \, {\rm cont}} (t) \right] ,
\\[2ex]
I_{\Delta 1, \, \Delta 2} &\equiv & -i \int\frac{d^4 k}{(2\pi )^4}
\frac{N_{\Delta 1, \, \Delta 2}}{(k_2^2 - M_\pi^2 + i0)
(k_1^2 - M_\pi^2 + i0) (l^2 - M_\Delta^2 + i0)} ,
\\[2ex]
N_{\Delta 1} &=& \frac{kP}{P^2} M_N \, F
\; + \; \frac{1}{P^2} \left\{
-\frac{t}{8} \left[ k^2 - \frac{(k\Delta)^2}{\Delta^2} \right]
\; + \; \left( M_N^2 + \frac{t}{8} \right) 
\frac{(kP)^2}{P^2} \right\} \, G ,
\\[2ex]
N_{\Delta 2} &=& - \frac{kP}{P^2} M_N \, F
\; - \; \frac{M_N^2}{P^2} \left[ -k^2 + 3 \frac{(kP)^2}{P^2}
+ \frac{(k\Delta)^2}{\Delta^2} \right] \frac{G}{2} ,
\\[2ex]
I_{\Delta 1 \, {\rm cont}, \; \Delta 2 \, {\rm cont}} 
&=& -i \int\frac{d^4 k}{(2\pi )^4}
\frac{N_{\Delta 1 \, {\rm cont}, \; \Delta 2 \, {\rm cont}}}
{(k_2^2 - M_\pi^2 + i0)(k_1^2 - M_\pi^2 + i0)} ,
\\[2ex]
N_{\Delta 1 \, {\rm cont}} &=& \frac{1}{3 M_\Delta^2}
\left[ k^2 - \frac{(k\Delta)^2}{\Delta^2} \right] 
\left[ -\frac{k^2}{2} - \frac{1}{6} (M_N + M_\Delta)^2 
- \frac{t}{24} \right] ,
\\[2ex]
N_{\Delta 2 \, {\rm cont}} &=& \frac{M_N^2}{9 M_\Delta^2}
\left[ k^2 - \frac{(k\Delta)^2}{\Delta^2} \right] .
\ee
The imaginary part on the two--pion cut can now be computed using the
$t$--channel cutting rule of Appendix~\ref{app:cutting} in the same
way as in the intermediate nucleon case. The virtuality of the 
intermediate $\Delta$ is
\be
l^2 - M_\Delta^2 &=&  -A_\Delta + i B \cos\theta ,
\\[1ex]
A_\Delta &\equiv & t/2 - M_\pi^2 + M_\Delta^2 - M_N^2 ,
\\[1ex]
B &\equiv&  2 k_{\rm cm} \sqrt{P^2} .
\ee
We obtain the spectral functions as
\be
\frac{1}{\pi} \, \textrm{Im} \, F_1^V (t)
&=& 
\frac{g_{\pi N \Delta}^2}{24 \pi^2 M_N^2 \sqrt{P^2} \sqrt{t}}
\left\{ -\frac{A_\Delta M_N F}{2 P^2} (x_\Delta - \arctan x_\Delta)
\right.
\nonumber \\[2ex]
&+& \left. \frac{A_\Delta^2 G}{4 (P^2)^2} 
\left[ -\frac{t}{8} x_\Delta^2 \arctan x_\Delta
+ \left( M_N^2 + \frac{t}{8} \right) (x_\Delta - \arctan x_\Delta)
\right] \right\} 
\label{Im_F1_delta}
\\[2ex]
&+& \frac{g_{\pi N \Delta}^2 k_{\rm cm}^3}{36 \pi^2 M_N^2 M_\Delta^2 \sqrt{t}}
\left[ - \frac{k_{\rm cm}^2}{2} + \frac{1}{6} (M_N + M_\Delta)^2 
+ \frac{t}{24} \right] ,
\label{Im_F1_delta_contact}
\\[2ex]
\frac{1}{\pi} \, \textrm{Im} \, F_2^V (t)
&=& \frac{g_{\pi N \Delta}^2}{24 \pi^2 M_N^2 \sqrt{P^2} \sqrt{t}}
\left\{ \frac{M_N A_\Delta F}{2 P^2} (x_\Delta - \arctan x_\Delta)
\right.
\nonumber \\[2ex]
&+& \left. \frac{M_N^2 A_\Delta^2 G}{8 (P^2)^2} 
\left[ (x_\Delta^2 + 3) \arctan x_\Delta - 3 x_\Delta
\right] \right\} 
\label{Im_F2_delta}
\\[2ex]
&-& \frac{g_{\pi N \Delta}^2 k_{\rm cm}^3}{108 \pi^2 M_\Delta^2 \sqrt{t}} ,
\label{Im_F2_delta_contact}
\\[2ex]
x_\Delta &\equiv & \frac{B}{A_\Delta}
\;\; = \;\; \frac{2 \sqrt{t/4 - M_\pi^2} \sqrt{M_N^2 - t/4}}
{t/2 - M_\pi^2 + M_\Delta^2 - M_N^2} ,
\ee
where $F$ and $G$ now denote the functions of Eq.~(\ref{f0_f1_def}) on
the pion mass shell,
\be
F &\equiv& F(t, \; k_1^2 = M_\pi^2, \; k_2^2 = M_\pi^2)
\nonumber
\\[2ex]
&=& \left[ \frac{t}{2} - M_N^2 + 
\frac{(M_\Delta^2 + M_N^2 - M_\pi^2)^2}{4 M_\Delta^2} \right]
(M_N + M_\Delta) 
- \frac{1}{3} 
\left( M_N + \frac{M_\Delta^2 + M_N^2 - M_\pi^2}{2 M_\Delta}\right)^2
(M_N - M_\Delta ) ,
\\[3ex]
G &\equiv& G(t, \; k_1^2 = M_\pi^2, \; k_2^2 = M_\pi^2)
\nonumber
\\[2ex]
&=& -\left[ \frac{t}{2} - M_N^2 + 
\frac{(M_\Delta^2 + M_N^2 - M_\pi^2)^2}{4 M_\Delta^2} \right]
+ \frac{1}{3} 
\left( M_N + \frac{M_\Delta^2 + M_N^2 - M_\pi^2}{2 M_\Delta}\right)^2 .
\label{f1_cut}
\ee

For further analysis it will be convenient to quote simplified expressions
according to the uniform approximation, 
Eqs.~(\ref{Im_F1_simplified})--(\ref{x_1}), in which we neglect terms 
$t/M_N^2$ without altering the analytic structure near threshold.
In the contact term Eq.~(\ref{Im_F1_delta_contact}) we can also neglect 
terms of order $t/M_{N, \Delta}^2$ and $M_\pi^2/M_{N, \Delta}^2$.
With these approximations we obtain
\be
\frac{1}{\pi} \, \textrm{Im} \, F_1^V (t)
&=& \frac{g_{\pi N \Delta}^2 \, A_\Delta 
(- 2 M_N F + A_\Delta G)}{96 \pi^2 M_N^5 \sqrt{t}}
(x_{\Delta 1} - \arctan x_{\Delta 1})
\label{Im_F1_delta_simplified}
\\[2ex]
&+& \frac{g_{\pi N \Delta}^2 (M_N + M_\Delta)^2 
k_{\rm cm}^3}{216 \pi^2 M_N^2 M_\Delta^2 \sqrt{t}} ,
\label{Im_F1_delta_contact_simplified}
\\[2ex]
\frac{1}{\pi} \, \textrm{Im} \, F_2^V (t)
&=& \frac{g_{\pi N \Delta}^2 \, A_\Delta }{192 \pi^2 M_N^5 \sqrt{t}}
\left[ (4 M_N F - 3 A_\Delta G) \, (x_{\Delta 1} - \arctan x_{\Delta 1})
\; + \; A_\Delta G \, x_{\Delta 1}^2 \arctan x_{\Delta 1} \right] 
\label{Im_F2_delta_simplified}
\\[2ex]
&-& \frac{g_{\pi N \Delta}^2 k_{\rm cm}^3}{108 \pi^2 M_\Delta^2 \sqrt{t}} ,
\label{Im_F2_delta_contact_simplified}
\\[2ex]
x_{\Delta 1} &\equiv & x_{\Delta 1}(t) \;\; \equiv \;\;
\frac{2  M_N k_{\rm cm}}{A_\Delta} 
\;\; = \;\;
\frac{2 M_N \sqrt{t/4 - M_\pi^2}}{t/2 - M_\pi^2 + M_\Delta^2 - M_N^2} .
\label{x_Delta_1}
\ee
Equations~(\ref{Im_F1_delta_simplified})--(\ref{x_Delta_1})
provide a compact and completely adequate representation of the
two--pion spectral functions resulting from $\Delta$ intermediate states.

%
% FIGURE
%
\begin{figure}
\begin{tabular}{ll}
\includegraphics[width=.45\textwidth]{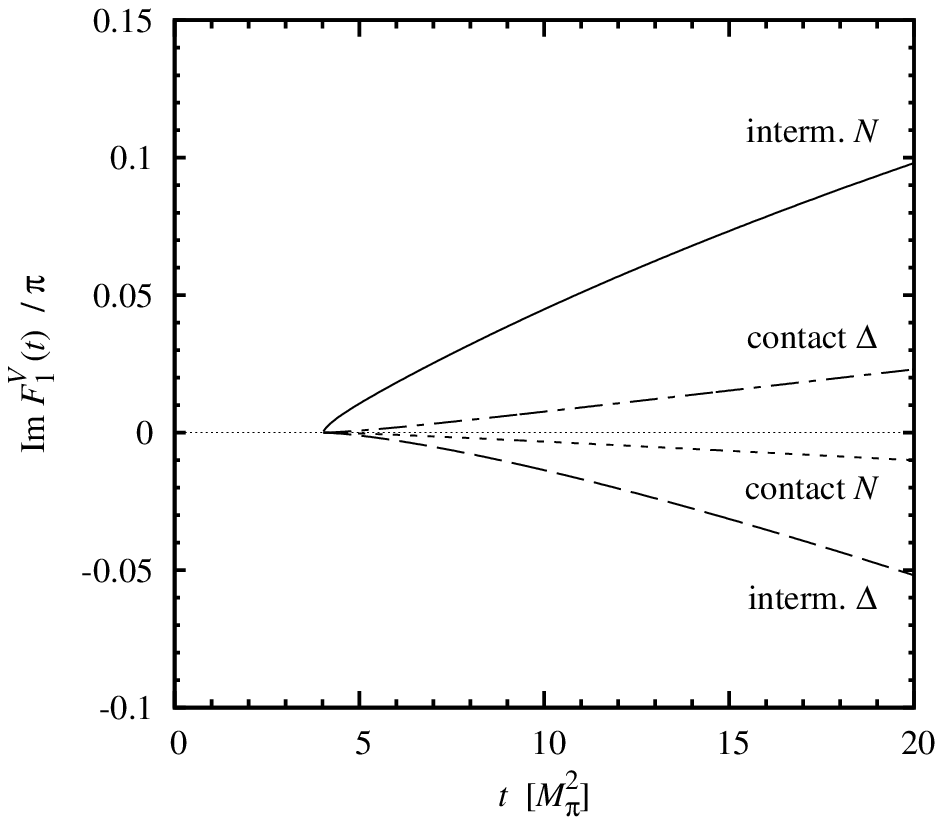}
&
\includegraphics[width=.45\textwidth]{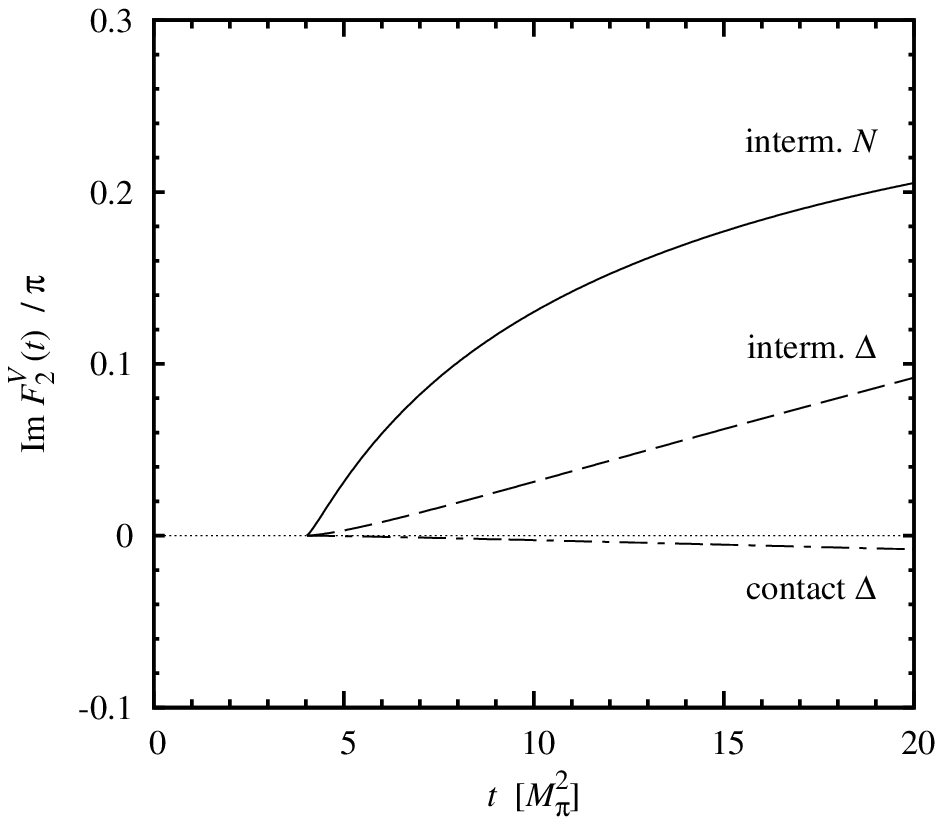}
\\[-2ex]
(a) & (b)
\end{tabular}
\caption[]{Comparison of chiral and $\Delta$ contributions to the
spectral functions of the isovector Dirac form factor
$\textrm{Im} \, F_1^V (t)/\pi$ [panel (a)] and 
Pauli form factor $\textrm{Im} \, F_2^V (t)/\pi$ [panel (b)].
Solid lines: Chiral component, intermediate $N$, 
Eq.~(\ref{Im_F1_noncontact}). Dotted line: Chiral component,
contact term Eq.~(\ref{Im_F1_contact}) [in the Pauli form factor
in plot (b) this term is absent]. Dashed lines: Contribution 
from intermediate $\Delta$, Eq.~(\ref{Im_F1_delta}). 
Dashed--dotted line: $\Delta$ contact term.}
\label{fig:ft_delta}
\end{figure}
The numerical results for the two--pion isovector spectral functions 
resulting from $\Delta$ intermediate states, 
Eqs.~(\ref{Im_F1_delta})--(\ref{f1_cut}),
are shown in Fig.~\ref{fig:ft_delta}, together with those from $N$ 
intermediate states, Eqs.~(\ref{Im_F1_noncontact})--(\ref{x}). Several
features are worth noting. First, in the Dirac spectral function
in Fig.~\ref{fig:ft_delta}a the $N$ and $\Delta$ contributions have 
opposite sign, both in the contact and non--contact terms, such that 
they partly cancel each other. In the Pauli spectral function
in Fig.~\ref{fig:ft_delta}b, in contrast, the $N$ and $\Delta$
contributions have the same sign. Both findings can naturally
be explained in the large--$N_c$ limit of QCD, where model--independent
relations between the $N$ and $\Delta$ spectral functions can
be derived (see Sec.~\ref{subsec:chiral_large_nc}).
Second, the $\Delta$ contributions do not show the strong rise
near threshold observed in the intermediate $N$ contributions to the
Dirac and Pauli spectral functions. This is because in the intermediate 
$\Delta$ case the subthreshold singularity is removed from threshold 
by a much larger distance, see Eq.~(\ref{t_sub_delta}). 
It implies that the transverse densities at large distances are 
dominated by the contribution from intermediate $N$ states,
as expected.

A comment is in order regarding the interpretation of the contact 
terms in the $\Delta$ contribution to the Dirac spectral function. 
Equation~(\ref{Im_F1_delta_contact}) represents the contact term 
as it comes out of the diagram of Fig.~\ref{fig:diag_delta}. 
It is seen from Fig.~\ref{fig:ft_delta}a
that numerically this term is considerably larger than the net contact
term in the chiral EFT result with nucleons only, Eq.~(\ref{Im_F1_contact}). 
The latter is the sum of the explicit contact term in the chiral Lagrangian 
entering in diagram Fig.~\ref{fig:diag}b, and the ``non--propagating'' 
piece of the diagram of Fig.~\ref{fig:diag}a, with substantial 
cancellations between the two, cf.\ Eq.~(\ref{contact_combined}). 
From a physical point of view, inclusion of the 
$\Delta$ as an explicit degree of freedom in the chiral Lagrangian 
amounts to a change of the short--distance structure of the nucleon, 
which could manifest itself in the appearance of a ``new'' $\pi\pi NN$ 
contact term, or, effectively, a renormalization of the old contact 
term in the chiral Lagrangian. Unlike the case of the Lagrangian with
$N$ only, the strength of this new contact term is not fixed by 
chiral symmetry, and we presently have no way to constrain it 
theoretically. It is likely that this new contact term would partly
cancel the contact term coming out of the $\Delta$ diagram of 
Fig.~\ref{fig:diag_delta}. Because we cannot determine the coefficient
of the total contact term from general principles, and because the 
$\Delta$ effects are altogether rather unimportant at large distances, 
we shall treat the $\Delta$ contact term in the Dirac spectral function
as a theoretical uncertainty (see below).

A contact term is also found in the $\Delta$ contribution to the Pauli 
spectral function, Eq.~(\ref{Im_F2_delta_contact}). However, its 
contribution is extremely small, see Fig.~\ref{fig:ft_delta}b, and 
we shall nexlect the theoretical uncertainty associated with it.
Note that in this channel there is no contact term in the chiral EFT 
result with nucleons only, Eq.~(\ref{Im_F2}). 

%
% FIGURE
%
\begin{figure}
\begin{tabular}{ll}
\includegraphics[width=.45\textwidth]{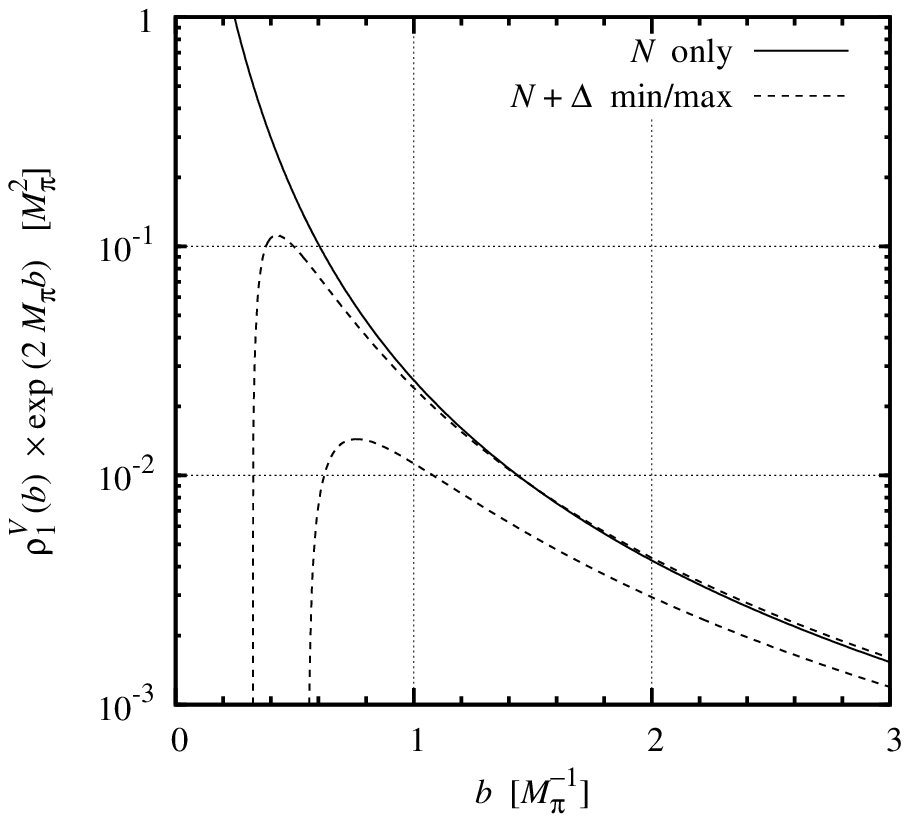}
&
\includegraphics[width=.45\textwidth]{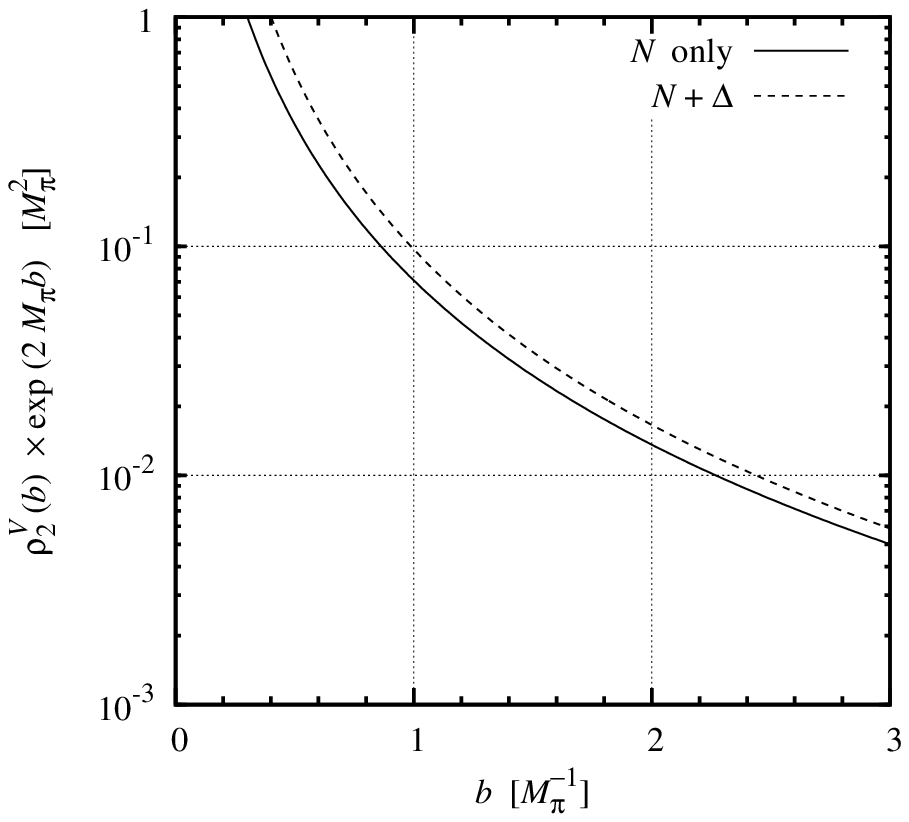}
\\[-2ex]
(a) & (b)
\end{tabular}
\caption[]{Effect of $\Delta$ excitation on the peripheral isovector 
transverse charge and magnetization densities of the nucleon. 
The plots shows the densities $\rho_{1, 2}^V (b)$ with the exponential 
factor $\exp (-2 M_\pi b)$ extracted (cf.\ Fig.~\ref{fig:rho_b}b).
The distance $b$ is given in units of $M_\pi^{-1}$, the densities in 
units of $M_\pi^2$. (a) Transverse charge density $\rho_1 (b)$. 
Solid line: Chiral component from intermediate $N$ states and the $N$
contact term, Eqs.~(\ref{Im_F1_noncontact}) and
Eqs.~(\ref{Im_F1_contact}).
Dotted lines: Sum of chiral component (solid line) and $\Delta$ contribution, 
Eqs.~(\ref{Im_F1_delta_contact})--(\ref{Im_F1_delta}). The 
two curves show the results obtained with the $\Delta$ contact 
term, Eq.~(\ref{Im_F1_delta_contact}), multiplied by 0 and 2, respectively;
their difference is an estimate of the theoretical uncertainty (details 
see text). (b) Transverse magnetization density $\rho_2 (b)$. 
Solid line: Chiral component from intermediate $N$ states, 
Eq.~(\ref{Im_F2}). Dotted line: Sum of chiral and $\Delta$ contributions, 
Eqs.~(\ref{Im_F2_delta_contact})--(\ref{Im_F2_delta}). 
The uncertainty resulting from the $\Delta$ contact term is negligible 
on the scale of the plot and not shown on the figure.}
\label{fig:rho_delta}
\end{figure}
Using the results for the spectral functions we can now calculate the 
peripheral transverse densities resulting from $\Delta$ intermediate states.
Figure~\ref{fig:rho_delta}a summarizes the numerical effect of the $\Delta$
on the transverse charge density. The solid line shows the chiral
component with intermediate $N$ states only (see Fig.~\ref{fig:rho_b}a). 
The dotted lines show the density after adding the $\Delta$ contribution.
The lower curve is obtained when including the contact term 
Eq.~(\ref{Im_F1_delta_contact}) with full strength, the lower curve
when setting it to zero, as would correspond to complete cancellation 
by an explicit new $\pi\pi NN$ contact term in the Lagrangian
(cf.\ the discussion of the theoretical uncertainty above).
One sees that at $b \lesssim 1 \, M_\pi^{-1}$ there are very substantial
cancellations between the $N$ and $\Delta$ contributions, causing the
total two--pion isovector density from intermediate $N$ and $\Delta$ 
to become negative at small $b$. [Note that the physical density at 
such small values of $b$ is dominated by vector meson singularity 
of the spectral function (see Sec.~\ref{sec:region}), and that 
Fig.~\ref{fig:rho_delta}a is only intended to illustrate the relative 
magnitude of the calculated $N$ and $\Delta$ contributions.]
At distances $b \gtrsim 2\, M_\pi^{-1}$ the $\Delta$ contribution
becomes a small correction of $\sim 20\%$, as expected from a 
``heavy'' degree of freedom. The theoretical uncertainty associated
with the inclusion of the $\Delta$ therefore does not affect our
numerical estimates of the chiral component of the transverse 
charge density.

Figure~\ref{fig:rho_delta}b shows the effect of the $\Delta$ on
the transverse magnetization density. At distances 
$b \gtrsim 2\, M_\pi^{-1}$ the $\Delta$ increases the intermediate $N$
result by $\sim 20\%$. One sees that the effect of the $\Delta$ has 
opposite sign in the charge and magnetization densities, as already
noted in relation to the spectral functions. This pattern is 
naturally explained by the relations between the intermediate $N$
and $\Delta$ contributions emerging in the large--$N_c$ limit of QCD.
\subsection{Transverse densities in large--$N_c$ QCD}
\label{subsec:large_nc_qcd}
The limit of a large number of colors in QCD, $N_c \rightarrow \infty$, 
is a powerful theoretical tool for studying properties of mesons and 
baryons and relating them to the microscopic theory of strong interactions.
While even in the large--$N_c$ limit QCD remains
a complex dynamical system that cannot be solved exactly, the scaling
behavior of meson and baryon properties with $N_c$ can be established
on general grounds and provides constraints for EFTs
or phenomenological models. In this subsection we want to
establish the $N_c$--scaling behavior of the transverse charge and 
magnetization densities on general grounds. In the following subsection we 
then show that the two--pion components of the peripheral charge
and magnetization densities obey these general scaling laws and discuss 
the essential role of $\Delta$ isobar excitation in bringing about 
this result.

The $N_c$--scaling of meson and baryon masses in QCD, their interactions,
and various current matrix elements, can be established using the 
classic techniques described in Ref.~\cite{Witten:1979kh}. It is
found that the low--lying meson and baryon masses [i.e., with 
spin and isospin of $O(N_c^0)$] scale as
\beq
M_{\textrm{\scriptsize meson}} \;\; = \;\; O(N_c^0), 
\hspace{2em}
M_{\textrm{\scriptsize baryon}} \;\; = \;\; O(N_c), 
\label{masses_nc}
\eeq
while their basic hadronic sizes scale as
\beq
R_{\textrm{\scriptsize meson}}, \,  R_{\textrm{\scriptsize baryon}} 
\;\; = \;\; O(N_c^0) .
\label{R_nc}
\eeq
Baryons in large--$N_c$ QCD thus are heavy systems of fixed spatial size. 
Their overall momentum and spin--isospin degrees of freedom can be described 
as the classical motion of a heavy body characterized by a mass and moment
of inertia of order $O(N_c)$. In particular, the $N$ and $\Delta$ are 
obtained as rotational states with $S = T = 1/2$ and $S = T = 3/2$, and 
their mass splitting is $M_\Delta - M_N = O(N_c^{-1})$. This description 
can be extended to transition matrix elements of current operators, 
which generally involve new parameters characterizing the internal 
structure of the classical rotor, and has been formalized using 
group--theoretical methods \cite{Dashen:1993jt,Dashen:1994qi}.

Turning to the transverse charge and magnetization densities, we are 
interested in their general $N_c$--scaling behavior at non--exceptional 
distances of the order
\beq
b \;\; = \;\; O(N_c^0),
\eeq
i.e., distances of the same order as the basic hadronic size of the 
large--$N_c$ nucleon, Eq.~(\ref{R_nc}). [Below we shall see that the
chiral component is contained in this parametric region, as it involves
distances of the order $b = O(M_\pi^{-1})$ and the pion mass scales
as $M_\pi = O(N_c^0)$.] For such distances the $N_c$--scaling behavior
of the density can be inferred from that of the corresponding total
charge, given by the integral of the density over $b$. The isovector 
densities are normalized, respectively, to the total isovector charge 
and anomalous magnetic moment of the nucleon, which scale as
\be
\int d^2 b \; \rho_1^V (b) &=& \; \frac{1}{2}
\;\; = \;\; O(N_c^0),
\label{int_largen_1}
\\
\int d^2 b \; \frac{\rho_2^V (b)}{M_N} 
&=& \frac{\kappa_p - \kappa_n}{2 M_N} 
\;\; = \;\; O(N_c) .
\label{int_largen_2}
\ee
Here we assume that the large--$N_c$ limit is taken at fixed 
spin and isospin of the baryon states, $\{ S, T \} = O(N_c^0)$, 
which is the domain usually considered in large--$N_c$ 
phenomenology \footnote{One can also consider the large--$N_c$ limit 
for baryon states whose spin and isospin scales as $\{S, T \} = O(N_c)$, 
which is a different parametric regime and leads to different scaling 
relations for current matrix elements. In the chiral soliton picture 
of large--$N_c$ baryons such states correspond to high--lying rotational
excitations with angular momentum $O(N_c)$ and with angular velocity 
$O(N_c^0)$.}. The scaling behavior of the isospin difference 
Eq.~(\ref{int_largen_1}) is immediately obvious. The scaling
behavior of the isovector anomalous magnetic moment, 
Eq.~(\ref{int_largen_2}), can be established in various ways, e.g., 
by explicitly constructing the spin--flavor wave functions of the 
nonrelativistic quark model at large $N_c$ \cite{Karl:1984cz,Jackson:1985bn}.
It is important to realize that the $N_c$ scaling thus obtained applies
to the \textit{dimensionful} isovector magnetic moment of the nucleon,
while the \textit{dimensionless} quantity $\kappa_p - \kappa_n$ measures 
the isovector anomalous magnetic moment in units of the nuclear 
magneton $e/(2 M_N)$; one therefore needs to explicitly include the 
factors $1/M_N = O(N_c^{-1})$ 
in the $N_c$ scaling relation, as done in Eq.~(\ref{int_largen_2}).
Because the range of the $b$--integration remains stable in the
large--$N_c$ limit, the scaling behavior of the densities follows
that of the charges, and we conclude that
\beq
\rho_1^V (b) \;\; = \;\; O(N_c^0),
\hspace{2em}
\frac{\rho_2^V (b)}{M_N} \;\; = \;\; O(N_c)
\hspace{2em}
[b = O(N_c^0)].
\label{rho_largen_general}
\eeq
Equation~(\ref{rho_largen_general}) represents the general scaling behavior 
of the isovector densities at non--exceptional distances in large--$N_c$ QCD.
One sees that the physical isovector magnetization density (including the 
factor $1/M_N$) is parametrically larger than the isovector charge density.
This is a consequence of the spin--flavor symmetry of the large--$N_c$
nucleon, which implies that the spin--dependent matrix elements of isovector
quark operators are larger than the spin--independent ones by one order 
in $N_c$, and represents a general pattern that is found also in matrix 
elements of other operators. We note that the $N_c$--scaling relations
Eq.~(\ref{rho_largen_general}) for the transverse densities could also 
be obtained from the more general scaling relations for the 
isovector nucleon GPDs $H^{u-d}(x, \xi, t)$ and $E^{u-d}(x, \xi, t)$ 
described in Ref.~\cite{Goeke:2001tz}, by integrating the latter over 
the quark momentum fraction $x = O(N_c^{-1})$, setting $\xi = 0$, and 
performing the transverse Fourier transform as in Eq.~(\ref{rho_fourier}) 
with $t = -\Delta_T^2 = O(N_c^0)$. 
\subsection{Two--pion component in large--$N_c$ limit}
\label{subsec:chiral_large_nc}
We now want to examine the $N_c$--scaling of the two--pion component of 
the transverse densities calculated in Secs.~\ref{subsec:two_pion}
and \ref{subsec:delta}. This exercise explains the interplay between the 
$N$ and $\Delta$ contributions observed in Sec.~\ref{subsec:delta} and 
provides a powerful check on the calculations. More generally, it shows 
that the two--pion component calculated using EFT methods obeys the 
large--$N_c$ scaling laws required by QCD.

Some general comments are in order regarding the compatibility of 
the large--$N_c$ limit of QCD with our identification of the chiral
component based on the spatial picture of nucleon structure. First, in our
approach we are interested in the transverse densities at distances 
$b \sim M_\pi^{-1}$, where $M_\pi^{-1}$ is assumed to be parametrically
large compared to the nucleon's non--chiral size but we do not actually
take the limit $M_\pi \rightarrow 0$. Since the pion mass scales as 
$M_\pi = O(N_c^0)$ this region of distances remains stable in the 
large--$N_c$ limit, 
\beq
b \;\; \sim \;\; M_\pi^{-1} \;\; = O(N_c^0),
\eeq
as does the nucleon's non--chiral hadronic size,
Eq.~(\ref{R_nc}). As a result, the basic proportion of the non-chiral 
and chiral regions of nucleon structure does not change in the 
large--$N_c$ limit, and the latter is naturally compatible with our 
spatial picture. Second, in the large--$N_c$ limit both the $N$ and 
the $\Delta$ become heavy, so that this limit corresponds to the
heavy--baryon expansion of the densities. All the findings of 
Sec.~\ref{subsec:heavy_baryon}, in particular the various consequences 
of the vanishing distance of the subthreshold singularity from the
physical threshold, can be carried over to the discussion of the
large--$N_c$ limit.

Using the explicit expressions for the leading--order chiral EFT
result for the two--pion spectral functions 
Eqs.~(\ref{Im_F1_noncontact})--(\ref{x}), we can determine the
$N_c$--scaling of the corresponding components of the transverse densities.
With the general scaling relations for the couplings
\beq
g_A \; = \; O(N_c), 
\hspace{2em}
F_\pi \; = \; O(N_c^{1/2}), 
\eeq
and the masses, Eq.~(\ref{masses_nc}), we find that for 
$t = O(M_\pi^2) = O(N_c^0)$ the spectral functions scale as
\be
\textrm{Im}\, F_1^V (t)_N &=& O(N_c) , 
\\[1ex]
\frac{\textrm{Im}\, F_2^V (t)_N}{M_N} &=& O(N_c)
\hspace{2em}
[t = O(N_c^0)] .
\ee
The subscript $N$ here indicates that these are the results obtained
from the chiral EFT with nucleons only (including the contributions
from intermediate $N$ states and the contact term) and distinguishes
them from the $\Delta$ contribution considered below.
In the dispersion integral Eq.~(\ref{rho_dispersion}) this implies 
that
\be
\rho_1^V (b)_N &=& O(N_c) , 
\label{rho1_largen_chiral}
\\[1ex]
\frac{\rho_2^V (b)_N}{M_N} &=& O(N_c) 
\hspace{2em}
[b = O(N_c^0)].
\label{rho2_largen_chiral}
\ee
These results can also be obtained directly from the heavy--baryon expansion
of the densities in the chiral region $b = O(M_\pi^{-1})$, 
Eqs.~(\ref{rho_1_heavy_beta}) and (\ref{rho_2_heavy_beta}), as in this
region the heavy--baryon limit $M_N \gg M_\pi$ effectively coincides with 
the large--$N_c$ limit. We now discuss the implications of 
Eq.~(\ref{rho1_largen_chiral}) and (\ref{rho2_largen_chiral}),
and the effect of including $\Delta$ intermediate states, 
separately for the charge and magnetization densities.

\textit{Charge density.} In the transverse charge density
the chiral component from nucleons only, Eq.~(\ref{rho1_largen_chiral}),
is \textit{larger} by a power of $N_c$ than what is allowed by the 
general $N_c$ scaling relation Eq.~(\ref{rho_largen_general}).
It shows that the chiral component from nucleons alone as an 
approximation to the peripheral isovector transverse densities would 
not be consistent with the large--$N_c$ limit of QCD. However, in 
Sec.~\ref{sec:densities} we argued on general grounds that the 
large--distance behavior of the isovector densities in the region 
$b = O(M_\pi^{-1})$ is governed by the two--pion spectral function 
near threshold, which should be true even in large--$N_c$ QCD. 
The paradox is resolved when one includes the $\Delta$ contribution
to the two--pion spectral function. In the large--$N_c$ limit the 
$N$ and $\Delta$ are degenerate,
\beq
M_N, M_\Delta \;\; = \;\; O(N_c),
\hspace{2em}
M_\Delta -M_N \;\; = \;\; O(N_c^{-1}) ,
\eeq
and the $\pi NN$ and $\pi N\Delta$ coupling constants are related as
[cf.\ Eq.~(\ref{axial_pseudoscalar})]
\beq
g_{\pi N\Delta} \;\; = \;\; \frac{3}{2} g_{\pi NN} ,
\hspace{2em}
g_{\pi NN} \; \equiv \; \frac{g_A M_N}{F_\pi} .
\eeq
Using these relations it is easy to see that for $t = O(N_c^0)$ 
the $N$ and $\Delta$ two--pion spectral functions given by
Eqs.~(\ref{Im_F1_simplified})--(\ref{x_1}) and
Eqs.~(\ref{Im_F1_delta_contact_simplified})--(\ref{x_Delta_1})
become equal and opposite at $O(N_c)$,
\be
\textrm{Im} \, F_1^V (t)_\Delta &=& -\textrm{Im} \, F_1^V (t)_N
+ O(N_c^{0}) 
\hspace{2em} [t = O(N_c^0)] .
\label{largen_F1}
\ee
The same applies to the corresponding transverse densities,
\be
\rho_1^V (b)_N &=& - \rho_1^V (b)_\Delta + O(N_c^{0}) 
\hspace{2em} [b = O(N_c^0)] ,
\label{largen_rho1_N_Delta}
\ee
so that adding the $N$ and $\Delta$ contribution we obtain
\be
\rho_1^V (b)_N \; + \; \rho_1^V (b)_\Delta &=& O(N_c^{0}) 
\hspace{2em} [b = O(N_c^0)] ,
\label{largen_rho1}
\ee
which is consistent with the general $N_c$ scaling of the transverse
densities, Eq.~(\ref{rho_largen_general}). Thus, we see that the inclusion
of the $\Delta$ cancels the leading $O(N_c)$ part of the $N$ contribution
and restores the proper $N_c$ scaling of the two--pion component of the 
transverse densities. 

Two circumstances are important in bringing about the remarkable
result of Eq.~(\ref{largen_F1}). First, the large--$N_c$ limit corresponds 
to the heavy--baryon limit of the spectral functions, in which the 
results for both intermediate $N$ and $\Delta$ are given by the 
leading terms in the $M_N/M_\pi$ and $M_\Delta/M_\pi$ expansion, 
respectively, which are not sensitive to the position of the 
subthreshold singularities. [These are the $x_1$ term in 
Eq.~(\ref{Im_F1_simplified}), and 
the $x_{1, \Delta}$ term in Eq.~(\ref{Im_F1_delta_simplified}); 
only the arctan terms in these expressions contain the subthreshold 
singularity.] We recall that the distances of the subthreshold 
singularities from the threshold, given by
Eqs.~(\ref{t_sub}) and (\ref{t_sub_delta}), are
\beq
\begin{array}{rrclcl}
N: & t_{\rm sub} - 4 M_\pi^2 &=& \displaystyle 
\frac{M_\pi^4}{M_N^2} &=& O(N_c^{-2}), 
\\[3ex]
\Delta: & \hspace{1em} t_{{\rm sub}, \Delta} - 4 M_\pi^2 
&=& \displaystyle \frac{(M_\Delta^2 - M_N^2 + M_\pi^2)^2}{M_\Delta^2}
&=& O(N_c^{-2}) . 
\end{array}
\eeq
They are of order $O(N_c^{-2})$ for both $N$ and $\Delta$. However, 
their magnitude (i.e., the coefficient of $N_c^{-2}$ in the scaling law) 
is different for $N$ and $\Delta$, because the term 
$M_\Delta^2 - M_N^2 = (M_\Delta - M_N)(M_\Delta + M_N) = O(N_c^0)$
in the $\Delta$ expression is of the same order as 
$M_\pi^2 = O(N_c^0)$. Thus, the subthreshold
branch points for the $N$ and $\Delta$ approach the threshold with
different speed as $N_c\rightarrow\infty$. The higher--order terms
in the large--$N_c$ expansion of the $N$ and $\Delta$ spectral functions 
are sensitive to this speed will in general not show a simple relation
in the large--$N_c$ limit; rather, their relation will depend on the 
ratio $M_\pi^2 / (M_\Delta^2 - M_N^2) = O(N_c^0)$ which remains
non--trivial in the large--$N_c$ limit.

Second, also the contact terms resulting from the graphs with 
intermediate $N$ and $\Delta$ states become equal and opposite.
Here it is important that in the large--$N_c$ limit the 
explicit contact term in the chiral Lagrangian can be neglected compared
to the contact term resulting from the $N$ triangle graph,
cf.\ Eq.~(\ref{contact_combined}), because the former has 
coefficient $1 = O(N_c^0)$ while the latter has $g_A^2 = O(N_c^2)$.
It is the $g_A^2$ term from the $N$ triangle graph,
Eq.~(\ref{Im_F1_contact}), which is matched by the corresponding
term from the $\Delta$ graph, Eq.~(\ref{Im_F1_delta_contact});
there is no explicit Lagrangian contact term in the $\Delta$ case.
Incidentally, this argument shows that introduction of an explicit
``new'' $\pi\pi NN$ contact term together with the $\Delta$ is
not required by the large--$N_c$ limit, supporting our treatment
of this term in Sec.~\ref{subsec:delta}.

\textit{Magnetization density.} In the transverse magnetization density 
the two--pion component obtained with intermediate $N$ only, 
Eq.~(\ref{rho2_largen_chiral}), shows the $N_c$--scaling 
behavior expected on general grounds, Eq.~(\ref{rho_largen_general}). 
The situation is thus very different from the charge density, and 
cancellation between $N$ and $\Delta$ is not required to ensure
the correct $N_c$ scaling of the magnetization density. Indeed, we see 
that the chiral dynamics exploits this freedom and produces $N$ 
and $\Delta$ contributions in a non--trivial ratio.
Using the $N_c$--scaling relations for the couplings and
masses as above, and the expressions for the spectral functions
Eqs.~(\ref{Im_F2_simplified})--(\ref{x_1}) and 
Eqs.~(\ref{Im_F2_delta_simplified})--(\ref{x_Delta_1}),
it is straightforward to show that in the large--$N_c$ limit
\be
\frac{\textrm{Im} \, F_2^V (t)_\Delta}{M_N} 
&=& \frac{1}{2} \, \frac{\textrm{Im} \, F_2^V (t)_N}{M_N}
\; + \; O(N_c^0) 
\hspace{2em} [t = O(N_c^0)] ,
\label{largen_F2}
\ee
and thus
\be
\frac{\rho_2^V (b)_\Delta}{M_N} 
&=& \frac{1}{2} \, \frac{\rho_2^V (b)_N}{M_N}
\; + \; O(N_c^0) 
\hspace{2em} [b = O(N_c^0)] .
\label{largen_rho2}
\ee
Combining the $N$ and $\Delta$ contributions one gets a density that
is $3/2$ times the original density from $N$ only,
\be
\frac{\rho_2^V (b)_N + \rho_2^V (b)_\Delta}{M_N} 
&=& \frac{3}{2} \, \frac{\rho_2^V (b)_N}{M_N}
\; + \; O(N_c^0) 
\hspace{2em} [b = O(N_c^0)] .
\label{largen_rho2_combined}
\ee
Such an enhancement by a factor of $3/2$ from including the $\Delta$
is typically found in the chiral component of matrix elements of 
isovector--vector operators; for example, the same factor was obtained for 
the chiral divergence of the 3--dimensional isovector magnetic radius 
of the nucleon \cite{Cohen:1992uy}; see Ref.~\cite{Cohen:1996zz} for a review. 
It can also be seen by comparing the chiral EFT predictions for the leading 
non--analytic dependence of nucleon matrix elements in the limit 
$M_\pi \rightarrow 0$ with those of chiral soliton models, which
naturally include the contributions from intermediate 
$\Delta$ states \footnote{The equivalence
of the chiral soliton model and two--pion exchange with intermediate
$N$ and $\Delta$ in the case of isoscalar peripheral partonic structure
is discussed in Ref.~\cite{Strikman:2003gz}.}. It appears very natural
that our result for the two--pion component of the nucleon's peripheral 
transverse magnetization density follow the same pattern.

In sum, we find that the two--pion components of the nucleon's isovector 
transverse charge and magnetization densities obey the general large--$N_c$ 
scaling behavior when the contributions from intermediate $\Delta$ 
states are included. In the charge density the $\Delta$ is ``required''
to cancel the wrong leading term in the intermediate $N$ result and
restore the proper $N_c$--scaling; in the magnetization density it is
``optional'' and results in a factor 3/2 enhancement in the large--$N_c$
limit. These theoretical results explain the numerical relation
between $N$ and $\Delta$ contributions observed in 
Sec.~\ref{subsec:delta} (see Fig.~\ref{fig:ft_delta}a and b, and
Fig.~\ref{fig:rho_delta}a and b). More importantly, our findings 
allow us to place the chiral EFT approach to peripheral nucleon structure
firmly in the context of large--$N_c$ QCD.

A more formal approach to combining the $1/N_c$ and chiral expansions
in nucleon structure was proposed recently in Ref.~\cite{CalleCordon:2012xz}
and applied to static nucleon properties. If this approach could be
extended to the near-threshold spectral functions, it could be used
to study peripheral transverse nucleon structure with the help of the
dispersion representation described in Sec.~\ref{subsec:dispersion}.
\section{Spatial region of chiral dynamics}
\label{sec:region}
\subsection{Spectral functions from vector mesons}
\label{subsec:vector}
The chiral EFT methods described in Sec.~\ref{sec:chiral} allow us
to calculate the transverse densities in the nucleon at distances
of the parametric order $b = O(M_\pi^{-1})$, i.e., distances that 
scale as $\textrm{const} \times M_\pi^{-1}$ when
the pion mass is considered small compared to the non--chiral 
mass scales. An important question is at what numerical values of
$b$ the chiral component dominates the non--chiral contributions
and thus represents a good approximation to the overall peripheral
densities in the nucleon. In the space--time picture in the nucleon 
rest frame of Sec.~\ref{subsec:charge_vs_current}, 
this defines the region of distances where one can truly think of 
the system as a ``bare'' nucleon and a peripheral pion, outside of 
the range of interaction defined by the intrinsic (or non--chiral) size 
of the bare nucleon. In the context of scattering processes,
it defines the region of impact parameters where the probe interacts
predominantly with the chiral component of the nucleon.

%
% FIGURE
%
\begin{figure}
\includegraphics[width=.38\textwidth]{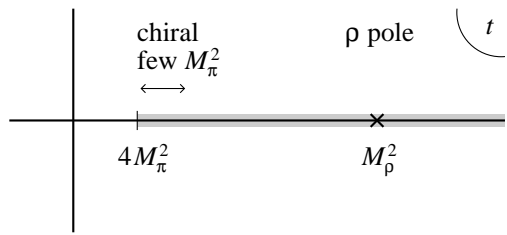}
\caption[]{Chiral and non--chiral contributions to the spectral functions 
of the isovector nucleon form factors.
In the near--threshold region $t - 4 M_\pi^2 \sim \textrm{few} \, M_\pi^2$
the spectral functions are governed by chiral dynamics and approximated
by the chiral EFT results Eqs.~(\ref{Im_F1_noncontact})--(\ref{x}). 
In the region 
$t \lesssim 1\, \textrm{GeV}^2$ they are dominated by the $\rho$ meson
resonance and approximated by the pole form Eqs.~(\ref{vm_spectral_1})
and (\ref{vm_spectral_2}). High--mass states at $t > 1\, \textrm{GeV}$
give negligible contribution to peripheral densities.}
\label{fig:tplane_rho}
\end{figure}
The dispersion representation of the transverse densities described
in Sec.~\ref{subsec:dispersion}, Eq.~(\ref{rho_dispersion}), allows
us to answer this question in a natural way (see Fig.~\ref{fig:tplane_rho}). 
The ``chiral'' component of the isovector transverse densities results from 
the near--threshold region $t = 4 M_\pi^2 + \textrm{few} \; M_\pi^2$, 
where the spectral functions are governed by chiral dynamics and are
well approximated by the chiral EFT expressions 
Eqs.~(\ref{Im_F1_noncontact})--(\ref{x}). ``Non--chiral'' densities 
are generated by higher--mass states in the spectral function, which 
include the prominent $\rho$ meson resonance in the two--pion channel
\cite{Hohler:1976ax} and a continuum of higher--mass 
hadronic states \cite{Belushkin:2006qa}.
By comparing the ``chiral'' and ``non--chiral'' densities defined 
in this sense we can quantify at what 
peripheral distances the chiral component becomes numerically dominant
and in this way identify the spatial region where the overall densities
are governed by chiral dynamics. Note that the dispersion representation 
enables us to perform this comparison model--independently and without
without double--counting. 

The non--chiral isovector transverse densities at distances 
$b \gtrsim 1 \, \textrm{fm}$ are overwhelmingly due to the $\rho$ 
resonance in the two--pion channel \cite{Miller:2011du}. States with 
masses $t > 1 \, \textrm{GeV}^2$ in the spectral function play a direct 
role only at small distances $b < 0.5\, \textrm{fm}$, 
which we are not interested in here. For the purpose of our comparison 
between chiral and non--chiral densities in the nucleon's periphery
it will be sufficient to consider only the non--chiral density
generated by the $\rho$ meson mass region of the spectral function.
We parametrize the distribution of strength in this region by a 
simple pole at the $\rho$ meson mass $M_\rho = 0.77 \, \textrm{GeV}$,
\be
\frac{1}{\pi} \, \textrm{Im} \, F_1^V (t)_{\rho}
&=& c_{1\rho} \, M_\rho^2 \delta (t - M_\rho^2) ,
\label{vm_spectral_1}
\\[1ex]
\frac{1}{\pi} \, \textrm{Im} \, F_2^V (t)_{\rho}
&=& c_{2\rho} \, M_\rho^2 \delta (t - M_\rho^2) ,
\label{vm_spectral_2}
\ee
where $c_{1\rho}$ and $c_{2 \rho}$ are parameters determined by
empirical information. In the Dirac
spectral function Eq.~(\ref{vm_spectral_1}), the vector meson 
dominance (or VMD) model, in which the entire isovector charge of the nucleon 
is carried by $\rho$ meson exchange, $F_1^V(0)_\rho = 1/2$, would 
correspond to
\beq
c_{1\rho} \;\; = \;\; \textstyle{\frac{1}{2}}
\hspace{2em} \textrm{(VMD)}.
\label{c_rho_1_vmd}
\eeq
A more realistic value is obtained using the empirical $\rho NN$
coupling from meson exchange parametrizations of the $NN$ interaction, 
$g_{\rho NN} = 3.25$ \cite{Machleidt:1987hj,Machleidt:2000ge}, and 
the $\rho$ meson coupling to the electromagnetic
current, $f_\rho = 5.01$, as extracted from the $\rho \rightarrow e^+e^-$
partial decay width $\Gamma (\rho \rightarrow e^+e^-) =  (\alpha M_\rho /3)
(e/f_\rho)^2 = 6.9 \, \textrm{keV}$ \cite{Nakamura:2010zzi}, where 
$\alpha = e^2/(4 \pi) \approx 1/137$ is the fine structure constant,
\beq
c_{1\rho} \;\; = \;\; g_{\rho NN}/f_\rho \;\; = \;\; 0.65
\hspace{2em} \textrm{(empirical couplings)}.
\label{c_rho_1_empirical}
\eeq
This value is $\sim 30\%$ larger than the simple VMD
result, Eq.~(\ref{c_rho_1_vmd}). The explanation is that in the 
full spectral function the ``excess'' isovector charge from $\rho$ 
exchange is compensated by a negative contribution from other states
above $\sim 1 \, \textrm{GeV}^2$. This is related to the
$1/t^{2}$ asymptotic power behavior of the spacelike form factor
in QCD for $|t| \rightarrow \infty$ (up to logarithmic corrections), 
which requires vanishing of the coefficient of $1/t$ in the
asymptotic series, or
\beq
\int\limits_{4M_\pi^2}^\infty dt \; \textrm{Im}\; F_{1, 2}^V (t)
\;\; = \;\; 0,
\label{absence_monopole}
\eeq
and is consistently seen in empirical fits to nucleon form 
factor data \cite{Belushkin:2006qa}. 
It can also be demonstrated in a two--pole model of the 
spectral density, in which the strength of the higher--mass states above 
the $\rho$ is parametrized by a second pole with negative residue
such that Eq.~(\ref{absence_monopole}) is satisfied; if the 
mass of that second pole is taken to be that of the first $\rho'$ 
resonance established in $e^+e^-$ annihilation experiments, 
$M_{\rho'} = 1.47 \, \textrm{GeV}$, one obtains
$c_{1\rho} = \frac{1}{2} M_{\rho'}^2/(M_{\rho'}^2 - M_\rho^2)
= 0.70$, in reasonable agreement with Eq.~(\ref{c_rho_1_empirical}). 
We shall use the empirical value
Eq.~(\ref{c_rho_1_empirical}) in our numerical estimates below.

The parameter $c_{2\rho}$ in Eq.~(\ref{vm_spectral_2}) determines the 
strength of the Pauli spectral function in the $\rho$ mass region.
In the simple VMD model (or the two--pole extension) it would be fixed 
by the isovector anomalous magnetic moment, namely
\beq
c_{2\rho} / c_{1\rho} \;\; = \;\; \kappa_p - \kappa_n \;\; = \;\; 3.7
\hspace{2em} \textrm{(VMD)} .
\label{c2_c1_vmd}
\eeq
In meson exchange phenomenology the ratio Eq.~(\ref{c2_c1_vmd}) is 
directly given by the ratio of the helicity--flip and non--flip 
$\rho NN$ couplings. The value obtained with the empirical 
couplings used in the parametrization of the $NN$ 
interaction \cite{Machleidt:1987hj,Machleidt:2000ge} 
is substantially larger,
\beq
c_{2\rho} / c_{1\rho} \;\; = \;\; 6.1
\hspace{2em} \textrm{(empirical couplings)} .
\label{c2_c1_empirical}
\eeq
Inspection of the full dispersion--theoretical result for the low--mass
spectral functions \cite{Belushkin:2005ds} shows that the ratio 
$\textrm{Im}\, F^V_2(t) / \textrm{Im}\, F^V_1(t)$ varies over 
the region $t \leq 1\, \textrm{GeV}^2$, roughly in the range between 
the values of Eqs.~(\ref{c2_c1_vmd}) and (\ref{c2_c1_empirical}),
and particularly fast near the $\rho$ resonance mass. With the simple
parametrization Eq.~(\ref{vm_spectral_2}) we are clearly not able to 
express such details. Rather, we shall use Eq.~(\ref{vm_spectral_2})
with the empirical value of the couplings Eq.~(\ref{c2_c1_empirical})
and treat the discrepancy between the values of Eqs.~(\ref{c2_c1_vmd}) 
and (\ref{c2_c1_empirical}) as a measure of the theoretical uncertainty
of our parametrization. For the numerical estimates performed in the
following this turns out to be fully sufficient. 
\subsection{Chiral vs.\ nonchiral densities}
\label{subsec:chiral_nonchiral}
%
% FIGURE
%
\begin{figure}
\begin{tabular}{ll}
\includegraphics[width=.45\textwidth]{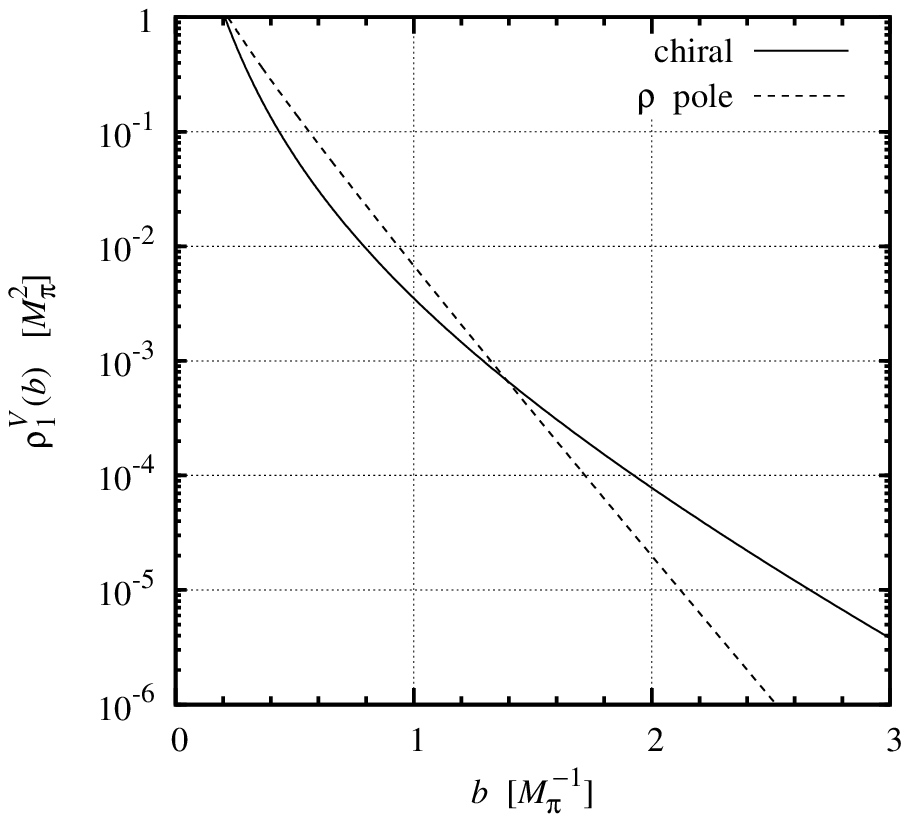}
&
\includegraphics[width=.45\textwidth]{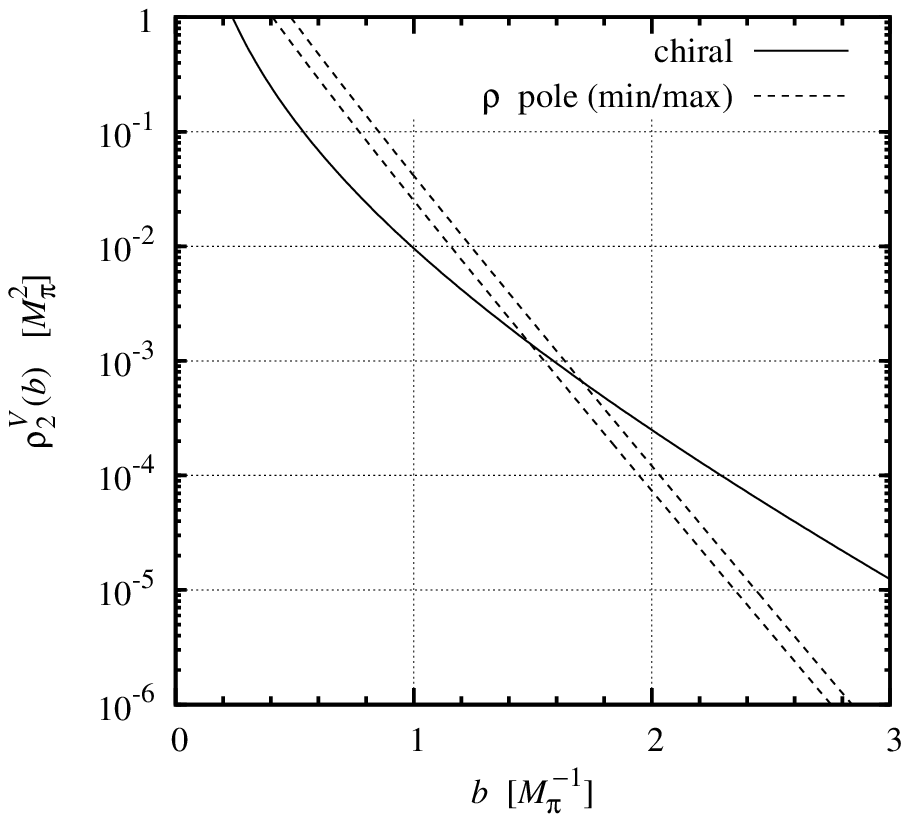}
\\[-1ex]
(a) & (b)
\end{tabular}
\caption[]{Comparison of the chiral component of the transverse charge
and magnetization densities from Eqs.~(\ref{Im_F1_noncontact})--(\ref{x}) 
(cf.~Fig.~\ref{fig:rho_b}) with the non--chiral densities parametrized 
by a $\rho$--meson pole, Eq.~(\ref{rho12_vm}).
(a) Charge density $\rho_1(b)$. The coefficient of the $\rho$ pole
parametrization is given by Eq.~(\ref{c_rho_1_empirical}).
(b) Magnetization density $\rho_2 (b)$. The two curves for the $\rho$ pole
parametrization correspond to the coefficients of 
Eqs.~(\ref{c2_c1_vmd}) and (\ref{c2_c1_empirical}) and reflect the
uncertainty of the empirical parametrization.}
\label{fig:vm}
\end{figure}
With the higher--mass spectral functions parametrized by 
Eqs.~(\ref{vm_spectral_1}) and (\ref{vm_spectral_2}), and the
parameters given by Eq.~(\ref{c_rho_1_empirical}) and 
(\ref{c2_c1_empirical}), we can now quantitatively compare
the ``chiral'' and ``non--chiral'' components of the nucleon's
peripheral transverse densities in the sense specified above.
The transverse densities generated by the $\rho$--pole spectral 
functions of Eqs.~(\ref{vm_spectral_1}) and (\ref{vm_spectral_2})
via the dispersion integral Eq.~(\ref{rho_dispersion}) are
\be
\rho^V_{1, 2} (b)_{\rho} &=& c_{1\rho, \, 2\rho} \, M_\rho^2
\; \frac{K_0 (M_\rho b)}{2\pi} 
\;\; \sim \;\; c_{1\rho, \, 2\rho} \, M_\rho^2
\; \frac{e^{-M_\rho b}}{\sqrt{8\pi M_\rho b}}
\hspace{2em} (b \rightarrow \infty).
\label{rho12_vm}
\ee
The asymptotic form given by the last expression describes the exact 
density with an accuracy better than $10\%$ already
for $M_\rho b > 1$ and can be used for calculational purposes.
In Fig.~\ref{fig:vm} we compare the ``chiral'' component of the 
densities obtained from the chiral EFT spectral functions,
Eqs.~(\ref{Im_F1_noncontact})--(\ref{x}), with the ``non--chiral'' 
densities from the $\rho$ meson pole parametrization, Eq.~(\ref{rho12_vm}). 
For clarity we plot here the chiral two--pion densities without the 
intermediate $\Delta$ contribution; the latter is numerically small 
at large $b$ (see Fig.~\ref{fig:rho_delta})
and does not substantially affect our conclusions. 
Figure~\ref{fig:vm} shows that in the limit $b \rightarrow \infty$ 
the chiral components of the charge and magnetization density indeed 
dominate, because their exponential fall--off is governed by the 
scale $2 M_\pi$ rather than $M_\rho$ (the deviations from exponential 
behavior due to the pre-exponential factor of the chiral component
were discussed in Sec.~\ref{subsec:chiral_densities}; 
see Fig.~\ref{fig:rho_b}). However, the numerical values of $b$
required for the chiral component to become substantially larger 
than the non--chiral one are surprisingly large. In both the charge 
and the magnetization density one has to go to 
$b \gtrsim 2\, M_\pi^{-1}$ for the chiral charge density to 
become 3--4 times larger than the one from the $\rho$ pole 
parametrizations. It is only at these distances that the theoretical
expectation based on exponential asymptotics is borne out by the
actual numerical values of the densities.

Some remarks are in order regarding the aim and significance of the 
numerical studies done here. First, the purpose of the numerical 
comparison of the chiral component with the $\rho$ pole parametrization 
in Fig.~\ref{fig:vm} is only to determine \textit{at what distances} 
the chiral component becomes numerically dominant.
This is to be understood in the sense of large--$b$ asymptotics: 
we compare the result of the ``theoretically leading'' singularity at 
$t \sim 4 M_\pi^2$ with a model of the ``theoretically subleading''
higher--mass singularities, summarized by a pole at $t = M_\rho^2$.
We do not advocate to add the chiral component and the $\rho$ pole
and construct in this way a model of the full spectral functions.
Excellent dispersion--theoretical parametrizations of the full spectral 
functions are available which serve that purpose
\cite{Belushkin:2006qa,Belushkin:2005ds,Ditsche:2012fv}. 
These parametrizations are fully consistent with the chiral EFT results 
at $t \lesssim 10 \, M_\pi^2$ (where the chiral expansion converges)
and embed them in an interpolating description that extends up to
$t \sim 1\, \textrm{GeV}^2$. A spectral analysis of the transverse
charge density based on the full spectral function \cite{Miller:2011du} 
arrives at practically the same conclusion regarding the region of 
distances associated with the ``chiral'' component as the estimate 
presented here. Second, for the stated purpose we only need to compare 
the ``chiral'' and ``non--chiral'' densities \textit{on a logarithmic scale,}
as shown in Fig.~\ref{fig:vm}, and roughly determine at what distances 
the chiral component becomes dominant. For this purpose the 
simple parametrizations of the non--chiral density described 
in Sec.~\ref{subsec:vector} are fully adequate,
and the uncertainties in the parameters do not affect our 
conclusions [see Fig.~\ref{fig:vm}b for the magnetization density obtained
with the parameters of Eqs.~(\ref{c2_c1_vmd}) and (\ref{c2_c1_empirical}),
which differ by a factor 1.6]. Likewise, it was shown in 
Ref.~\cite{Miller:2011du} that account of the finite width of the 
$\rho$ meson resonance increases the peripheral densities generated
by the $\rho$ only moderately in the region of interest 
(by $\sim 40\%$ at $b = 2\, \textrm{fm}$) and does not substantially 
change our conclusions regarding the region of dominance of the 
chiral component.

The results of Fig.~\ref{fig:vm} have interesting implications for our 
general understanding of nucleon structure. First, they invalidate the
naive picture of the nucleon's spatial structure as a ``core'' of size
$\sim 1\, \textrm{fm}$ surrounded by a ``pion cloud'' generated by chiral 
dynamics. The numerical results show that the density associated with 
the $\rho$ meson region of the spectral function, which is \textit{not}
associated with chiral dynamics, dominates up to much larger distances,
and that one has to go to $b \gtrsim 2\, M_\pi^{-1} \sim 3 \, \textrm{fm}$
to clearly see the component due to chiral dynamics. Only at
such transverse distances can one think of the relevant configurations 
in nucleon's light--cone wave function
(see Secs.~\ref{subsec:definition} and \ref{subsec:charge_vs_current}) 
as a nucleon--like core and a peripheral pion interacting through the 
physical $\pi N$ coupling. Second, dominance of the chiral component 
starts at roughly the
same distances $b \gtrsim 2\, M_\pi^{-1}$ in the charge and magnetization
densities. Again, this runs counter to the expectation that the
``pion cloud'' should be more prominent in the magnetization density
(see also Sec.~\ref{subsec:charge_vs_current}). In absolute terms the 
chiral component is indeed larger in $\rho_2(b)$ than in $\rho_1(b)$, 
but the same is true for the non--chiral density
parametrized by the $\rho$ pole, so that the proportion remains roughly
the same. Note that this conclusion changes when taking into
account the $\Delta$ contribution, as it diminishes the change density and
enhances the magnetization density (see Sec.~\ref{sec:delta}); however,
with the $\Delta$ one is leaving the domain of strict chiral dynamics,
so that the comparison with the ``non--chiral'' density modeled by
$\rho$ pole becomes less meaningful.

Our findings do not imply that chiral dynamics plays no role in
transverse nucleon structure at $b \lesssim 2 M_\pi^{-1}$. We only
find that as such distances the behavior of the transverse densities 
will always be essentially influenced by the nucleon's ``intrinsic'' 
size, i.e., a distance scale other than $M_\pi^{-1}$, represented by the 
$\rho$ meson mass in the example discussed here. It is only at larger
distances that the densities lose the memory of this intrinsic size
of the nucleon and the latter can be thought of as a structureless source
coupling to soft pions. Chiral symmetry still plays an important role 
in nucleon structure at smaller distances, as a constraint on the 
long--distance behavior of the overall effective dynamics. Dynamical 
models have been formulated which ``interpolate'' between the universal
chiral dynamics at distances $M_\pi^{-1}$, summarized by the chiral 
Lagrangian, and dynamics at shorter distance scales giving rise to the
nucleon's non--chiral intrinsic size. One such class of models
is the skyrmion, which describes the nucleon as a soliton of a 
non--linear chiral Lagrangian coupled to vector meson 
fields (or, equivalently, a Lagrangian with higher--derivative terms 
resulting from integrating out the vector mesons); here the intrinsic size 
of the nucleon is determined by the vector meson mass and the inherent 
non--linearity of the dynamics; see Refs.~\cite{Zahed:1986qz,Meissner:1987ge} 
for a review. Another example is the chiral quark--soliton
model \cite{Kahana:1984be,Diakonov:1987ty}, 
which uses constituent quarks coupled to the pion field as
effective degrees of freedom; here the short--distance scale governing 
the intrinsic size of the nucleon is the constituent quark mass.
Both models are explicit realizations of the generic soliton picture of
baryons in the large--$N_c$ limit of QCD \cite{Witten:1979kh}
and therefore include the 
equivalent of $\Delta$ intermediate states in chiral processes.
They give rise to a successful phenomenology of nucleon form factors
at intermediate momentum transfers $|t| \lesssim 1\, \textrm{GeV}$
(see Refs.~\cite{Meissner:1987ge,Christov:1995vm} for reviews),
which testifies to the proper implementation of the nucleon's non--chiral
intrinsic size. They can therefore be used to model the nucleon's
transverse densities over a wide range of distances 
$b \gtrsim 0.3\, \textrm{fm}$ in a manner that matches with
chiral dynamics at large distances $b \gtrsim 3\, \textrm{fm}$
(including intermediate $\Delta$).
\section{Moments and chiral divergences}
\label{sec:moments}
\subsection{Moments of transverse densities}
In traditional applications of chiral EFT one studies the dependence of 
nucleon observables such as the vector and axial charges, charge radii,
etc., on the pion mass in the limit $M_\pi \rightarrow 0$. Of particular
interest is the leading non--analytic behavior of these quantities
(``chiral singularities''), which can be traced back to universal
characteristics of the effective chiral dynamics. In the context of
the spatial representation of nucleon structure (see 
Sec.~\ref{subsec:definition}) these quantities appear as weighted 
integrals of the transverse densities over $b$. To conclude our
study we want to show how the chiral components of the transverse
charge and magnetization densities at distances $b = O(M_\pi^{-1})$ 
computed in Sec.~\ref{sec:chiral} are related to the well--known chiral 
singularities in the nucleon charge and magnetic radii. This serves 
as a further test of the formalism developed here and offers
new insights into the spatial support of the chiral divergences.
More generally, it explains the connection between the traditional
usage of chiral EFT for bulk quantities and the spatial picture of 
nucleon structure employed here.

For theoretical analysis it is convenient to consider ``truncated'' 
moments of the transverse charge and magnetization densities, defined as
\be
M_{1, 2}(n, b_0) \;\; \equiv \;\; \int d^2 b \, \Theta (b > b_0) \; 
b^{2 n} \; \rho_{1, 2} (b)
\hspace{2em} (n = 0, 1, 2, \ldots).
\label{moment_def}
\ee
The theta function restricts the integration to distances $b > b_0$.
For $b_0 = 0$ the truncated moments coincide with the usual moments
of the densities. In particular, the moments with $n = 0$ reproduce
the form factors at $t = 0$,
\be
M_1(n = 0, \, b_0 = 0) \; = \; F_1(t = 0),
\hspace{2em}
M_2(n = 0, \, b_0 = 0) \; = \; F_2(t = 0);
\ee
their values for the isoscalar and isovector combinations in our
convention are given in Eq.~(\ref{S_V_def}). More generally, for
any integer $n \geq 0$ the moment with $b_0 = 0$ is proportional
to the $n$'th derivative of the form factor at $t = 0$,
\beq
M_{1, 2}(n, \, b_0 = 0) \;\; = \;\; 2^{2 n} n! \; 
\frac{d^n F_{1, 2}}{dt^n}  (t = 0) .
\label{moment_derivative}
\eeq
The coefficient can be determined by repeated differentiation of
the Fourier representation of the form factor, Eq.~(\ref{rho_def}),
with respect to the vector $\bm{\Delta}_T$, or more elegantly
by comparing the dispersion integral for the moments given
below, Eq.~(\ref{moment_dispersion}), with the dispersion integral
for the derivatives of the form factor obtained by 
differentiation of Eq.~(\ref{dispersion}) with respect to $t$.
The normalized averages of powers of $b^2$ over the transverse 
densities are obtained as
\beq
\langle b^{2n} \rangle_{1, 2}
\;\; \equiv \;\; \frac{\int d^2 b \; b^{2 n} \; \rho_{1, 2}(b)}
{\int d^2 b \; \rho_{1, 2}(b)}
\;\; = \;\; \frac{M_{1,2}(n, \, b_0 = 0)}{M_{1,2}(0, \, b_0 = 0)} 
\eeq
(with the explicit normalization factor in the denominator, the expressions 
are valid for any normalization convention of the form factors).

The analytic properties of the form factor guarantee that the transverse
densities decay exponentially at $b \rightarrow \infty$; in the case of
the isovector densities the exponential decay is $\sim \exp (- 2 M_\pi b)$
(cf.\ Sec.~\ref{subsec:dispersion}). The $b$--integral in 
Eq.~(\ref{moment_def}) therefore converges for any $n \geq 0$, and 
the series of moments provides an alternative representation of the
information contained in the transverse densities. This can also be
deduced from the fact that the form factor is analytic near $t = 0$, 
and that the the moments are proportional to its derivatives,
cf.\ Eq.~(\ref{moment_derivative}). 

From the dispersion representation of the transverse densities, 
Eq.~(\ref{rho_dispersion}), we can now derive a dispersion representation
of the truncated moments defined by Eq.~(\ref{moment_def}). 
Multiplying Eq.~(\ref{rho_dispersion}) by $b^{2n}$ and integrating 
over $b$ we obtain
\be
M_{1, 2} (n, b_0) &=& \int\limits_{4 M_\pi^2}^\infty dt \; 
\left[ \, \int\limits_{b_0}^\infty db \; b^{2 n + 1} \; K_0 (\sqrt{t} b) 
\right]
\; \frac{\textrm{Im} \, F_{1, 2}(t + i0)}{\pi}
\\[1ex]
&=& 2^{2 n} (n!)^2
\int\limits_{4 M_\pi^2}^\infty \frac{dt}{t^{n + 1}} 
\; R (n, \sqrt{t} b_0)
\; \frac{\textrm{Im} \, F_{1, 2}(t + i0)}{\pi} .
\label{moment_dispersion}
\ee
The function $R$ introduced in the last step is defined 
as the dimensionless integral (here $z \equiv \sqrt{t} b$ and
$z_0 \equiv \sqrt{t} b_0$)
\beq
R(n, z_0) \;\; \equiv \;\; \frac{1}{2^{2 n} (n!)^2}
\int_{z_0}^\infty dz \, z^{2 n + 1} \, K_0 (z)
\eeq
and has the properties that, for any $n \geq 0$, it is normalized to 
unity at zero argument and vanishes exponentially at large values
\be
R(n, 0) &=& 1, 
\\
R(n, z_0) &\sim& \sqrt{\frac{\pi}{2}} \; z_0^{2 n + 1/2} \; e^{-z_0} 
\hspace{2em} (z_0 \rightarrow \infty) .
\label{R_def}
\ee
It thus acts as an ultraviolet cutoff in the dispersion integral
for the truncated moment, Eq.~(\ref{moment_dispersion}), which
has no effect on masses $\sqrt{t} \ll 1/b_0$ but exponentially 
suppresses the contributions from masses $\sqrt{t} \gg 1/b_0$.
Note that for $b_0 = 0$ the function $R$ in Eq.~(\ref{moment_dispersion}) 
is identically equal to unity, and the dispersion integral reverts
to that for the usual moments (or, up to a factor, the derivatives
of the form factor), where large masses are not suppressed.
Thus we see that the elimination of small transverse distances
in the truncated moments Eq.~(\ref{moment_def}) implements a very
natural ultraviolet cutoff in the dispersion integral and renders
it exponentially convergent for all $n$. In a sense, the truncated
moments Eq.~(\ref{moment_def}) can be regarded as a coordinate--space
based regularization of the derivatives of the form factor.
\subsection{Chiral divergence of moments}
We now want to demonstrate that the chiral component of the isovector 
charge and magnetization densities at $b \sim M_\pi^{-1}$, derived in 
Sec.~\ref{subsec:chiral_densities}, reproduces the well--known chiral 
divergences of the nucleon's isovector charge and magnetic radius.
This will establish the connection between our spatial identification
of the nucleon's chiral component and the pion mass dependence of 
traditional chiral EFT and reveal what physical distances are 
involved in the chiral divergences of these quantities.

To this end we consider the truncated $n = 1$ 
moments of the isovector charge and magnetization densities,
\beq
M^V_{1, 2}(1, \, b_0) \;\; = \;\; \int d^2 b \; \Theta (b > b_0) \; 
b^2 \; \rho^V_{1, 2} (b) ,
\eeq
which for $b_0 = 0$ are, up to a factor, equal to the first derivatives 
of the form factors,
\beq
M_{1, 2}^V(1, \, b_0 = 0) \;\; = \;\; 4 \, \frac{dF^V_{1, 2}}{dt}(t = 0)
\label{moment_1_derivative} .
\eeq
Their dispersion representation is provided by 
Eq.~(\ref{moment_dispersion}). Changing the integration variable to 
the dimensionless variable $u \equiv \sqrt{t}/(2 M_\pi)$, such that 
the threshold $t = 4 M_\pi^2$ corresponds to $u = 1$, cf.\ Eq.~(\ref{u_def}),
the dispersion integral becomes
\beq
M^V_{1, 2}(1, \, b_0) \;\; = \;\; \frac{2}{M_\pi^2} \int\limits_1^\infty
\frac{du}{u^3} \; R (1, \, 2 M_\pi b_0 u) \; 
\frac{\textrm{Im} F_{1, 2}^V(t)}{\pi} 
\hspace{2em} (t = 4 M_\pi^2 u^2),
\label{moment_1_dispersion_u}
\eeq
where the kernel $R$ is defined in Eq.~(\ref{R_def}). We want to evaluate
this integral with the leading--order chiral result for the isovector 
spectral functions at $t = O(M_\pi^2)$ quoted in 
Sec.~\ref{subsec:chiral_densities}, and extract the leading
asymptotic behavior of the truncated moment at $M_\pi \rightarrow 0$.
The cutoff $b_0$ in the moment is regarded as a non--chiral scale, i.e., 
it is \textit{not} of order $O(M_\pi^{-1})$ and remains finite in the limit 
$M_\pi \rightarrow 0$. The leading chiral singularities of
the moments can be obtained from the leading term in the the heavy--baryon 
expansion of the spectral functions derived in Appendix~\ref{app:heavy},
Eqs.~(\ref{Im_F1_heavy_u}) and (\ref{Im_F2_heavy_u}); one can easily 
show that higher--order terms in the heavy--baryon expansion of the 
spectral function do not modify the leading asymptotic behavior for 
$M_\pi \rightarrow 0$. In leading order of $M_\pi / M_N$ the explicit 
expressions given by Eqs.~(\ref{Im_F1_heavy_u}) and (\ref{Im_F2_heavy_u}) 
are (combining the intermediate nucleon and contact terms in the 
Dirac spectral function)
\be
\frac{1}{\pi} \, \textrm{Im} \, F_1^V (t)
&=& \frac{M_\pi^2}{(4\pi F_\pi)^2} \; \frac{\sqrt{u^2 - 1}}{u}
\left[ \frac{5 g_A^2 + 1}{3} \, (u^2 - 1)  \; + \;  2 g_A^2 \right]
\; + \; O\left(\frac{M_\pi^3}{M_N}\right) ,
\label{F_1_heavy_moment}
\\[1ex]
\frac{1}{\pi} \, \textrm{Im} \, F_2^V (t)
&=& \frac{\pi g_A^2 M_\pi M_N}{(4\pi F_\pi)^2} \; \frac{u^2 - 1}{u} 
\; + \; O(M_\pi^2)
\hspace{2em} (t = 4 M_\pi^2 u^2).
\label{F_2_heavy_moment}
\ee
In the representation of Eq.~(\ref{moment_1_dispersion_u}) the chiral
singularities of the moments arise from the $u \rightarrow \infty$
region of the integral, where the behavior of the integrand depends
on both the spectral functions and the cutoff $R$. This behavior needs
to be discussed separately for the moment of the charge and magnetization
density.

\textit{Moment of charge density (``charge radius'').} The spectral function 
of the isovector Dirac form factor in leading--order heavy--baryon expansion,
Eq.~(\ref{F_1_heavy_moment}), behaves at large $u$ as
\beq
\frac{1}{\pi} \, \textrm{Im} \, F_1^V (t)
\;\; \sim \;\; \frac{M_\pi^2}{(4\pi F_\pi)^2} \; 
 \frac{5 g_A^2 + 1}{3} \, u^2 
\hspace{2em} (u \rightarrow \infty, \; t = 4 M_\pi^2 u^2).
\eeq
The integrand in Eq.~(\ref{moment_1_dispersion_u}), excluding the
factor $R$, effectively behaves as $u^{-1}$ at large $u$. The integral
thus has a would--be logarithmic divergence that is regulated by 
the function $R$, which restricts the integration to values
$u \lesssim 1/(M_\pi b_0)$. The truncated moment of the transverse 
charge density acquires a logarithmic chiral singularity of the form
\beq
M^V_1(1, \, b_0) \;\; \sim \;\; \frac{2(5 g_A^2 + 1)}{3 (4\pi F_\pi)^2}
\; \log (M_\pi b_0) \;  + \; O(M_\pi^0) .
\label{moment_1_chiral_log}
\eeq
That the argument of the logarithm of the truncated moment involves the 
combination $M_\pi b_0$ shows explicitly that the minimum transverse
distance $b_0$ plays the role of an ultraviolet cutoff here.
The coefficient of the chiral logarithm of the truncated moment, 
Eq.~(\ref{moment_1_chiral_log}), agrees with the well--known chiral 
logarithm of the nucleon charge radius obtained in standard chiral EFT 
calculations of the nucleon form factors with other regularization schemes,
such as dimensional 
regularization \cite{Gasser:1987rb,Bernard:1992qa,Kubis:2000zd}.

It is natural to ask what physical distances are responsible for the
chiral logarithm of the charge radius within our spatial picture.
This question can be answered by considering the case that $1/M_\pi$ 
is very much larger than $b_0$, while $b_0$ itself is of the order 
of the nucleon's non--chiral size. In this case there is a broad range of 
distances $b_0 \ll b \ll 1/M_\pi$. The chiral logarithm is the result
of the integration over this broad range. That the coefficient of the
logarithm is the same as that obtained with other regularization
schemes shows that the approximations made in calculating the transverse
charge density at large $b$ are sufficiently accurate to permit 
integration down to $b \sim b_0$ with logarithmic accuracy.
The picture sketched here follows the general pattern by which 
``large logarithms'' in quantum field theory arise from integrating 
over modes with wavelengths in a range limited by two widely 
different scales.

\textit{Moment of magnetization density (``magnetic radius'').} 
The spectral function of the isovector Pauli form factor in leading--order 
heavy--baryon expansion, Eq.~(\ref{F_2_heavy_moment}), behaves as
$u$ in the limit $u \rightarrow \infty$. The integrand in
Eq.~(\ref{moment_1_dispersion_u}) (excluding the factor $R$) therefore 
drops as $u^{-2}$ at large $u$, and the integral converges without
the ultraviolet cutoff by the function $R$. The leading power behavior
of the moment in $M_\pi$ is obtained by simply setting $M_\pi = 0$ in 
the integral, whence the function $R$ becomes unity,
\be
M^V_2(1, \, b_0) &\sim& 
\frac{\pi g_A^2 M_N}{(4\pi F_\pi)^2 M_\pi} \; \int\limits_1^\infty \! du 
\; \frac{u^2 - 1}{u^4} \; + \; O(M_\pi^0) 
\\
&=& \frac{2 \pi g_A^2 M_N }{3 (4\pi F_\pi)^2 M_\pi} \; + \; O(M_\pi^0).
\label{moment_2_chiral_div}
\ee
The $b^2$--moment of the transverse magnetization density diverges 
as $M_\pi^{-1}$ in the chiral limit. This result agrees in power and 
coefficient with the well--known chiral divergence of the slope of the 
isovector Pauli form factor (or the nucleon's isovector magnetic radius) 
obtained in standard chiral 
EFT \cite{Gasser:1987rb,Bernard:1992qa,Kubis:2000zd}.

The power--like chiral singularity of the truncated $b^2$--moment of the 
transverse magnetization density, Eq.~(\ref{moment_2_chiral_div}), does not 
depend on the value of the short--distance cutoff $b_0$. It shows that 
this chiral singularity really arises from the integration over distances 
$b \sim M_\pi^{-1}$. The situation is different from the moment of the 
charge density, where integration over a broad range 
$b_0 \ll b \ll M_\pi^{-1}$, extending down to non--chiral distances, 
is required to bring about the logarithmic singularity. In this sense, 
the power--like divergence of the magnetization 
density moment represents a purer chiral long--distance effect than
the logarithmic divergence of the charge density moment.

In sum, our investigation confirms that the $b^2$--weighted moments
of the chiral component of the transverse charge and magnetization density 
reproduce the well--known chiral singularities of the isovector charge 
and magnetic radius in the limit $M_\pi \rightarrow 0$. It shows that 
the approximations made in our calculation of the peripheral densities
in chiral EFT are sufficient to permit integration over $b$ 
with the necessary accuracy. In the logarithmic singularity of the 
charge density moment there is (necessarily) a residual dependence
on the short--distance cutoff $b_0$, while the power--like singularity
of the magnetization density moment is completely independent of $b_0$
and represents a pure large--distance effect. In higher moments of the
densities ($b^4, b^6$ etc.) short--distance contributions are even more
suppressed; these moments exhibit power--like divergences that can
likewise be obtained by integrating the chiral result for the 
peripheral densities at $b \sim M_\pi^{-1}$.

A comment is in order regarding the $n = 0$ truncated moment of the 
densities, which gives the total isovector charge and anomalous magnetic 
moment located at transverse distances $b > b_0$,
\beq
M^V_{1, 2}(0, \, b_0) \;\; = \;\; \int d^2 b \; \Theta (b > b_0) \; 
\rho^V_{1, 2} (b) .
\label{moment_total}
\eeq
As there is no factor $b^2$, contributions from short distances
are not suppressed in this integral. We therefore cannot evaluate
the moment Eq.~(\ref{moment_total}) using only the peripheral 
densities at $b = O(M_\pi^{-1})$ computed in 
Sec.~\ref{subsec:chiral_densities} [more precisely, we could do so 
only for $b_0 = O(M_\pi^{-1})$, which would not be interesting].
To calculate the $M_\pi$ dependence of the moment Eq.~(\ref{moment_total}) 
we would need also the chiral contributions to the transverse
densities at ``non--chiral'' distances. The latter arise from 
chiral EFT processes in which the current operator couples to the
nucleon field. Those diagrams could also be computed within our
dispersion approach and give either contributions to the density at
finite distances $b \sim O(M_N^{-1})$ (the loop diagram with a two--nucleon 
cut in the $t$--channel) or delta functions $\delta^{(2)}(\bm{b})$ 
(loop diagrams with no $t$--channel cut). Such a calculation could show 
in what sense, and with what accuracy, the chiral EFT result approximates
the empirical central (non--peripheral) charge and magnetization densities 
in the nucleon, which are known to dominated by the $\rho$ meson region of 
the spectral exchange over a wide region of distances; see 
Ref.~\cite{Miller:2011du} and Sec.~\ref{sec:region}.
In the present work the focus is on peripheral densities, and we leave 
this exercise to a future study.
\section{Summary and outlook}
\label{sec:summary}
\subsection{Specific results}
In this work we have studied the transverse charge and magnetization 
densities in the nucleon's chiral periphery using methods of dispersion 
analysis and chiral effective field theory. Our investigation has provided 
many new insights into the behavior of the transverse densities and the 
merits of the theoretical methods employed in their calculation.
In the following we summarize the specific results for the transverse 
densities, the methodological aspects of broader relevance,
and possible experimental tests of the structures found here.
We also discuss possible future extensions and applications of
the methods developed here. 

Our study has produced the following specific results regarding the 
structure of the nucleon's peripheral transverse charge and magnetization 
densities:
\begin{itemize}
\item \textit{Exponential vs.\ pre-exponential dependence.}
The transverse densities show a very strong $b$--dependence beyond the
exponential fall--off $\sim \exp (-2 M_\pi b)$ required by the position
of the two--pion threshold. It reflects the non--trivial structure of 
the $\pi N$ scattering amplitude near threshold, particularly the 
subthreshold nucleon singularity, which brings in the small 
parameter $\epsilon = M_\pi/M_N$.
\item \textit{Charge vs.\ magnetization density.} In the chiral region
$b = O(M_\pi^{-1})$ the spin--independent and --dependent components
of the 4--vector current density, $\rho_1^V(b)$ and $\widetilde\rho_2^V(b)$, 
are of the same order in the chiral expansion. A mechanical explanation can be 
provided in the rest frame, where a peripheral pion in the 
nucleon's light--cone wave function at distances $O(M_\pi^{-1})$ 
has velocity $v = O(1)$ and generates charge and current 
densities of the same order. In the region of molecular distances 
$b = O(M_N^2/M_\pi^3)$ the pion velocity is $v = O(M_\pi / M_N)$, 
and the current density is suppressed compared to the charge density.
\item \textit{Role of $\Delta$ excitation.} The inclusion of intermediate 
$\Delta$ isobars in the $\pi N$ amplitude diminishes the peripheral
charge density but enhances the magnetization density. 
The pattern is explained 
by the large--$N_c$ limit of QCD, which requires that the $N$ and $\Delta$ 
contributions to the charge density cancel in leading order of $N_c^{-1}$, 
while in the magnetization density they add to give $3/2$ times 
the $N$ result.
\item \textit{Spatial region of chiral component.} The chiral component
of the charge and magnetization densities becomes numerically dominant only 
at very large transverse distances $b \gtrsim 2 \, M_\pi^{-1}$. 
At smaller distances the densities are generated mostly by the $\rho$ mass 
region of the spectral functions. The spatial representation of nucleon
structure afforded by the transverse densities gives a precise meaning 
to the notion of the ``pion cloud.''
\item \textit{Spatial support of chiral divergences.} The integrals of
the chiral long--distance component of the densities reproduce the 
well--known chiral divergences of the isovector charge radius and the 
magnetic moment in the limit $M_\pi \rightarrow 0$. The chiral logarithm
of the charge radius results from the integral over a broad range of 
distances $b_0 \ll b \ll 1/M_\pi$, while the power--like divergence
of the magnetic moment results from $b \sim M_\pi^{-1}$.
\end{itemize}
The calculations reported here could easily be extended to study other 
elements of the nucleon's transverse structure. One obvious extension
are the nucleon's transverse axial and pseudoscalar charge densities,
which are the Fourier transforms of the form factors of the axial 
current operator (or, equivalently, the $x$--integral of the 
axial vector--type GPDs $\tilde H$ and $\tilde E$  \cite{Goeke:2001tz}). 
Of particular
interest is that the spectral function of the pseudoscalar form factor 
contains a pion pole at $t = M_\pi^2$, so that the corresponding density
should represent the longest--range transverse structure accessible
through matrix elements of local current operators. Another extension
are the transverse densities corresponding to the form factors of the
nucleon's energy--momentum tensor \cite{Jaffe:1989jz,Ji:1995sv,Ji:1996ek}, 
which describe the spatial distributions
of matter, momentum, and stress (or forces) in the nucleon. The study of
the peripheral chiral component of these densities would be of fundamental 
interest and possibly offer a new perspective on the partonic interpretation
of orbital angular momentum in the nucleon's periphery. Last, the methods
described here could be extended to study the spatial structure of 
higher $x$--moments of the nucleon GPDs in the chiral region and perhaps
provide new insight into their chiral extrapolation properties.
\subsection{Methodological aspects}
\textit{Usefulness of spatial representation.}
The spatial representation of nucleon structure in the light--front 
formulation offers a natural framework for identifying and calculating the 
chiral component of nucleon structure. The transverse distance $b$ acts as
a natural parameter for the chiral expansion, and the expansion of the
peripheral densities at $b = O(M_\pi^{-1})$ provides much more theoretical 
control than that of the total charges. The formulation also allows one to 
combine chiral and non--chiral contributions in a consistent fashion. 
Even the inclusion of $\Delta$ intermediate states with the additional
scale $M_\Delta - M_N$ and the implementation of the large--$N_c$ limit of 
QCD can be accomplished easily when focusing on the peripheral 
$b$--dependent densities. When developed further, the transverse spatial 
representation could become a valuable tool for the interpretation
of chiral EFT calculations of nucleon structure. Its role can be
compared to that of the coordinate--space potential in summarizing
the properties of the low--energy $NN$ interaction in chiral 
EFT \cite{Epelbaum:2005pn}.

\textit{Invariant vs.\ time--ordered formulation of chiral EFT.}
In the present work we have used the dispersion representation of the
transverse densities to study their behavior in the chiral periphery
$b = O(M_\pi^{-1})$. This has allowed us to employ chiral EFT in its 
Lorentz--invariant relativistic formulation to describe the spectral
functions of the form factors --- an efficient and safe approach, 
particularly when higher--spin particles such as the $\Delta$ are involved.
In this way the ``transverse'' context is hidden in the structure of 
the kernel of the dispersion integral, while the calculations
are performed in invariant perturbation theory. Alternatively, one 
could study the chiral processes in peripheral transverse nucleon 
structure directly in time--ordered perturbation theory, where they
take the form of emission and absorption of soft pions by the nucleon
[momentum $k = O(M_\pi)$ in the nucleon rest frame]. 
This can be done using either the 
infinite--momentum frame, where one considers a nucleon moving with 
momentum $P \gg R^{-1}$ ($R$ represents the non--chiral nucleon size), 
or light--front time--ordered perturbation theory, where one studies
the time evolution of the chiral $\pi N$ system in $x^+ = x^0 + z$.
(In both formulations a careful limiting procedure is required to 
correctly account for the $\pi\pi NN$  contact terms representing 
the effect of high--mass intermediate states on nucleon structure.)
The time--ordered formulation obtained in this way offers many 
interesting new insights. The nucleon in chiral EFT is characterized
by a light--front wave function with $\pi N, \pi\pi N$ etc components, 
which is calculable from the chiral Lagrangian and provides a 
particle--based first--quantized description of chiral nucleon structure
(Fock expansion). It gives a precise meaning to the orbital angular
momentum of the chiral $\pi N$ configuration and allows one to make
contact with the non--relativistic Schr\"odinger wave function 
description. The calculation of the peripheral densities in the 
``time--ordered'' formulation, and the demonstration of its equivalence 
with the ``invariant'' formulation based on dispersion relations, 
will be the subject of a subsequent article \cite{inprep}. 
Several aspects of nucleon structure at zero momentum transfer 
(self--energies, pion momentum distribution,
electromagnetic couplings) in light--front chiral dynamics 
have been studied in Refs.~\cite{Ji:2009jc,Burkardt:2012hk,Ji:2013bca}. 
Other recent work has focused on expressing the consequences of dynamical 
chiral symmetry breaking in the light--front formulation of QCD 
at a more abstract level \cite{Beane:2013ksa}.

\textit{Longitudinal structure and resummation.}
In the study reported here we have focused on the transverse charge
and magnetization densities, which are integrals of the nucleon
GPDs over $x$. The chiral component was identified only through the 
transverse distance $b = O(M_\pi^{-1})$, and the leading--order 
chiral EFT result was used to evaluate the densities. Much more structure 
becomes available, of course, when one considers the chiral component of
the GPDs as a function of both longitudinal momentum and transverse distance. 
The longitudinal properties of the chiral expansion were studied 
extensively for the pion, where it was shown that the fixed--order
chiral expansion breaks down at pion light--cone momentum fractions 
of the order $y = O[M_\pi^2/(4 \pi F_\pi)^2]$, and an all--order resummation
was proposed for this regime \cite{Kivel:2007jj,Kivel:2008ry}. 
It was shown that the transverse radius of the pion grows with
decreasing $y$ as a result of chiral dynamics, and that this ``inflation''
is consistent with the well--known chiral divergence of the pion 
charge radius \cite{Perevalova:2011qi}. What these findings imply for the 
nucleon GPDs and transverse densities deserves further study. 
Generally, the leading--order chiral component of the nucleon corresponds 
to pion light--cone momentum fractions of the order 
$y = O(M_\pi / M_N)$ \cite{Strikman:2009bd,Strikman:2003gz}, and 
the integral of the pion distributions over this parametric domain 
reproduces the leading--order transverse densities described in the 
work here \cite{Strikman:2010pu} (for a discussion of the role of 
$y = 0$ modes see the quoted article). Interesting questions are how much 
a regime of exceptionally small ``chiral'' momentum fractions 
(i.e., parametrically smaller than the natural value $M_\pi / M_N$) 
would contribute to the $y$--integral of the transverse densities; 
how such a chiral contribution to the densities could be reconciled with 
known chiral behavior of the spectral functions near threshold, 
to which they are related by a dispersion relation; and how the 
subthreshold singularity and the ``molecular regime'' described 
in Sec.~\ref{subsec:parametric} would manifest themselves in 
that formulation. The $y$--dependence of the chiral component of
nucleon GPDs was recently studied heavy--baryon chiral 
EFT \cite{Moiseeva:2012zi}. A discussion of the relevance of chiral
resummation in the nucleon's partonic structure from a phenomenological 
perspective can be found in Ref.~\cite{Strikman:2009bd}.

\textit{Importance of analyticity.}
Analyticity plays a central role in the study of peripheral nucleon 
structure. The peripheral densities are the dispersive image of the spectral 
functions of the form factors near threshold and embody their full 
complexity. In fact, the $b$--representation represents the mathematically
cleanest way of displaying the analytic structure of the form 
factors \cite{Miller:2011du} and might well have been invented for that 
purpose had it not been known for its physical meaning. The relativistic 
formulation of chiral EFT produces amplitudes with the correct analytic 
properties and can safely be used to study peripheral nucleon structure. 
Heavy--baryon formulations can be employed to the extent that their 
results can be represented as approximations to the spectral functions of 
the form factors with an analytic structure based on exact 
kinematics \cite{Bernard:1996cc,Kaiser:2003qp}. 
While we have used chiral EFT to obtain
explicit approximations to the peripheral densities, many of the results
presented here could be obtained in a more general amplitude analysis
based on analyticity and dispersion relations. The properties of the 
chiral component of the densities studied in Sec.~\ref{sec:chiral}
could be deduced from the two--pion cut of the form factor using the 
general $\pi N$ scattering amplitude and its analytic properties.
Likewise, the $\rho$ meson contribution to the densities computed
in Sec.~\ref{sec:region} could be obtained from a dispersion analysis
of the isovector spectral function, using elastic unitarity below the
$4\pi$ threshold \cite{Ditsche:2012fv}. It would be interesting to extend 
this general amplitude analysis to other elements of peripheral nucleon 
structure, e.g.\ the nucleon's $x$--dependent parton densities 
at $b = O(M_\pi^{-1})$.
\subsection{Experimental tests}
To conclude our discussion we briefly want to comment on observable
effects and possible experimental tests of the structures described here. 
The aim of the present study has been to calculate the peripheral 
transverse charge and magnetization densities in chiral EFT and 
understand their mechanical properties. The chiral large--distance 
components of the densities described here represent model--independent 
elements of the nucleon's light--front (or partonic) structure.
They could be implemented as constraints (limiting cases) in empirical 
parametrizations of the nucleon's transverse densities. Using the
methods developed in Refs.~\cite{Strikman:2009bd,Strikman:2003gz},
this approach could easily be extended to the $x$--dependent peripheral 
transverse densities of quarks, antiquarks and gluons in the nucleon 
(GPDs).

\textit{Chiral component in empirical densities.}
An obvious question is whether the chiral component of the transverse 
densities could be ``seen'' in empirical $b$--dependent densities 
extracted from form nucleon factor data. A detailed study of this
complex problem remains beyond the scope of the present article, 
and we limit ourselves to a few comments here. The results of 
Sec.~\ref{sec:region} show that the chiral components dominate the
overall charge and magnetization densities only at very large distances 
$b \gtrsim 2\, M_\pi^{-1} \approx 3 \, \textrm{fm}$. To probe the 
chiral component directly one therefore has to extract the empirical 
densities at such large distances, where they are exponentially small. 
This is possible \textit{only} with form factor parametrizations that 
respect the exact analytic structure of the form factor in the
complex $t$--plane (principal cut starting at $4 M_\pi^2$, absence
of spurious singularities), as are provided by dispersion fits
\cite{Belushkin:2006qa}. Form factor parametrizations based on rational 
approximations \cite{Venkat:2010by,Vanderhaeghen:2010nd,Cates:2011pz}
generally produce singularities at unphysical complex values of $t$ 
and are \textit{principally not adequate} for extracting densities in 
the region of their leading exponential fall--off; cf.\ the
discussion in Ref.~\cite{Miller:2011du}. It follows that traditional
dispersion--type fits to form factors are the only mathematically 
reliable method to extract the peripheral densities and identify the
chiral component. Moreover, the dispersion--theoretical spectral functions 
of Refs.~\cite{Hohler:1976ax,Belushkin:2006qa} incorporate the 
full chiral structure of the form factor near the threshold 
(see Sec.~\ref{subsec:spectral}) as obtained also in chiral EFT and
expressed in the peripheral densities described here. There is thus 
no need to fundamentally change these parametrizations in order to 
``see'' the chiral component of the densities. Efforts should rather 
concentrate on studying the sensitivity of the form factor data to
small variations of the spectral functions near threshold, consistent
with their overall analytic structure; such variations will in turn
change the peripheral transverse densities and thus establish their
sensitivity to the form factor data. Based on the results of 
Ref.~\cite{Miller:2011du} we expect that present experimental uncertainties 
in the form factor data are much larger than the theoretical uncertainty
with which the spectral functions near threshold can be calculated 
using dispersion theory or chiral EFT.

\textit{Higher derivatives of form factors.}
The chiral components do reveal themselves clearly in the $b^{2n}$--weighted
moments of the transverse densities with $n \geq 2$, which govern
the $n$'th derivatives of the form factors at $t = 0$; 
cf.~Eq.~(\ref{moment_derivative}). The estimate of 
Ref.~\cite{Strikman:2010pu} shows that the $b^2$--weighted integral
of the chiral component of the transverse charge density [over the range
$b > b_0 = 2 \, M_\rho^{-1} \approx 0.4\, M_\pi^{-1}$] gives
$\sim 20\%$ of the experimental value $\langle b^2\rangle_{1, \text{exp}}$ 
extracted from the Dirac form factor slope. The $b^4$--weighted integral 
calculated with the chiral component is $\sim 1.5 \times 
\langle b^2 \rangle^2_{1, \text{exp}}$; i.e., the contribution of the 
chiral component alone is as large as the ``natural'' expectation for
this moment based on the empirical charge radius. This suggests that the 
chiral components should manifest themselves in an ``unnatural'' behavior 
of the second and higher derivatives of the isovector form factors 
$F^V_{1, 2}(t)$ at $t = 0$. Such behavior could be tested experimentally 
by comparing fits to low--$|t|$ spacelike form factor data with 
the slope (first derivative) obtained from the proton charge radius 
measured in atomic physics experiments. 
Again it is necessary to use dispersion--based parametrizations of the 
form factors with the correct analytic properties, as the collective 
behavior of higher derivatives at $t = 0$ is sensitive to the 
singularities of the form factor in the complex plane (this is 
equivalent to the statement that the peripheral densities
are sensitive to the singularities). A recent dispersion fit
\cite{Lorenz:2012tm}, which updates Ref.~\cite{Belushkin:2006qa}
and incorporates new form factor data, found that the charge form 
factor consistently extrapolates to the charge radius obtained in
atomic physics experiments. As already noted, the spectral functions 
used in this dispersion fit incorporate the full chiral structure of 
the form factor near the threshold, which generates the peripheral
transverse densities discussed here. With precise form factor 
data at $|t| \lesssim 10^{-2}\, \textrm{GeV}^2$ 
(see Ref.~\cite{Bernauer:2013tpr} for a recent update) one might be 
able to observe the predicted unnatural higher derivatives and in this way 
conclusively establish the presence of the chiral component
in the nucleon electromagnetic form factors. However, such 
measurements are extremely challenging, as the relevant observable
is not the cross section itself but its small deviation from the 
value at $t = 0$.

\textit{Peripheral high--energy processes.}
More direct experimental tests of the nucleon's chiral component are 
possible through measurements of peripheral processes in high--energy 
$eN, \pi N$ or $NN$ scattering. In scattering at moderately 
large center--of--mass energies $W \gtrsim 10 \, \textrm{GeV}$
and impact parameters $\sim \textrm{few} \, M_\pi^{-1}$ certain
final states are predominantly produced by scattering from a 
peripheral pion in the nucleon's light--cone wave function, while 
the nucleon--like system at the center remains a spectator.
The amplitude for such reactions can be expressed in terms of
the light--cone wave functions of the peripheral $\pi N$ 
system, which are calculable from the chiral Lagrangian \cite{inprep}, 
and the pertinent short--distance structure of the pion probed in the 
high--energy subprocess. The challenge lies in selecting final
states where the probability for scattering on a peripheral pion 
is maximal while other competing mechanisms are suppressed.
One possibility are exclusive processes in which a pion is observed 
in the final state with a moderately large transverse momentum 
$p_{T, \pi} \sim 1-2 \, \textrm{GeV}$, while the forward nucleon 
emerges with a small $p_{T, N} \lesssim 100\, \textrm{MeV}$.
In $eN$ scattering this could be realized with hard 
exclusive processes such as vector meson production 
$\gamma^\ast N \rightarrow V + \pi + N$ and deeply--virtual 
Compton scattering $\gamma^\ast N \rightarrow \gamma + \pi + N$,
in which the $\gamma^\ast\pi$ subprocess probes the GPDs of the 
peripheral pion \cite{Strikman:2003gz,Amrath:2008vx}.
Because the typical light--cone momentum fractions of peripheral pions
in the nucleon are $y \sim M_\pi / M_N \sim 0.1$, one needs to measure
at values of the Bjorken variable $x \ll 0.1$ to enable peripheral
scattering \cite{Strikman:2009bd,Strikman:2003gz}. 
Such processes could be measured at a future Electron--Ion 
Collider with appropriate forward 
detection capabilities \cite{Accardi:2011mz,Accardi:2012hwp}.
In $\pi N$ or $N N$ scattering one could select processes in which
the incoming hadron scatters with a large momentum transfer 
from a peripheral pion, which is then observed in the final state.
Generally, processes in which the participating peripheral pion is 
``knocked out'' and identified in the final state offer much better 
prospects for probing the chiral component than purely elastic scattering, 
where the only option is to reconstruct the transverse densities 
at very large $b$.
\section*{Acknowledgments}
Our study has benefited greatly from collaboration with M.~Strikman 
on other aspects of chiral dynamics in partonic 
structure \cite{Strikman:2010pu,Strikman:2009bd,Strikman:2003gz}.
We are indebted to S.~J.~Brodsky, A.~Calle Cordon, E.~Epelbaum, J.~Goity, 
F.~Gross, Peng Guo, B.~Kubis, U.--G.~Meissner, W.~Melnitchouk, 
G.~A.~Miller, and M.~V.~Polyakov for discussions and helpful suggestions.

Notice: Authored by Jefferson Science Associates, 
LLC under U.S.\ DOE Contract No.~DE-AC05-06OR23177. The U.S.\ Government 
retains a non--exclusive, paid--up, irrevocable, world--wide license to 
publish or reproduce this manuscript for U.S.\ Government purposes.
\appendix
\section{Cutting rule for $t$--channel discontinuity}
\label{app:cutting}
In the dispersion approach to the chiral contribution to the spectral 
functions of the nucleon form factors near threshold one needs to 
calculate the imaginary part of Feynman integrals resulting from 
processes of the type of Fig.~\ref{fig:diag} and
Fig.~\ref{fig:diag_delta},
describing two--particle exchange in the $t$--channel. This can be
done very efficiently using a modified version of the Cutkosky rules
(see Ref.~\cite{LLIV} for an introduction). 
Consider a Feynman integral of the form
\beq
I(t) \;\; \equiv \;\; -i \int\frac{d^4 k}{(2\pi)^4}
\frac{\Phi (k, \ldots)}{(k_2^2 - M_\pi^2 + i0)(k_1^2 - M_\pi^2 + i0)} ,
\label{I_cutting}
\eeq
where $k_{1,2} \equiv k \mp \Delta / 2, t \equiv \Delta^2$,
and the function $\Phi$ generally depends on the integration variable
$k$ as well as other external 4--vectors. The integral has a cut for 
$t > 4 M_\pi^2$, and we aim to evaluate the discontinuity
\beq
\Delta I(t) \;\; \equiv \;\; I(t + i0) - I(t - i0) 
\;\; = \;\; 2 i \, \textrm{Im} \, I(t + i0) .
\eeq
We assume that the function $\Phi$ has no singularities in the
region of $t$ considered here, so that the discontinuity is entirely 
due to the pion propagators in Eq.~(\ref{I_cutting}). To calculate it,
we go to the $t$--channel CM frame, Eq.~(\ref{Delta_cm}), where
\beq
\Delta^\mu \;\; = \;\; (\sqrt{t}, \bm{0}) ,
\eeq
and apply the Cutkosky rules in analogy to the calculation of
$s$--channel two--particle cuts. Replacing the propagators by
delta functions,
\be
\frac{1}{k_{1, 2}^2 - M_\pi^2 + i0}
&\rightarrow& -2\pi i \, \delta (k_{1, 2}^2 - M_\pi^2) ,
\ee
we obtain the constraints
\beq
k_{1, 2}^2 - M_\pi^2 \;\; = \;\; (k^0 \mp \sqrt{t}/2)^2 
- \bm{k}^2 - M_\pi^2 \;\; = \;\; 0,
\label{k_constraint}
\eeq
whose solution for $t > 4 M_\pi^2$ is [cf.\ Eq.~(\ref{k_cm})]
\be
k^0 &=& 0, \\
|\bm{k}| &=& {\textstyle\sqrt{t/4 - M_\pi^2}}
\;\; = \;\; k_{\rm cm} .
\ee
Including the Jacobian factors, the product of 
delta functions in the integral becomes
\be
\delta (k_1^2 - M_\pi^2) \, \delta (k_2^2 - M_\pi^2)
&=& \frac{\delta (k^0) \, \delta (|\bm{k}| - k_{\rm cm})}
{4\sqrt{t} k_{\rm cm}} .
\ee
We thus obtain the discontinuity and the imaginary part as
\be
\frac{1}{\pi} \, \textrm{Im} \, I(t + i0)
&=& \frac{\Delta I}{2\pi i}
\;\; = \;\;
\frac{k_{\rm cm}}{32 \pi^3 \sqrt{t}} \; \int d\Omega 
\; \Phi (k, \ldots)_{k^0 = 0, |\bm{k}| = k_{\rm cm}} ,
\label{cutkosky_omega}
\ee
where $\Omega$ denotes the solid angle of $\bm{k}$ in the $t$--channel
CM frame. The actual form of the integrand is determined by the 
external vectors on which the function $\Phi$ 
depends in the given case. The components of 
these external vectors also have to be analytically continued to the 
$t$--channel CM frame, in such a way as to preserve the values of the 
other invariants besides $t$ on which the integral depends, and may 
take imaginary values in this frame. In the case that $\Phi$ depends
only on a single external vector that is chosen to point in the
$3$--direction, the integrand in Eq.~(\ref{cutkosky_omega}) 
becomes independent of the azimuthal angle of $\bm{k}$,
and the integral reduces to
\be
\frac{1}{\pi} \, \textrm{Im} \, I(t + i0)
&=& \frac{k_{\rm cm}}{16 \pi^2 \sqrt{t}} \; \int_{-1}^1 d\cos\theta
\; \Phi (k, \ldots)_{k^0 = 0, |\bm{k}| = k_{\rm cm}} ,
\label{cutkosky_theta}
\ee
where $\theta$ is the polar angle of $\bm{k}$.
\section{Dispersion integral in heavy--baryon expansion}
\label{app:heavy}
With the heavy--baryon expansion of the spectral functions in the
chiral region, Eqs.~(\ref{F_1_heavy}) and (\ref{F_2_heavy}), the 
dispersion integrals for the transverse densities can be evaluated 
analytically. In this appendix we present the relevant formulas and results. 
In the region of distances $b = O(M_\pi^{-1})$ one has 
$t - 4 M_\pi^2 = O(M_\pi^2)$ and $k_{\rm cm} = O(M_\pi)$.
It is convenient to introduce dimensionless scaling variables for 
$\sqrt{t}$ and the CM momentum as
\be
u &\equiv& \sqrt{t}/(2 M_\pi), 
\label{u_def}
\\
\kappa &\equiv& k_{\rm cm}/M_\pi \;\; = \;\; \sqrt{u^2 - 1} ,
\ee
so that the threshold $t = 4 M_\pi^2$ corresponds to $u = 1$. The result of 
the heavy--baryon expansion of the spectral functions, 
Eqs.~(\ref{F_1_heavy}) and (\ref{F_2_heavy}), can then be stated as
\be
\frac{1}{\pi} \, \textrm{Im} \, F_1^V (u)
&=& \frac{g_A^2 M_\pi^2}{(4\pi F_\pi)^2 \, u}
\left[ f_0(u) - \frac{\pi \epsilon}{4} f_1(u)
+ \frac{\epsilon^2}{4} f_2(u)
- \frac{3\pi \epsilon^3}{4} f_3(u) + O(\epsilon^4) \right]
\nonumber
\\[1ex]
&+& 
\frac{(1 - g_A^2) M_\pi^2}{(4\pi F_\pi)^2 \, u} \; 
\frac{f_{\rm cont}(u)}{3} ,
\label{Im_F1_heavy_u}
\\[2ex]
\frac{1}{\pi} \, \textrm{Im} \, F_2^V (u)
&=& 
\frac{g_A^2 M_\pi^2}{(4\pi F_\pi)^2 \, u} \left[ 
\frac{\pi}{2 \epsilon} f_{-1} (u) - 2 f_0(u)
+ \frac{3\pi \epsilon}{8} f_1(u)
- \frac{\epsilon^2}{3} f_2(u)
+ \frac{15\pi \epsilon^3}{16} f_3(u) + O(\epsilon^4) \right] ,
\label{Im_F2_heavy_u}
\ee
where $\epsilon = M_\pi / M_N$, cf.\ Eq.~(\ref{epsilon}), and 
$f_{\rm cont}(u)$ and $f_n(u)$ denote rational functions of the 
dimensionless CM momentum,
\be
f_{\rm cont} &\equiv & \kappa^3 ,
\\ 
f_0 &\equiv & \kappa + 2 \kappa^3 ,
\\
f_2 &\equiv& \kappa^{-1} + 18 \kappa + 48 \kappa^3 + 32 \kappa^5 ;
\\[2ex] 
f_{-1} &\equiv& \kappa^2 ,
\label{f_m1}
\\
f_1 &\equiv& 1 + 6 \kappa^2 + 6 \kappa^4 ,
\label{f_1}
\\
f_3 &\equiv& 1 + 6\kappa^2 + 10 \kappa^4 + 5 \kappa^6 .
\label{f_3}
\ee
In terms of the dimensionless variable $u$ the dispersion integral 
for the density, Eq.~(\ref{rho_dispersion}), now becomes
\be
\rho_{1, 2} (b) &=& \frac{4 M_\pi^2}{\pi} \int\limits_1^\infty du \, u 
\; K_0(2 \beta u) \; \frac{\textrm{Im}\, F_{1, 2} (u)}{\pi} 
\hspace{2em} (\beta \equiv M_\pi b ) .
\label{rho_dispersion_u}
\ee
Substituting the expansion Eqs.~(\ref{Im_F1_heavy_u}) and 
(\ref{Im_F2_heavy_u}), the result can be expressed as
\be
\rho_1^V (b) &=& \frac{(1 - g_A^2) M_\pi^4}{(4\pi F_\pi)^2} \; 
\frac{4 \, R_{\rm cont}(\beta)}{3\pi}
\nonumber
\\[1ex]
&+& \frac{g_A^2 M_\pi^4}{(4\pi F_\pi)^2}
\left[ \frac{4}{\pi} R_0(\beta) - \epsilon R_1(\beta)
+ \frac{\epsilon^2}{\pi} R_2(\beta)
- 3 \epsilon^3 R_3(\beta) + O(\epsilon^4) \right] ,
\label{rho_1_heavy_beta}
\\[2ex]
\rho_2^V (b) &=& 
\frac{g_A^2 M_\pi^4}{(4\pi F_\pi)^2} \left[ 
\frac{2}{\epsilon} R_{-1} (\beta) - \frac{8}{\pi} R_0(\beta)
+ \frac{3 \epsilon}{2} R_1(\beta)
- \frac{4 \epsilon^2}{3\pi} R_2(\beta)
+ \frac{15 \epsilon^3}{4} R_3(\beta) + O(\epsilon^4) \right] ,
\label{rho_2_heavy_beta}
\ee
where $R_{\rm cont}(\beta)$ and $R_n(\beta)$ denote the basic integrals
\be
R_{\rm cont}(\beta ) &\equiv & \int\limits_1^\infty du 
\; K_0(2 \beta u) \; f_{\rm cont}(u) , 
\\
R_n(\beta ) &\equiv & \int\limits_1^\infty du \; K_0(2 \beta u) \;  f_n (u) .
\label{R_n}
\ee
The integrand in $R_{\rm cont}, R_0$ and $R_2$ involves odd powers of 
$\kappa$ and has a branch cut singularity 
at $u = 1$. These integrals can be reduced to standard integrals of the type
\beq
\int\limits_1^\infty du \; K_0(2 \beta u) \; (u^2 - 1)^{m/2}
\; = \; \int\limits_0^\infty dv \; K_0(2 \beta \cosh v) 
\; (\sinh v)^{m + 1}
\hspace{2em} (m = -1, 1, 3, \ldots ) ,
\eeq
which can be expressed in closed form in terms of products of modified 
Bessel functions. We obtain
\be
R_{\rm cont} &=& \frac{1}{16} \left\{ 
[K_2(\beta)]^2 - 4 [K_1(\beta)]^2 + 3 [K_0(\beta)]^2
\right\} ,
\label{R_cont_res}
\\
R_0 &=& 
\frac{1}{8} \left\{ [K_2(\beta)]^2 - 2 [K_1(\beta)]^2 + [K_0(\beta)]^2
\right\} ,
\label{R_0_res}
\\
R_2 &=& \frac{1}{2} [K_3(\beta)]^2 .
\label{R_2_res}
\ee
Asymptotic expansions of the integrals for large argument $\beta$ can 
be obtained by substituting the known asymptotic expansion of the modified 
Bessel functions,
\be
R_{\rm cont} &=& \frac{3\pi e^{-2 \beta}}{16 \beta^3} 
\left( 1 + \frac{1}{4 \beta} \right) ,
\label{R_cont_asymp}
\\
R_0 &=& 
\frac{\pi e^{-2 \beta}}{8 \beta^2} 
\left( 1 + \frac{11}{4 \beta} + \frac{33}{32 \beta^2} \right) ,
\label{R_0_asymp}
\\
R_2 &=&  \frac{\pi e^{-2 \beta}}{4 \beta} 
\left( 1 
+ \frac{35}{4 \, \beta}
+ \frac{1085}{32 \, \beta^2}
+ \frac{9135}{128 \, \beta^3} 
+ \frac{166635}{2048 \, \beta^4} 
+ \frac{336105}{8192 \, \beta^5} 
\right) .
\label{R_2_asymp}
\ee
The expressions here quote the asymptotic expansion to the order which, 
respectively, gives the best numerical approximation at $\beta = 1$.
In the region $\beta > 1$ the series Eq.~(\ref{R_cont_asymp}) 
describes $R_{\rm cont}$ with an accuracy of $< 7\%$; 
Eq.~(\ref{R_0_asymp}) describes $R_0$ with an accuracy of $< 3\%$, 
and Eq.~(\ref{R_2_asymp}) describes $R_2$ with an accuracy of 
$< 1\%$. As can be seen from the magnitude of the coefficients, the series 
differ widely in their convergence properties at fixed $\beta$, and care
is required when using them for numerical evaluation; it is necessary 
to include the inverse power terms exactly as quoted here to obtain 
an approximation with the stated accuracy.

In the integrals $R_{-1}, R_1$ and $R_3$ the integrand involves
even powers of the CM momentum, cf.~Eqs.~(\ref{f_m1})--(\ref{f_1}); 
it is therefore polynomial in $u$ and not singular at $u = 1$. 
These integrals cannot be expressed in closed form in terms of 
Bessel functions. However, excellent approximations can be obtained 
by substituting the modified Bessel function $K_0$ under the 
integral Eq.~(\ref{R_n}) by its asymptotic expansion for large argument $u$,
\beq
K_0 (2\beta u) \;\; = \;\; \frac{\sqrt{\pi} \, 
e^{-2\beta u}}{2 (\beta u)^{1/2}} 
\left( 1 - \frac{1}{16 \, \beta u} 
+ \ldots
\right) .
\label{K_u_expansion}
\eeq
The approximation is justified because we are interested in values
$\beta \gtrsim 1$ and the functions $f_n (u) \, (n = -1, 1, 3)$ 
emphasize large values $u \gg 1$ in the integral, where the expansion
converges very fast. With the substitution $u = w^2$ the integral 
then becomes an incomplete Gaussian integral, which can be expressed 
in terms of the error function. Keeping the first two terms of the 
expansion Eq.~(\ref{K_u_expansion}) we obtain
\be
R_{-1} &=& \frac{\pi \; \textrm{erfc}\left[ (2 \beta)^{1/2} \right]}
{\sqrt{2} \, \beta}
\left( -\frac{5}{8} + \frac{11}{128 \, \beta^2} \right)
\nonumber 
\\
&+& \frac{\sqrt{\pi} \; e^{-2\beta}}{\beta^{3/2}}
\left( \frac{5}{16} + \frac{11}{64\, \beta} \right) ,
\label{R_m1_app}
\\[2ex]
R_1 &=& \frac{\pi \; \textrm{erfc}\left[ (2 \beta)^{1/2} \right]}
{\sqrt{2} \, \beta}
\left( \frac{5}{8} - \frac{33}{64 \, \beta^2} + \frac{1215}{1024 \, \beta^4} 
\right) 
\nonumber 
\\
&+& \frac{\sqrt{\pi} \; e^{-2\beta}}{\beta^{3/2}}
\left( -\frac{1}{16} + \frac{3}{2\, \beta} + \frac{405}{128\, \beta^2} 
+ \frac{1215}{512\, \beta^3} \right) ,
\label{R_1_app}
\\[2ex]
R_3 &=& \frac{\pi \; \textrm{erfc}\left[ (2 \beta)^{1/2} \right]}
{\sqrt{2} \, \beta}
\left( \frac{11}{128 \beta^2} - \frac{2025}{2048 \, \beta^4} + 
\frac{203175}{32768 \, \beta^6} 
\right) ,
\nonumber 
\\
&+& \frac{\sqrt{\pi} \; e^{-2\beta}}{\beta^{3/2}}
\left( \frac{1}{4} + \frac{91}{64\, \beta} + \frac{315}{64\, \beta^2} 
+ \frac{45}{4\, \beta^3} 
+ \frac{67725}{4096\, \beta^4} 
+ \frac{203175}{16384\, \beta^5} 
\right) .
\label{R_3_app}
\ee
These expressions approximate the exact integrals with an accuracy far 
better than $1\%$ at all $\beta > 1$. When substituting the asymptotic
expansion of the error function complement,
\beq
\textrm{erfc}[ (2 \beta)^{1/2} ]
\;\; \sim \;\; \frac{e^{-2\beta}}{\sqrt{2 \pi \beta}}
\left( 1 - \frac{1}{4\beta} + \ldots \right) ,
\eeq
Eqs.~(\ref{R_m1_app})--(\ref{R_3_app}) reproduce the leading asymptotic
behavior of the integrals at large $\beta$,
\beq
R_{-1} \; \sim \; \frac{\sqrt{\pi} \; e^{-2\beta}}{4 \beta^{5/2}} ,
\hspace{2em}
R_{1, 3} \; \sim \; \frac{\sqrt{\pi} \; e^{-2\beta}}{4 \beta^{3/2}} 
\hspace{2em}
(\beta \rightarrow \infty ).
\eeq
Higher powers in the asymptotic expansion of $R_{-1}, R_1$ and $R_3$ at
large $\beta$ could be calculated by expanding the $K_0$ function in the
integral to higher order, cf.\ Eq.~(\ref{K_u_expansion}); however, 
the resulting series are poorly convergent for $\beta \sim 1$.
For numerical evaluation it is better to use the full expressions 
in terms of the error function, Eqs.~(\ref{R_m1_app})--(\ref{R_3_app}),
than the asymptotic series.

In sum, evaluating Eqs.~(\ref{rho_1_heavy_beta}) and (\ref{rho_2_heavy_beta})
with the integrals given in Eqs.~(\ref{R_cont_res})--(\ref{R_2_res})
and Eqs.~(\ref{R_m1_app})--(\ref{R_3_app}) one readily obtains the
numerical values of the transverse charge and magnetization densities 
in the heavy--baryon expansion at all distances of practical interest
$\beta = M_\pi b \gtrsim 1$. The accuracy of the heavy--baryon expansion
as an approximation to the full leading--order chiral component is
discussed in Sec.~\ref{subsec:heavy_baryon} (see Fig.~\ref{fig:heavy}). 
We note that the methods presented here can be applied also to integrals 
appearing in the heavy--baryon expansion of other transverse densities, 
such as the matter and momentum densities (form factors of the 
energy--momentum tensor) or the $x$--moments of generalized parton 
distributions (generalized form factors).
\end{document}